%% file: ms.tex
\documentclass[review,authoryear,5p,twocolumn]{elsarticle}

\usepackage{lineno}
\usepackage[bookmarks=false]{hyperref}

\usepackage{amsmath,amssymb,mathtools}
\usepackage{stmaryrd} 
\usepackage{physics} 

\let\originalleft\left
\let\originalright\right
\def\left#1{\mathopen{}\originalleft#1}
\def\right#1{\originalright#1\mathclose{}}

\newcommand{\vect}[1]{\mathbf{#1}}
\newcommand{\matr}[1]{\mathbf{#1}}
\newcommand{\tens}[1]{\boldsymbol{\mathcal{#1}}}
\newcommand{\kr}{\odot}

\newcommand{\outprod}{\circ}

\newcommand{\hadam}{\ast}

\newcommand{\pnorm}[3]{\norm{#3}^{#1}_{{}_{#2}}}
\newcommand{\transp}{^{\text{T}}}

\usepackage{xcolor}
\definecolor{entropyblue}{RGB}{0,176,240}
\definecolor{textgray}{rgb}{0.3,0.3,0.3}

\usepackage{epstopdf}

\usepackage{graphicx}
\graphicspath{{figures/}}
\usepackage{subcaption}

\newcommand{\subfigcap}[1]{\textbf{(#1)}}

\usepackage{pdflscape}

\usepackage{booktabs}
\usepackage{multirow} 
\usepackage{array} 

\newlength{\mycellwidth}

\usepackage[labelsep=quad,skip=3pt]{caption}
\usepackage{soul}

\journal{NeuroImage}

\usepackage{etoolbox}
\patchcmd{\emailauthor}{(#2)}{}{}{} 

\biboptions{authoryear}

\begin{document}

\begin{frontmatter}

\title{Augmenting interictal mapping with neurovascular coupling biomarkers by structured factorization of epileptic EEG and fMRI data}

\author[kul]{Simon Van Eyndhoven}\corref{mainauthor}
\ead{simon.vaneyndhoven@kuleuven.be, simon.vaneyndhoven@gmail.com}

\author[labcogn,lbi]{Patrick Dupont}

\author[kempenhaeghe]{Simon Tousseyn}

\author[kul]{Nico Vervliet}

\author[ler,deptneur]{Wim Van Paesschen}

\author[kul]{Sabine Van Huffel}
\author[tudelft]{Borb\'{a}la Hunyadi}

\address[kul]{KU Leuven, Department of Electrical Engineering (ESAT), STADIUS Center for Dynamical Systems, Signal Processing and Data Analytics}
\address[labcogn]{Laboratory for Cognitive Neurology, Department of Neurosciences, KU Leuven, Leuven, Belgium}
\address[lbi]{Leuven Brain Institute, Leuven, Belgium}

\address[kempenhaeghe]{Academic Center for Epileptology, Kempenhaeghe and Maastricht UMC+, Heeze, The Netherlands}

\address[ler]{Laboratory for Epilepsy Research, KU Leuven, Leuven, Belgium}
\address[deptneur]{Department of Neurology, University Hospitals Leuven, Leuven, Belgium}

\address[tudelft]{Circuits and Systems Group (CAS), Department of Microelectronics, Delft University of Technology, Delft, the Netherlands}

\cortext[mainauthor]{Corresponding author}

\begin{abstract}
EEG-correlated fMRI analysis is widely used to detect regional blood oxygen level dependent fluctuations that are significantly synchronized to interictal epileptic discharges, which can provide evidence for localizing the ictal onset zone. 
However, such an asymmetrical, mass-univariate approach cannot capture the inherent, higher order structure in the EEG data, nor multivariate relations in the fMRI data, and it is nontrivial to accurately handle varying neurovascular coupling over patients and brain regions.
We aim to overcome these drawbacks in a data-driven manner by means of a novel structured matrix-tensor factorization: the single-subject EEG data (represented as a third-order spectrogram tensor) and fMRI data (represented as a spatiotemporal BOLD signal matrix) are jointly decomposed into a superposition of several sources, characterized by space-time-frequency profiles. 
In the shared temporal mode, Toeplitz-structured factors account for a spatially specific, neurovascular `bridge' between the EEG and fMRI temporal fluctuations, capturing the hemodynamic response's variability over brain regions. 
We show that the extracted source signatures provide a sensitive localization of the ictal onset zone, and, moreover, that complementary localizing information can be derived from the spatial variation of the hemodynamic response. 
Hence, this multivariate, multimodal factorization provides two useful sets of EEG-fMRI biomarkers, which can inform the presurgical evaluation of epilepsy.
We make all code required to perform the computations available.

\end{abstract}

\begin{keyword}
EEG-fMRI\sep blind source separation\sep tensor factorization \sep interictal epileptic discharge \sep neurovascular coupling \sep hemodynamic response function
\end{keyword}

\end{frontmatter}


\input{sections/introduction}

\input{sections/methods}
\input{sections/results}
\input{sections/discussion}
\input{sections/acknowledgment}

\input{sections/declaration}

\bibliography{ms}

\input{sections/appendix-optimization}

\input{sections/appendix-standardization}

\input{sections/appendix-entropy}

\end{document}


\begin{frontmatter}
		
		\title{Augmenting interictal mapping with neurovascular coupling biomarkers by structured factorization of epileptic EEG and fMRI data}
		
		
		
		
		\author[kul]{Simon Van Eyndhoven}\corref{mainauthor}
		\ead{simon.vaneyndhoven@kuleuven.be, simon.vaneyndhoven@gmail.com}
		
		\author[labcogn,lbi]{Patrick Dupont}
		
		\author[kempenhaeghe]{Simon Tousseyn}
		
		\author[kul]{Nico Vervliet}
		
		\author[ler,deptneur]{Wim Van Paesschen}
		
		\author[kul]{Sabine Van Huffel}
		\author[tudelft]{Borb\'{a}la Hunyadi}
		
		\address[kul]{KU Leuven, Department of Electrical Engineering (ESAT), STADIUS Center for Dynamical Systems, Signal Processing and Data Analytics}
		\address[labcogn]{Laboratory for Cognitive Neurology, Department of Neurosciences, KU Leuven, Leuven, Belgium}
		\address[lbi]{Leuven Brain Institute, Leuven, Belgium}
		
		\address[kempenhaeghe]{Academic Center for Epileptology, Kempenhaeghe and Maastricht UMC+, Heeze, The Netherlands}
		
		\address[ler]{Laboratory for Epilepsy Research, KU Leuven, Leuven, Belgium}
		\address[deptneur]{Department of Neurology, University Hospitals Leuven, Leuven, Belgium}
		
		\address[tudelft]{Circuits and Systems Group (CAS), Department of Microelectronics, Delft University of Technology, Delft, the Netherlands}

		\cortext[mainauthor]{Corresponding author}

	\end{frontmatter}
	
	\input{sections/suppfigures}

%% file: sections/introduction.tex
\section{Introduction}\label{sec:intro}

Refractory epilepsy is a neurological disorder suffered by 30\% of approximately 50 million epilepsy patients worldwide \citep{who2019epilepsy2}, in which seizures cannot adequately be controlled by anti-epileptic medication.
In the preparation of treatment via resective surgery, interictal epileptic discharges (IEDs) can be localized in the brain with simultaneous EEG-fMRI, which provides a good surrogate for mapping the seizure onset zone \citep{lemieux2001event,thornton2010eeg,vanhoudt2013eeg,grouiller2011or,zijlmans2007eeg,an2013electroencephalography,khoo2017hemodynamic}. 
This mapping is often conducted via EEG-correlated fMRI analysis, wherein a reference temporal representation of the IEDs is used to interrogate all brain regions' blood oxygen level dependent (BOLD) signals for significant correlations; voxels for which a statistical threshold is exceeded can then be considered part of the epileptic brain network, along which epileptic seizures are generated and propagated \citep{gotman2008epileptic,lemieux2001event,zijlmans2007eeg,thornton2010eeg,salek2003studying}.

The workhorse for conducting EEG-correlated fMRI analyis has been \citep{salek2006hemodynamic}---and will likely continue to be \citep{poline2012general}---the general linear model (GLM) framework \citep{friston1994statistical}. 
Over the past years, it has become clear that using the GLM comes with several hurdles, related to the many modeling assumptions, that may reduce its sensitivity or specificity (increasing Type I errors) when violated \citep{poline2012general,monti2011statistical,lindquist2009modeling}. Remedies for several of these issues are not yet widely applied, or are not yet available.

First of all, the adoption of a relevant representation of IED occurrences to construct a regressor for the design matrix has proven vital to the sensitivity. This aspect has been investigated in \citep{rosa2010estimating,murta2015electrophysiological,abreu2018eeg,vaneyndhoven2019semi}. In previous work \citep{vaneyndhoven2019semi}, we addressed this issue by pre-enhancing the EEG signals using a spatiotemporal filter that is tuned to maximize the signal-to-noise ratio (SNR) of IEDs with respect to the background EEG. We have shown that taking the time-varying power of the filtered EEG leads to a robust regressor, which is more performant than many other types of regressors, including those based on stick functions \citep{lemieux2001event,salek2006hemodynamic}, ICA \citep{formaggio2011integrating,abreu2016objective} or EEG synchronization \citep{abreu2018eeg}.

Model mismatch may occur due to the unknown neurovascular coupling from electrophysiological phenomena measured on the EEG to hemodynamic variations captured by the BOLD signals. In many papers on EEG-correlated fMRI, a canonical hemodynamic response function (HRF) based on two gamma density functions is used to translate IED-related temporal dynamics to BOLD fluctuations \citep{friston1998event}. However, there is insurmountable evidence that the HRF is not fixed, but varies substantially over subjects \citep{aguirre1998variability}, over brain regions \citep{handwerker2004variation}, with age \citep{jacobs2008variability}, or even with stress level \citep{elbau2018brain}. For the diseased brain, this issue may be even greater: i.e., additional variation, e.g. in brain areas involved in the epileptic network, has been observed compared to healthy controls \citep{vanhoudt2013eeg,benar2002bold,jacobs2009hemodynamic,lemieux2008noncanonical,grouiller2010characterization}. Plenty of previous research has shown that failing to account for this variability may lead to substantial bias and increased variance of the estimated activation, which in turn inflates Type I and/or Type II error rates \citep{lindquist2009modeling,lindquist2007validity,calhoun2004fmri,monti2011statistical}. 

Several methods have been devised to deal with this variability. A widely used approach is to model the HRF as a linear combination of several basis functions. Some popular choices for these bases, which are also supported by open source toolboxes like SPM are the `informed basis set' \citep{friston1998event}, consisting of the HRF plus its derivative w.r.t. time and its derivative w.r.t. the dispersion parameter (leading to a Taylor-like extension which can capture slight changes in peak onset and width), and the finite impulse reponse (FIR) basis set, in which every basis function fits exactly one sample of the HRF in every voxel \citep{glover1999deconvolution,aguirre1998variability}. Other researchers have aimed to find a basis set by computing a low-dimensional subspace of a large set of `reasonable' HRFs \citep{woolrich2004constrained} or by fitting nonlinear functions to given fMRI data  \citep{lindquist2007validity,vaneyndhoven2017flexible}. Alternatively, multiple copies of a standard HRF, which differ only in their peak latencies, can be used \citep{bagshaw2004eeg}. 
Finally, approaches exist that aim to be immune to differences in neurovascular coupling, such as those based on mutual information (MI), which does not rely on any predefined model or even linearity of the HRF \citep{ostwald2011information,caballero2013mapping}. Perhaps surprisingly, the authors of \citep{caballero2013mapping} found that the results based on MI were often very similar to those based on the informed basis set, leading to the conclusion that the assumption of a linear time-invariant system, as described by the convolution with an appropriate HRF, is sufficiently accurate. 
Instead, it may be useful to not make abstraction of the variable neurovascular coupling, but rather consider it as an additional biomarker to localize epileptogenic zones \citep{vanhoudt2013eeg}. 
Indeed, in several studies HRFs that deviate from the canonical model were found in regions of the epileptic network \citep{benar2002bold,lemieux2008noncanonical,hawco2007bold,pittau2011changes,jacobs2009hemodynamic,moeller2008changes,vanhoudt2013eeg}. Several hypotheses have been postulated to explain this varability, including altered autoregulation due to higher metabolic demand following (inter)ictal events \citep{schwartz2007neurovascular}, vascular reorganization near the epileptogenic region \citep{rigau2007angiogenesis}, or the existence of pre-spike changes in neuro-electrical activity which are not visible on EEG and which culminate in the IED \citep{jacobs2009hemodynamic}. It is thus an opportunity to map not only regions with statistically significant BOLD changes in response to IEDs, but also the spatial modulation of the HRF itself, in order to discover regions where an affected HRF shape may provide additional evidence towards the epileptic onset.

The previous considerations indicate that it is difficult to meet all assumptions in the general linear model, which may compromise inference power \citep{lindquist2009modeling,handwerker2004variation,monti2011statistical}. Data-driven alternatives may relieve this burden, since they adapt to the complexity of the data more easily compared to model-based approaches, and are especially suited for exploratory analyses \citep{mantini2007electrophysiological,marecek2016can}. Blind Source Separation (BSS) techniques consider EEG and/or fMRI data to be a superposition of several `sources' of physiological activity and nonphysiological influences. Based on the observed data alone, BSS techniques are used to estimate both the sources and the mixing system, by means of a factorization of the data into two (or more) factor matrices, holding sources or mixing profiles along the columns. They naturally allow a symmetrical treatment of EEG and fMRI data, enabling true fusion of both modalities \citep{valdes2009model,lahat2015multimodal,calhoun2009review}, which is in contrast to EEG-correlated fMRI, where EEG-derived IEDs inform the fMRI analysis. Furthermore, BSS techniques naturally accommodate higher-order representations of the data in the form of tensors or multiway arrays, which can capture the rich structure in the data. Indeed, measurements of brain activity inherently vary along several modes (subjects, EEG channels, frequency, time, ...), which cannot be represented using matrix-based techniques like ICA without loss of structure or information \citep{sidiropoulos2017tensor,lahat2015multimodal,kolda2009tensor,acar2007multiway}. 
Tensor-based BSS techniques have been used to mine unimodal EEG data by decomposing third-order spectrograms \mbox{(channels $\times$ time points $\times$ wavelet scales)} into several `atoms' (also coined `components' or `sources'), each with a distinct spatial, temporal and spectral profile/signature \citep{miwakeichi2004decomposing,morup2006parallel,marecek2016can}, with successful application in seizure EEG analysis \citep{acar2007multiway,devos2007canonical}. While a tensor extension of ICA for group fMRI data (in the form of \mbox{subjects $\times$ time points $\times$ voxels}) exists \citep{beckmann2005tensorial}, matrix representations of fMRI remain dominant for single-subject analyses. 
Coupled BSS techniques can estimate components which are shared between both modalities, providing a characterization in both domains \citep{hunyadi2017tensor}. For example, in \citep{acar2017acmtf,acar2019unraveling,hunyadi2016fusion,chatzichristos2018fusion}, multi-subject EEG and fMRI data have been analyzed using coupled matrix-tensor factorization (CMTF), wherein the `subjects' factor is shared between the EEG trilinear tensor decomposition and the fMRI matrix decomposition. In \citep{hunyadi2016fusion}, the resulting factor signatures revealed onset and propagation zones of an interictal epileptic network that was common over patients, as well as the modulation of the default-mode network (DMN) activity. 
Also single-subject data can be decomposed into distinct components,
using a shared temporal factor for EEG and fMRI. This requires the use of a model of the neurovascular coupling, to ensure temporal alignment of EEG and BOLD dynamics. In \citep{martinez2004concurrent}, a fixed canonical HRF was used, followed by multiway partial least squares to extract components with spatial, temporal, and spectral signatures. In previous work, we proposed an extension to this technique, where a subject-specific HRF is co-estimated from the available data, along with the components \citep{vaneyndhoven2017flexible}.

In this paper, we extend this latter technique in order to account not only for subject-wise variation of the HRF, but also capture variations over brain regions. This results in a highly structured CMTF (sCMTF) of the interictal multimodal data, in which HRF basis functions and spatial weighting coefficients are estimated along with spatial, spectral and temporal signatures of components. By preprocessing the EEG using the data-driven filters from \citep{vaneyndhoven2019semi}, we aim to maximize the sensitivity in mapping the interictal discharges. We analyze whether the estimated spatial modulation of the HRF is a viable biomarker when localizing the ictal onset zone, besides the BOLD spatial signatures themselves.

%% file: sections/methods.tex
\section{Methods and materials}\label{sec:methodology}

\subsection{Patient group}\label{subsec:method_patients}
We use data of twelve patients with refractory focal epilepsy, whom we previously studied in \citep{tousseyn2014sensitivity,tousseyn2014reliable,tousseyn2015correspondence,hunyadi2015prospective}. These patients were selected based on the following criteria: 1) they were adults which underwent presurgical evaluation using EEG-fMRI, and for which there was concordance of all the available clinical evidence regarding the epileptic focus; 2) subtraction ictal single-photon emission tomography (SPECT) coregistered to MRI (SISCOM) images were available for all patients, as well as post-surgery MRI scans when patients were seizure-free (international league against epilepsy (ILAE) outcome classification 1–3 (1, completely seizure-free; 2, only auras; 3, one to three seizure days per year $\pm$ auras; 4, four seizure days per year to 50\% reduction of baseline seizure days $\pm$ auras; 5, $<$50\% reduction of baseline seizure days to 100\% increase of baseline seizure days $\pm$ auras; 6,more than 100\% increase of baseline seizure days $\pm$ auras)); 3) interictal spikes were recorded during the EEG-fMRI recording session. 
This study was carried out in accordance with the recommendations of the International Conference on Harmonization guidelines on Good Clinical Practice with written informed consent from all subjects. All subjects gave written informed consent in accordance with the Declaration of Helsinki, for their data to be used in this study, but not to be made publicly available. The protocol was approved by the Medical Ethics Committee of the University Hospitals KU Leuven.
For the complete data on the patients' etiology, and the number of observed IEDs, we refer to Table~\ref{tab:clindata}.

\begin{table*}
		\caption{Clinical patient data. Abbreviations: F = female, M = male, L = left, R = right, CNS = central nervous system, DNET = dysembryoplastic neuroepihelial tumor, FCD = focal cortical dysplasia, HS = hippocampal sclerosis, cx = cortex.}
	\scalebox{0.8}{
		{\def\arraystretch{0.9}
			\begin{tabular}{ccllp{25mm} c c}
				\toprule
				patient & 
				gender & 
				\begin{tabular}[c]{@{}c@{}}ictal \\ onset zone \end{tabular} &
				etiology &
				surgery & 
				\begin{tabular}[c]{@{}c@{}}ILAE \\ outcome \end{tabular} &
				\begin{tabular}[c]{@{}c@{}}follow-up time \\ after surgery (y) \end{tabular}\\
				\midrule
				p01 & F & L temporal & HS & \begin{tabular}[c]{@{}l@{}}temporal lobe \\ resection\end{tabular} & 3 & 5\\[10pt]
				p02 & F & L parietal & FCD & \begin{tabular}[c]{@{}l@{}}partial\\ lesionectomy\end{tabular} & 4 & 5\\[10pt]
				p03 & F & \begin{tabular}[c]{@{}l@{}}R parieto- \\occipito-temporal \end{tabular} & Sturge-Weber &  & &\\[10pt]
				p04 & M & R temporal & unknown &  & &\\[10pt]
				p05 & F & \begin{tabular}[c]{@{}l@{}}L anterior \\temporal\end{tabular} & HS & \begin{tabular}[c]{@{}l@{}}temporal lobe \\ resection\end{tabular} & 1 & 8\\[10pt]
				p06 & F & R frontal & FCD & \begin{tabular}[c]{@{}l@{}}partial\\ lesionectomy\end{tabular} & 5 & 2\\[10pt]
				p07 & F & \begin{tabular}[c]{@{}l@{}}L anterior \\temporal\end{tabular} & DNET & \begin{tabular}[c]{@{}l@{}}temporal lobe \\ resection\end{tabular} & 1 & 4\\[10pt]
				p08 & M & \begin{tabular}[c]{@{}l@{}}L temporo- \\parietal\end{tabular} & unknown & \begin{tabular}[c]{@{}l@{}}overlap \\eloquent cx \end{tabular} &  & \\[10pt]
				p09 & F & L occipital & FCD & \begin{tabular}[c]{@{}l@{}}overlap \\ eloquent cx \end{tabular} & &\\[10pt]
				p10 & F & R temporal & HS & refused &  &\\[10pt]
				p11 & M & \begin{tabular}[c]{@{}l@{}}L anterior \\ temporal\end{tabular} & HS & \begin{tabular}[c]{@{}l@{}}temporal lobe \\ resection\end{tabular} & 1 & 6\\[10pt]
				p12 & F & R temporal & CNS infection & refused & & \\
				\bottomrule
			\end{tabular}
			\label{tab:clindata}
		}
	}
\end{table*}

\subsection{Data acquisition and preprocessing}\label{subsec:method_acquisition}
Functional MRI data were acquired on one of two 3T MR scanners (Achieva TX with a 32-channel head coil and Intera Achieva with an eight-channel head coil, Philips Medical Systems, Best, The Netherlands) with an echo time (TE) of 33 ms, a repetition time (TR) of either 2.2 or 2.5 s, and a voxel size of 2.6 $\times$ 3 $\times$ 2.6 mm\textsuperscript{3}. 
EEG data were recorded using MR-compatible caps with 30 to 64 electrodes, sampled at 5 kHz. The EEG signals were band-pass filtered offline between 1-50 Hz, gradient artifacts were removed and pulse artifacts were subtracted. The signal of every channel is divided by its standard deviation. Two neurologists subsequently inspected and annotated the EEG signals for IEDs. 

The fMRI images were realigned, slice-time corrected and normalized to MNI space, resampled to a voxel size of 2 $\times$ 2 $\times$ 2 mm\textsuperscript{3}, and smoothed using a Gaussian kernel of 6 mm full width at half maximum (FWHM). These processing steps were carried out using SPM8 (Functional Imaging Laboratory, Wellcome Center for Human Neuroimaging, University College London, UK) \citep{friston1994statistical}. We refer the reader to \citep{tousseyn2014sensitivity} for a detailed description of these preprocessing steps.

We regress out covariates of no interest from the fMRI data. These include the six motion-correction parameters, and the average time series in the white matter and the lateral ventricles (cerebrospinal fluid). If necessary, also boxcar regressors are added at moments of substantial scan-to-scan head movement (larger than 1 mm based on the translation parameters). 
To reduce remaining nuisance effects, we regress out the first five principal components of the BOLD time series within the cerebrospinal fluid and white matter regions \citep{behzadi2007component}.

Subsequently, the BOLD time series are band-pass filtered between 0.008--0.20 Hz using the CONN toolbox \citep{whitfield2012conn}.

The dimensionality of the fMRI data is reduced by means of an anatomical parcellation of the brain. The initial $79\times95\times68$ images are segmented into regions-of-interest (ROIs) according to the Brainnetome atlas, which consists of 246 parcels in the grey matter\citep{fan2016human}. 
For every ROI, one BOLD time series is constructed as the average of the time series of all voxels within the ROI. 

Further customized EEG preprocessing steps are treated in Sections \ref{subsec:method_mwf} and \ref{subsec:method_tensorization}.

If multiple acquisition runs (within the same recording session) had been done, the EEG and fMRI data of the different runs are temporally concatenated.

\subsection{Multi-channel Wiener filtering for spatio-temporal EEG enhancement}\label{subsec:method_mwf}
In previous work \citep{vaneyndhoven2019semi}, we have shown that pre-enhancing the EEG signals using a data-driven, spatiotemporal filter that is tuned to maximize the signal-to-noise ratio (SNR) of IEDs with respect to the background EEG and artifacts, leads to a BOLD predictor that is more performant than many other predictors, including those based on simple stick functions \citep{lemieux2001event,salek2006hemodynamic}, ICA \citep{formaggio2011integrating,abreu2016objective} or EEG synchronization \citep{abreu2018eeg}.
This pre-enhancement strategy based on multi-channel Wiener filters (MWF) has error-correcting capabilities and produces an IED representation that improves the localization sensitivity of EEG-correlated fMRI \citep{vaneyndhoven2019semi}.

In brief, the MWF is estimated by first performing time-delay embedding of the multi-channel EEG signals $\mathbf{x}[t]$, leading to an extended multi-channel, multi-lag signal
\begin{equation}
\tilde{\mathbf{x}} = \begin{bmatrix}
\mathbf{x}[t-\tau] \\
\vdots \\
\mathbf{x}[t-1] \\
\mathbf{x}[t] \\
\mathbf{x}[t+1] \\
\vdots \\
\mathbf{x}[t+\tau]
\end{bmatrix}
\end{equation}
and subsequently computing the filter coefficients as
\begin{equation}\label{eq:wienerfilter}
\hat{\mathbf{W}} = \mathbf{R}_{xx}^{-1}(\mathbf{R}_{xx}-\mathbf{R}_{nn})\;,
\end{equation}
where $\mathbf{R}_{xx} = \mathrm{E}\left\{\tilde{\mathbf{x}}\tilde{\mathbf{x}}\transp\middle\vert H=1\right\}$ is the covariance matrix of the EEG observed during annotated IED segments \mbox{($H=1$)}, and $\mathbf{R}_{nn} = \mathrm{E}\left\{\tilde{\mathbf{x}}\tilde{\mathbf{x}}\transp\middle\vert H=0\right\}$ is the covariance matrix of the EEG outside of IED segments \mbox{($H=0$)}. For the full derivation, we refer the reader to \citep{somers2018generic,vaneyndhoven2019semi}. The EEG signals are then filtered as $\hat{\mathbf{W}}\transp\tilde{\mathbf{x}}$. Due to the extension with lagged copies of the signals, channel-specific finite impulse response filters are found. Hence, $\hat{\mathbf{W}}\transp\tilde{\mathbf{x}}$ is a set of spatiotemporally filtered output signals, in which IED-like waveforms are preserved while other waveforms, which are not specific to epilepsy, are supressed\footnote{Subsampling of the rows of $\hat{\mathbf{W}}\transp\tilde{\mathbf{x}}$ is needed to reverse the time-delay embedding, collapsing it into a multi-channel output signal at time $t$ only.}. 

We train MWFs for each patient individually, using the same parameter settings as in \citep{vaneyndhoven2019semi}: we embed the EEG signals using $\tau = 4$ positive and negative lags and compute the final filter using the generalized eigenvalue decomposition, which ensures the positive definiteness property of the subtracted covariance matrix in \eqref{eq:wienerfilter}\citep{somers2018generic}.

\subsection{Higher-order data representation}\label{subsec:method_tensorization}
To preserve the intrinsic multiway nature of the data, we represent the preprocessed EEG and fMRI as a tensor and matrix respectively, which are subsequently factorized jointly. 
This approach differs from the mass-univariate treatment in the traditional GLM, where each voxel is treated individually, and only `flattened' EEG time courses can be entered as regressors. 
Since epilepsy is manifested with considerable variability between patients, we handle the multimodal data of each patient separately.

\subsubsection{Spatio-temporal-spectral tensor representation of EEG}\label{subsec:methods_eegtens}
We adopt a tensorization strategy based on time-frequency transformation of the EEG data to third-order spectrograms (time points $\times$ frequencies $\times$ channels).  
After the pre-enhancement step described in Section \ref{subsec:method_mwf}, we create a spectrogram using the Thomson multitaper method \citep{thomson1982spectrum}, applied on nonoverlapping EEG segments with a length equal to one repetition time (TR) of the fMRI acquisition. 
The squared Fourier magnitudes are averaged into 1~Hz bins, from 1~Hz to 40~Hz. Hence, for every EEG channel, we obtain a spectrogram which is synchronized to the fMRI time series. 
The time points $\times$ frequencies $\times$ channels spectrogram, \mbox{$\tens{X}\in \mathbb{R}^{I_s \times I_g \times I_m}$} is further normalized to equalize the influence of each channel and each frequency: the mean of each mode-1 fiber (holding one time series of squared amplitudes) is subtracted, and afterwards scale normalization is iteratively carried out over the second and third mode to ensure that each channel and each frequency bin contributed the same amount of variance to the data \citep{bro1997parafac}.

\subsubsection{Spatio-temporal matrix representation of fMRI}\label{subsec:methods_fmrimat}
The average BOLD time series are stacked in a time points $\times$ ROIs matrix \mbox{$\matr{Y}\in \mathbb{R}^{I_s \times I_v}$}, where $I_v=246$ ROIs. 
We normalize each ROI's time series by subtracting its mean and dividing by its standard deviation.

\begin{figure*}
	\centering
	\includegraphics[width=1.0\linewidth]{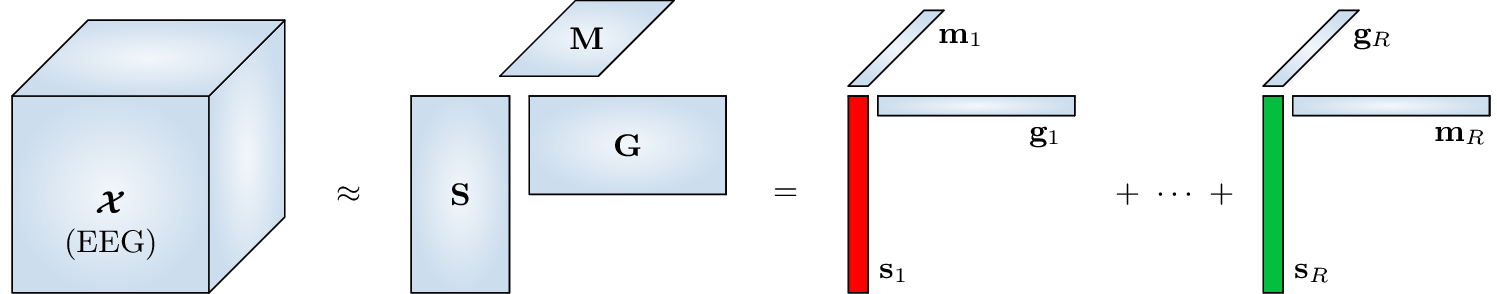} 
		\label{fig:Xdecomp}
	\includegraphics[width=1.0\linewidth]{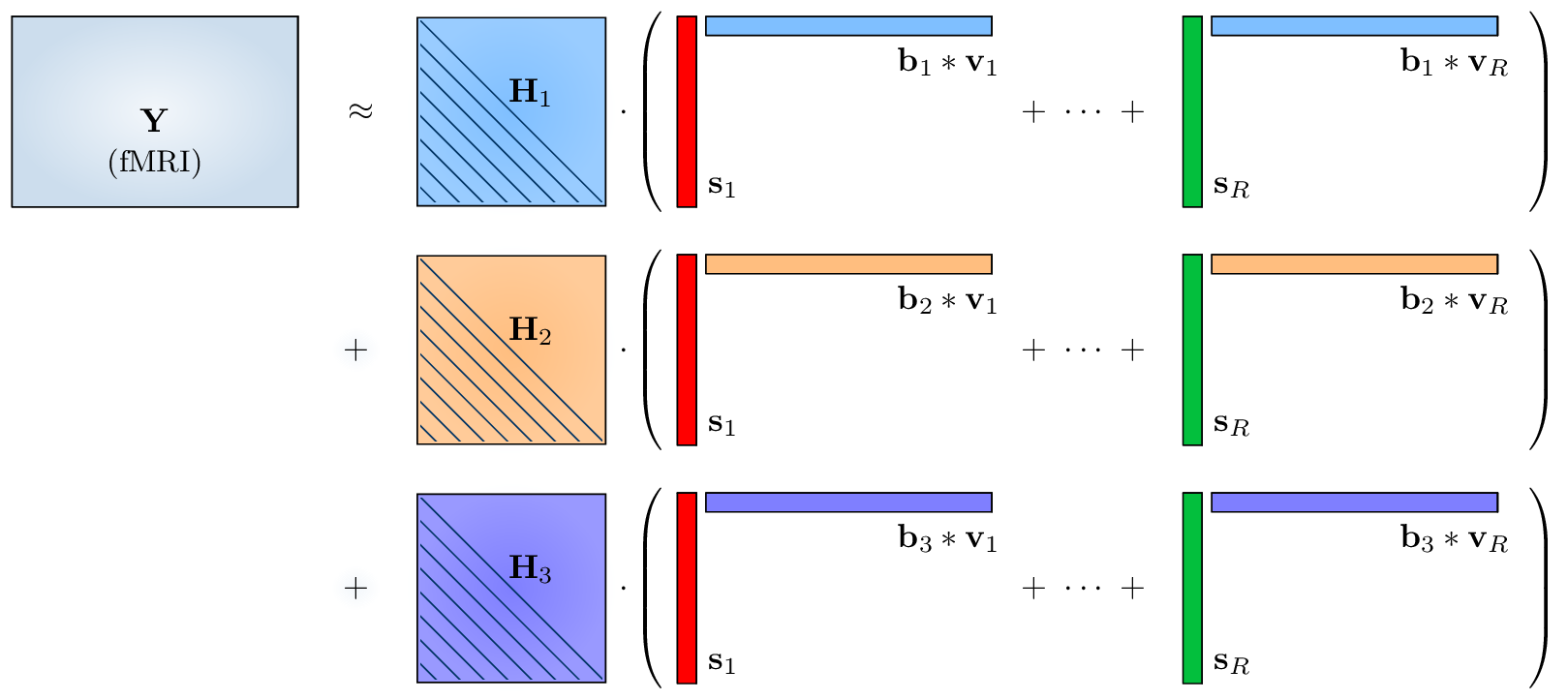} 
		\label{fig:Ydecomp}
	\caption{Structured coupled matrix-tensor factorization (sCMTF) of EEG and fMRI data can reveal neural sources that are encoded in both modalities, as well as capture the varying neurovascular coupling between the electrophysiological and BOLD changes. \subfigcap{(a)} The EEG signals vary over time points $\times$ frequencies $\times$ electrodes. The resulting third-order spectrogram tensor $\tens{X}$ is factorized according to \eqref{eq:cmtf_eegrank1} into $R$ rank-1 components, which each consist of a temporal signature $\vect{s}_r$, a spectral signature $\vect{g}_r$ and a spatial signature $\vect{m}_r$. \subfigcap{(b)} The fMRI data consist of the average BOLD signal in different brain parcels or regions of interest (ROIs), represented in a time points $\times$ ROI matrix $\matr{Y}$, and are factorized according to \eqref{eq:cmtf_fmrirank1}. Neurovascular coupling is modeled as a convolution of the EEG temporal dynamics with a ROI-specific hemodynamic response function (HRF), as in \eqref{eq:cmtf_fmrirank1}--\eqref{eq:cmtf_fmriflattened}. In this example, each local HRF is represented as a linear combination (encoded by coefficients $\vect{b}_k$) of $K=3$ optimized basis functions, each populating a Toeplitz matrix $\matr{H}_k$ which implements a convolution through matrix multiplication with the temporal signatures $\vect{s}_r$. Afterwards, each smoothed component $r$ is spatially weighted by a signature $\vect{v}_r$. This is accomplished by the elementwise product $\vect{b}_k \hadam \vect{v}_r$ of the HRF basis function-specific coefficients $\vect{b}_k$ and the component-specific amplitudes $\vect{v}_r$. }
	\label{fig:sCMTF}
\end{figure*}

\subsubsection{Neurovascular coupling in the temporal mode}\label{subsec:methods_coupling}
EEG and fMRI data are acquired simultaneously per subject, and are thus naturally coregistered along the `time' mode. 
This is captured in a temporal factor matrix that is common between the EEG factorization and the fMRI factorization.
However, the electrophysiological changes that are picked up by EEG vary on a much more rapid time scale than the sluggish BOLD fluctuations that (indirectly) correspond to the same neural process.
The neurovascular coupling that describes the relation between these two complementary signals can be described by a convolution with an HRF\footnote{In this paper, we use the term `neurovascular coupling' to describe the coupling characteristic between EEG and fMRI temporal dynamics, and make the silent assumption that this characteristic is a proxy/surrogate for `neurovascular coupling' as it is understood in neuroscience: the model that describes BOLD changes in response to electrical neural `events', which take the form of local field potentials at the synapses.}.

In previous work, we developed a CMTF model in which the HRF itself is parametrically estimated from the data \citep{vaneyndhoven2017flexible}, and a matrix multiplication with Toeplitz structure implements the HRF convolution, as proposed in \citep{valdes2009model}. 
In the same paper, we hinted towards an extension based on multiple basis functions to account for the variability of the HRF over brain regions.
In the following, we assume that the time course of each EEG source is convolved with an a priori unknown, ROI-specific HRF, which is a superposition of $K$ parametrized basis functions, which leads to a modelled contribution of this source to the ROI's BOLD signal.
Hence, in every ROI $i_v$, the modeled (unscaled) BOLD time course $\vect{z}^{(r)}_{i_v}$ of the $r$-th neural source is 
\begin{align}
\vect{z}^{(r)}_{i_v} 
&= \matr{H}_{i_v} \vect{s}_r \label{eq:convolution_ROItoep}\\
&= \sum_{k = 1}^{K} b_{k,i_v} \matr{H}_k \vect{s}_r \label{eq:convolution_sumofKtoep}\\
&= \sum_{k = 1}^{K} b_{k,i_v} \mathcal{T}\left(\vect{h}_k\right) \vect{s}_r \label{eq:convolution_sumofKoperator}\\
&= \sum_{k = 1}^{K} b_{k,i_v} \mathcal{T}\left(\mathcal{H}(\theta_k)\right) \vect{s}_r \label{eq:convolution_sumofKoperator2}
\;.
\end{align}
Here, $\vect{s}_r$ is a factor vector holding the time course of the $r$-th EEG source; $\mathcal{H}$ is an operator that transforms a set of parameters $\theta_k$ into a full HRF, represented as a vector $\vect{h}_k$; $\mathcal{T}$ is an operator that transforms $\vect{h}_k$ into a Toeplitz matrix $\matr{H}_k$ by populating the main and lower diagonals with the HRF samples (see also \ref{apx:opt_noncausal}); $b_{k,i_v}$ is the weight for the $k$-th HRF basis function in the $i_v$-th ROI; $\matr{H}_{i_v}$ is the Toeplitz matrix holding the total HRF in the $i_v$-th ROI\footnote{The HRF in every ROI does not depend on $r$, and is hence shared between all sources. In \citep{makni2008bayesian,vincent2010spatially,pedregosa2015data}, such a constraint has been used to promote robust estimation.}. 

This time course $\vect{z}^{(r)}_{i_v}$ is conceptually equivalent to a regressor in the GLM's design matrix. 
We treat the HRF parameter sets $\theta_k, k = 1\ldots K$ as unknown variables, which need to be fitted to the data at hand \citep{lindquist2007validity}.
By parametrizing each basis function, we embed protection against nonsensical HRF shapes, and against overfitting, since the number of parameters to be estimated is greatly reduced compared to the FIR basis in \citep{glover1999deconvolution,aguirre1998variability}.
We employ a double-gamma HRF, i.e., each HRF basis function is described by five parameters as $h(t) = f(t;\boldsymbol{\theta}) =$ \mbox{$\Gamma(\theta_1)^{-1}\cdot\theta_2^{\theta_1}t^{\theta_1-1}e^{-\theta_2t}-\theta_5\Gamma(\theta_3)^{-1}\cdot\theta_4^{\theta_3}t^{\theta_3-1}e^{-\theta_4t}$}.

\subsection{Coupled matrix-tensor factorization of EEG and fMRI}\label{subsec:methods_cmtf}
After tensorization, we jointly decompose the EEG tensor $\tens{X}$ and the fMRI matrix $\matr{Y}$ into a set of distinct sources.

The third-order EEG spectrogram is approximated by a sum of $R$ rank-1 terms according to the trilinear canonical polyadic decomposition (CPD) (also referred to as Parallel Factor Analysis (PARAFAC)) as in \citep{miwakeichi2004decomposing,marecek2016can,martinez2004concurrent,vaneyndhoven2017flexible}.
Each rank-1 term $\vect{s}_r\outprod\,\vect{g}_r\outprod\,\vect{m}_r$ describes a source (also called `component') in terms of an outer product ($\outprod$) of a 
temporal, spectral, and spatial signature, respectively. 
Unlike matrix decompositions, the decomposition of a higher-order tensor into a set of sources is unique, up to scaling and permutation ambiguities, without imposing constraints (under mild conditions).

The fMRI matrix is similarly approximated as a sum of rank-1 terms. Coupling arises from the temporal signatures $\vect{s}_r$, which are shared between the EEG and fMRI factorization. 
After processing through a hemodynamic system (as described in Section \ref{subsec:methods_coupling}), each source's BOLD temporal signature is weighted with a spatial signature $\vect{v_r}$. 

To accommodate additional structured variation in the fMRI data, that is not related to electrophysiological dynamics, we allow a low-rank term to the fMRI factorization which is not coupled with the EEG factorization.
We have empirically found that such a low-rank term can capture structured noise, preventing it from biasing the estimation of the parameters which are coupled with the EEG factorization.

The full sCMTF model is then described as:
\begin{align}\label{eq:cmtf}
\tens{X} 
&= \tens{\hat{X}} + \tens{E}_x \\
&= \sum_{r = 1}^{R} \vect{s}_r \outprod \,\vect{g}_r  \outprod \,\vect{m}_r + \tens{E}_x\label{eq:cmtf_eegrank1} \\
&= \left\llbracket\matr{S},\matr{G},\matr{M}\right\rrbracket + \tens{E}_x 
\label{eq:cmtf_eegcpdfactors}\\
\matr{Y}
&= \matr{\hat{Y}}  + \matr{E}_y \\
&= \sum_{r = 1}^{R} \sum_{k = 1}^{K} \left(\matr{H}_k \vect{s}_r\right)\outprod\left(\vect{b}_k \hadam \vect{v}_r\right) + \sum_{q = 1}^{Q}\vect{n}_q\outprod\vect{p}_q + \matr{E}_y\label{eq:cmtf_fmrirank1} \\
&= \sum_{k = 1}^{K} \left(\matr{H}_k\matr{S}\right)~\left(\vect{b}\transp_k \kr \matr{V}\transp\right) + \matr{N}\matr{P}\transp + \matr{E}_y \label{eq:cmtf_fmriblocks} \\
&= \begin{bmatrix}
\matr{H}_1\matr{S}~\ldots~\matr{H}_K\matr{S} 
\end{bmatrix}\cdot
\begin{bmatrix}
\matr{B}\transp \kr \matr{V}\transp
\end{bmatrix}  + \left\llbracket\matr{N},\matr{P}\right\rrbracket + \matr{E}_y
\;, \label{eq:cmtf_fmriflattened}
\end{align}
where $\tens{\hat{X}}$ and $\matr{\hat{Y}}$ are the low-rank approximations; $\tens{E}_x$ and $\matr{E}_y$ hold the residuals of both factorizations; $\left\llbracket\matr{S},\matr{G},\matr{M}\right\rrbracket$ describes the CPD model composed of factor matrices \mbox{$\matr{S}\in \mathbb{R}^{I_s \times R}$}, \mbox{$\matr{G}\in \mathbb{R}^{I_g \times R}$}, \mbox{$\matr{M}\in \mathbb{R}^{I_m \times R}$}, which hold the temporal, spectral and EEG spatial signatures in the columns; the HRF matrices $\matr{H}_k$ are constructed as in \eqref{eq:convolution_ROItoep}--\eqref{eq:convolution_sumofKoperator2}; \mbox{$\matr{V}\in \mathbb{R}^{I_v \times R}$} is the fMRI spatial factor matrix; \mbox{$\matr{B}\in \mathbb{R}^{I_v \times K}$} is the HRF basis coefficient matrix; $\left\llbracket\matr{N},\matr{P}\right\rrbracket$ is a \mbox{rank-$Q$} term to capture fMRI-only structured nuisance; $\hadam$ denotes the Hadamard or elementwise product; $\kr$ denotes the Khatri--Rao product \citep{kolda2009tensor}.

Note that the coupled part of $\matr{Y}$ is described by $RK$ nonindependent rank-1 terms, or equivalently, by $K$ rank-$R$ block terms.
Namely, each rank-1 term \mbox{$\left(\matr{H}_k \vect{s}_r\right)\outprod\left(\vect{b}_k\hadam \vect{v}_r\right)$} describes the convolution of the $r$-th source's temporal signature with the $k$-th basis function, after which a spatial loading with vector $\left(\vect{b}_k\hadam \vect{v}_r\right)$ is performed; in all ROIs, there is one source-nonspecific relative weight for the basis function (captured in $\vect{b}_k$), and source-specific amplitudes (captured in $\vect{v}_r$).
To limit the degrees of freedom, the Khatri--Rao product in \eqref{eq:cmtf_fmriblocks}--\eqref{eq:cmtf_fmriflattened} expresses that the HRF is shared among all sources,  which is a constraint that has earlier been used for this purpose \citep{pedregosa2015data}.
Hence, there are not $RKI_v$ spatial coefficients, but $(R+K)I_v$, i.e., $K$ basis function weights and $R$ source amplitudes in all $I_v$ ROIs\footnote{In this way, the Khatri--Rao structure also breaks the curse of dimensionality in the fMRI decomposition if either the number of sources $R$ or the number of basis functions $K$ is high (or both).}.
This model is depicted in Figure~\ref{fig:sCMTF}, omitting $\left\llbracket\matr{N},\matr{P}\right\rrbracket$ to not overload the diagram.

We estimate all parameters of the model in \eqref{eq:cmtf_eegrank1} and \eqref{eq:cmtf_fmrirank1} by iteratively minimizing the cost function $J$, composed of two data fitting terms and two regularization terms as in \citep{acar2014structure}:
\begin{equation}
\begin{split}\label{eq:costfunction}
J(\matr{S},\matr{G},\matr{M},\matr{B},\matr{V},\boldsymbol{\theta}) 
&= \; \beta_x\pnorm{2}{F}{\tens{X} - \tens{\hat{X}}} \\
&+ \; \beta_y\pnorm{2}{F}{\matr{Y} - \matr{\hat{Y}}} \\ 
&+\; \gamma_x\pnorm{}{1}{\boldsymbol{\lambda}_x} \; + \; \gamma_y\pnorm{}{1}{\boldsymbol{\lambda}_y}
\end{split}
\end{equation}
\begin{equation}
\begin{split}\label{eq:constraints}
\text{s.t.} \; 
&\matr{H}_k = \mathcal{T}\left(\vect{h}_k\right) = \mathcal{T}\left(\mathcal{H}(\boldsymbol{\theta}_k)\right)\\
&\boldsymbol{\lambda}_x = \begin{bmatrix}
\lambda_{x,1} & \ldots & \lambda_{x,R}
\end{bmatrix}\transp \\
&\lambda_{x,r} = \pnorm{}{2}{\vect{s}_r}\cdot\pnorm{}{2}{\vect{g}_r}\cdot\pnorm{}{2}{\vect{m}_r} \\
&\boldsymbol{\lambda}_y = \begin{bmatrix}
\lambda_{y,1} & \ldots & \lambda_{y,R}
\end{bmatrix}\transp \\
&\lambda_{y,r} = \sum_{k = 1}^{K} \pnorm{}{2}{\vect{b}_k \hadam \vect{v}_r}
\;,
\end{split}
\end{equation}
where the squared Frobenius norm $\pnorm{2}{F}{\tens{A}}$ of a tensor $\tens{A}$ is the sum of its squared elements; $\pnorm{}{2}{\vect{a}}$ and $\pnorm{}{1}{\vect{a}}$ denote the Euclidean or $\ell_2$-norm  and the $\ell_1$-norm or sum of the elements' absolute values of a vector $\vect{A}$, respectively; $\beta_x$, $\beta_y$, $\gamma_x$ and $\gamma_y$ are positive weights; $\boldsymbol{\lambda}_x$ and $\boldsymbol{\lambda}_y$ are vectors which hold the amplitudes with which each source is expressed in the EEG and fMRI data, respectively.
The squared Frobenius norms of the residuals promote a good fit of the low-rank approximations to the data, while the $\ell_1$-regularization terms penalize excessive source amplitudes and promote a parsimonious\footnote{The sparsity-promoting properties of the LASSO penalty are most useful in the context of an underdetermined system, with more coefficients than observations, e.g. in dictionary learning. Here, the problem is heavily overdetermined, and we do not expect that the amplitudes $\boldsymbol{\lambda}_x$ an $\boldsymbol{\lambda}_y$ go exactly to zero. However, the $\ell_1$-penalty is still a useful heuristic to avoid degenerate components in the EEG's CP decomposition.} model, similar to the group-LASSO method \citep{acar2014structure,yuan2006model}. 
At the same time, the latter penalty also tends to prevent the occurrence of degenerate terms \citep{bro1997parafac}.
We minimize \eqref{eq:costfunction} using the Structured Data Fusion framework in Tensorlab \citep{sorber2015structured,vervliet2016tensorlab}, using a quasi-Newton method based on a limited-memory BFGS algorithm, for 50 independent initializations (see \ref{apx:optimization} for details regarding the optimization procedure and parameters).
After convergence, each set of estimated factors needs to be calibrated to remove certain ambiguities, and model selection must be performed to pick the best solution, with an appropriate $R$ (see  \ref{apx:standardization} for details).

\subsection{Statistical inference}\label{subsec:method_inference} 
We create statistical nonparametric maps (SnPMs) of the obtained spatial signatures $\vect{v}_r$ to determine which ROIs sources are significantly (de)activated in relation to the found sources \citep{nichols2002nonparametric,waites2005reliable}.
To this end, we use permutation-based inference, in which the spatial signatures $\vect{v}_r$ are compared against their empirically derived distributions, which are obtained via resampling of the fMRI data while freezing the other estimated sCMTF factors\footnote{Under the null hypothesis of no significant BOLD effect related to the EEG dynamics, the fMRI data may be temporally reshuffled 
without a significant loss of fit to the EEG dynamics in $\vect{s}_r$.}.
To account for serial correlations in the fMRI time series, we use the robust wavelet-based resampling approach in \citep{bullmore2001colored} to ensure exchangeability and to preserve spatiotemporal correlation structure of the original data in the produced surrogate datasets. 
For each fMRI dataset and every sCMTF solution, we generate $L=250$ surrogate fMRI $\matr{\tilde{Y}}^{(l)}$ datasets using the procedure in \citep{bullmore2001colored}.
We resample only the adjusted data $\matr{Y}-\matr{N}\matr{P}\transp$, i.e., after removing the components which model variation specific to the fMRI data.
We perform inference on a pseudo t-statistic, which we compute for every ROI and for every source as follows:
\begin{enumerate}
	\item construct a local `design matrix' with all estimated temporal signatures as in~\eqref{eq:convolution_ROItoep}:~$\matr{D}_{i_v} = \begin{bmatrix}
	\vect{z}^{(1)}_{i_v}\ldots\vect{z}^{(R)}_{i_v}
	\end{bmatrix}\;,$
	\item find the new `betas' by solving $\boldsymbol{\beta}^{(l)}_{i_v} = \matr{D}^\dagger_{i_v}\vect{\tilde{y}}^{(l)}_{i_v}~,~\forall~l\;,$
	\item convert the betas to a t-statistic per source by dividing them by their estimated standard deviation (see \citep{friston1994statistical,poline2012general}).
\end{enumerate}
Through this procedure, we obtain $L$-point empirical null distributions for every source and every ROI.
We set the significance threshold as to control the familywise error (FWE) rate at $\alpha=0.05$, according to the maximum statistic procedure outlined in \citep{nichols2003controlling}.
That is, for every source $r$, we form the empirical distribution of the maximal t-statistic over all $I_v$ ROIs, and determine source-specific thresholds $T^{(r)}_{(1-\alpha)}$ as the 95\%-percentile (to test for activation) and $T^{(r)}_{(\alpha)}$ as the 5\%-percentile (to test for deactivation).
Finally, we obtain statistical maps for all sources $r$ by applying these thresholds to the original spatial signatures $\vect{v}_r$, which can be considered as the betas of the unshuffled data.

Furthermore, we create a map of the HRF variability over ROIs. 
For every ROI, we assess how `unusual' the local HRF is, by measuring its calibrated distance in HRF space to all other ROIs' HRFs.
We use two metrics to quantify this  (see \ref{apx:hrfentropy} for details on the computation).
\begin{enumerate}
	\item \textit{Extremity} is computed as one minus the average of the absolute values of the correlations between a HRF waveform and all other HRFs' waveforms.
	\item \textit{Entropy} of the HRF waveform is computed as the negative logarithm of the conditional probability of the HRF.
\end{enumerate}

Both for the pseudo t-maps as for the HRF extremity and entropy maps, we furthermore limit the inspection to the 20 ROIs with the highest values, if applicable.

\begin{figure*}
	\centering
	\includegraphics[width=0.85\linewidth,trim={0 0 0 2cm},clip]{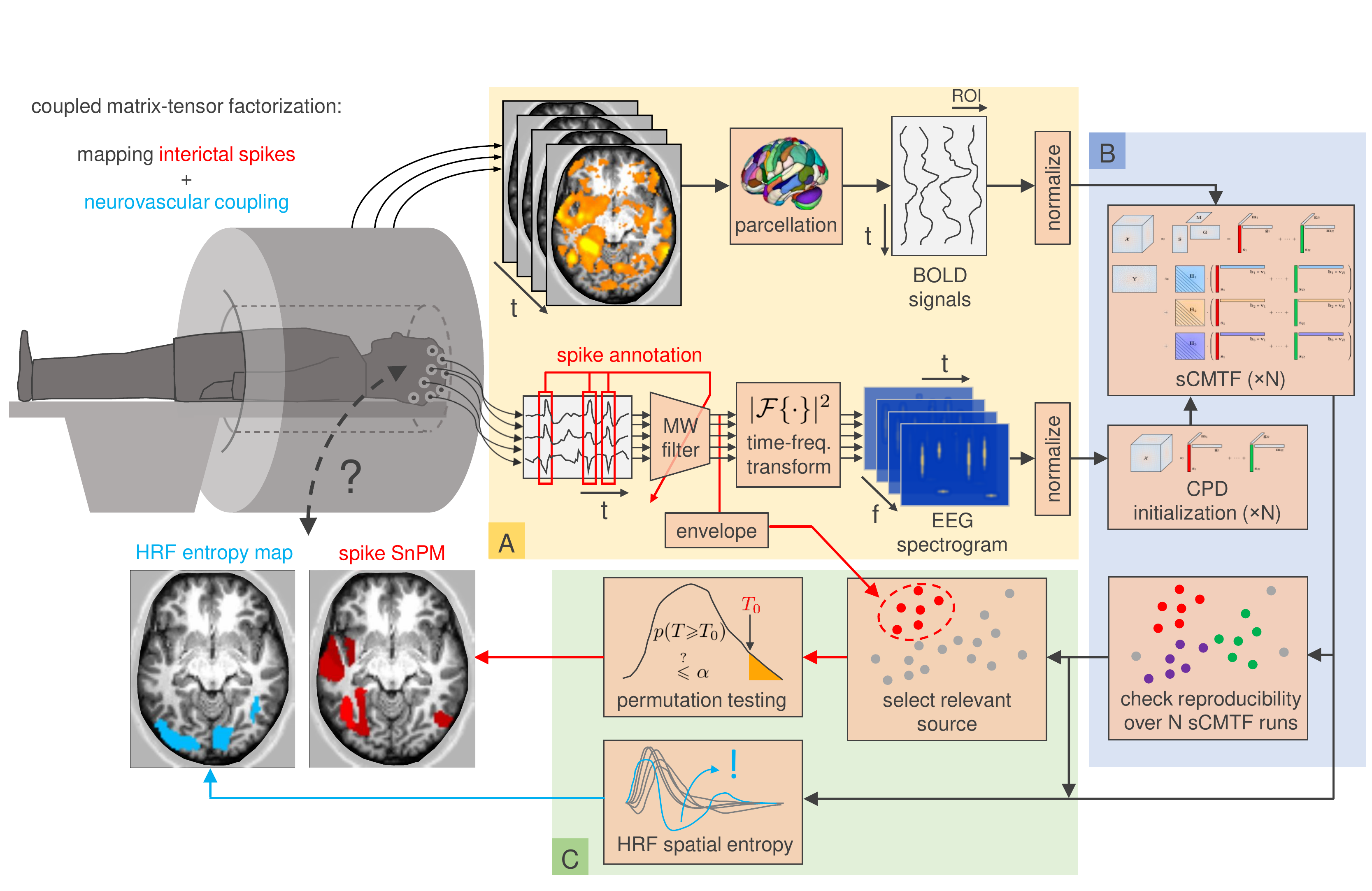} 
	\caption{Interictal EEG and fMRI data can be analyzed via structured coupled matrix-tensor factorizations (sCMTF), which reveals both \textcolor{red}{spatial localization of interictal discharges} (spikes), and also \textcolor{entropyblue}{localized deviations in neurovascular coupling} between electrical and BOLD fluctuations. \subfigcap{a} fMRI and EEG data are first separately preprocessed (yellow block). The fMRI data (top row) are structured as a time points $\times$ regions of interest (ROIs) matrix, after BOLD time courses are averaged within predefined or data-driven parcels. The EEG data (bottom row) are structured as a channels $\times$ time points $\times$ frequencies tensor, after the signals are enhanced via a multi-channel Wiener filter (MWF) which is calibrated based on spike annotations, and subsequently undergo a time-frequency transform. \subfigcap{b} The sCTMF of the EEG and fMRI data (blue block) reveals temporally, spatially and spectrally resolved components, and captures spatially varying hemodynamic response functions (HRFs) (cfr. Figure~\ref{fig:sCMTF}). We show the EEG temporal, spatial and spectral signatures in Figures~\ref{fig:p05eeg} and \ref{fig:p12eeg}, and the HRFs in Figures~\ref{fig:p05hrf} and \ref{fig:p12hrf}, for two selected patients. To initialize the sCMTF factors, first a canonical polyadic decomposition (CPD) of the EEG tensor is computed, from which the remaining fMRI factors are initialized. 
	The full sCMTF model is then computed $N$ times, from these $N$ different initializations, and the stability of the resulting factors over runs is assessed. \subfigcap{c} Statistical images are created for the patient's data and the corresponding sCMTF factors (green block). From the sCMTF factors, the spike-related component is picked as the one with the highest temporal correlation to the filtered EEG signals's broadband power envelope. A statistical nonparametric map (SnPM) of this interictal spike-related component is created, revealing co-activated ROIs in a pseudo-t-map (red). For every ROI, the entropy (and also the extremity) of the HRF is computed by assessing its likelihood under the distribution of all other ROIs' HRFs, and a map of this metric is constructed (blue) to reveal localized HRF abnormalities. Both maps can be used to form a hypothesis on the location of the epileptogenic zone, as we show in in Figures~\ref{fig:p05fmri} and \ref{fig:p12fmri} for the two selected patients. In this paper, we validated our technique on a set of patients for which the outcome is known.}
	\label{fig:blockdiagram}
\end{figure*}

An end-to-end overview of our pipeline, from data preprocessing up until statistical inference, is depicted in Figure~\ref{fig:blockdiagram}.

\subsection{Model performance}\label{subsec:method_performance}
We use several metrics to quantify the quality of the obtained sCMTF solutions. 

We compare the statistical maps with a ground truth delineation of the ictal onset zone (IOZ) to assess their concordance.
This ground truth is the manually delineated resection zone for patients that had undergone surgical treatment and that were seizure-free afterwards \citep{vanhoudt2013eeg,grouiller2011or,zijlmans2007eeg,an2013electroencephalography,thornton2010eeg}, or otherwise the hypothetical resection zone, based on concordant evidence from multiple modalities other than EEG-fMRI (cfr. Section \ref{subsec:method_patients}), for patients that were ineligible for or refused surgery \citep{tousseyn2014sensitivity}.
The sensitivity for detecting the IOZ is then computed as the fraction of `true positive' cases, which are determined by the presence or absence of significant activation clusters which overlap the IOZ in the spatial signatures $\vect{v}_r$. Following the reasoning in \citep{tousseyn2014sensitivity}, we do not consider significantly active voxels or regions outside of the delineated IOZ as false positives. Acknowledging epilepsy as a network disorder, such active regions might reflect seizure or IED propagation, despite not being involved in their generation.

Furthermore, we hypothesize that the spatial variation of the HRF over the brain might reveal additional localizing information regarding the IOZ, i.e., based on considerations explained in Section~\ref{sec:intro}, we assume that the HRF in or near the IOZ might be distorted compared to nonepileptic brain regions. 
We test this hypothesis by assessing whether those regions correspond to high values in the HRF entropy and HRF extremity  maps (see \ref{subsec:method_inference}).

Additionally, we inspect the spectral, spatial and temporal EEG signatures of the extracted sources, and we measure whether the spatial fMRI signatures bear any similarity to known networks of resting-state human brain activity \citep{shirer2012decoding}.

%% file: sections/results.tex
\section{Experiments}\label{sec:results}

\subsection{Patient-specific model selection}
Table~\ref{tab:Rselection} compiles the results of the steps described in \ref{apx:standardization}.
For each patient, we select the set of sCMTF factors of rank $\hat{R}$, which best fulfill the criteria.
In all cases, we found at least one such a solution, including an IED-related component within that solution.
Note that sometimes models with different $R$ might score well on different (subsets of) the criteria, so the selection of the rank is inevitably ambiguous.
In the next section, we analyze the individual set of results for each patient, based on the selected rank, and we analyze the sensitivity of the results to the choice of $R$.

\begin{figure}
	\centering
	\includegraphics[width=1\linewidth]{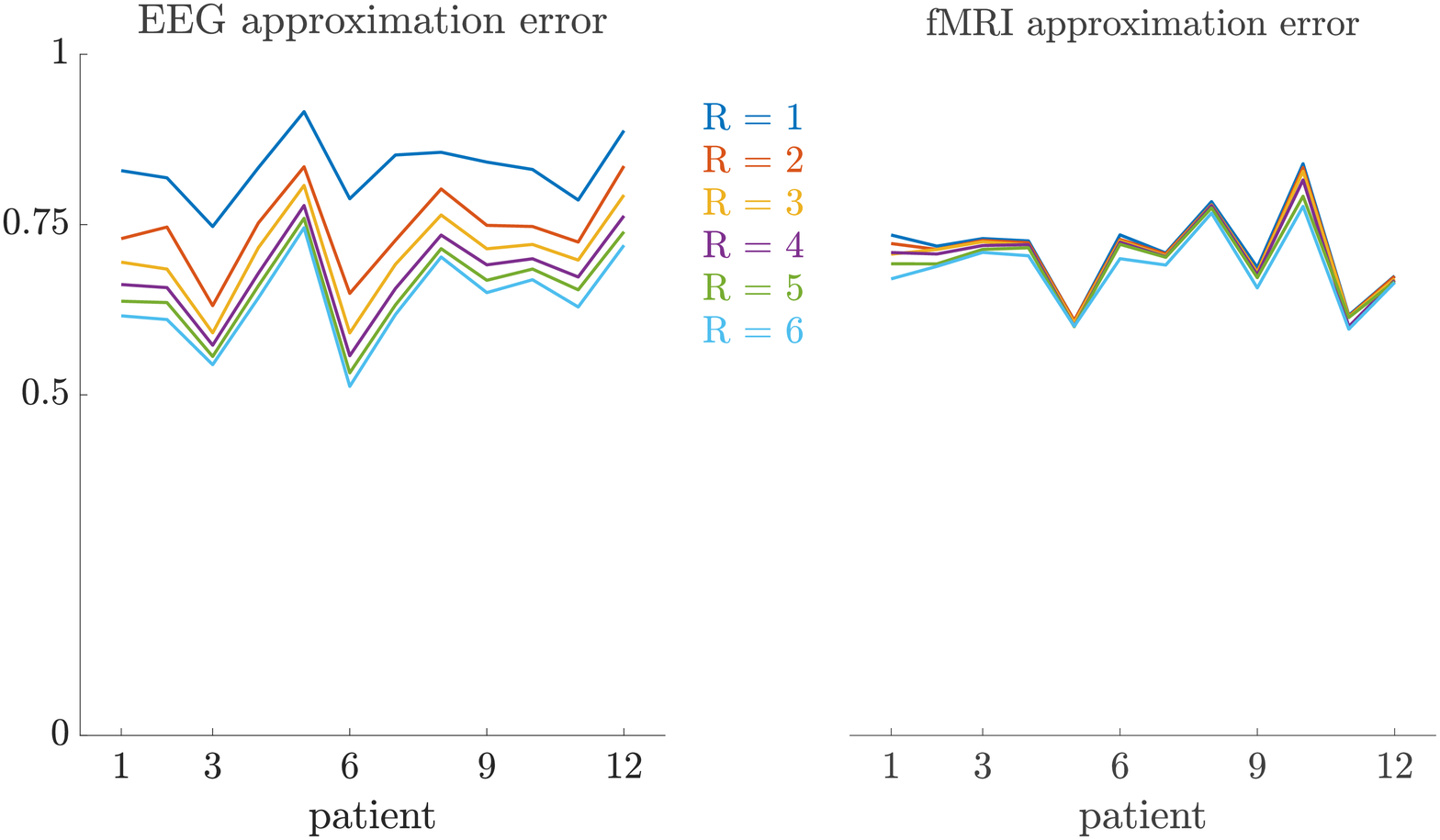}
	\caption{Goodness of fit of each patient's EEG tensor~$\tens{X}$ and fMRI matrix $\matr{Y}$, for varying choices of the rank $R$ in the sCMTF. Naturally, the EEG approximation error decreases monotonically for increasing rank (intra-patient). For the fMRI data, the fit already plateaus for very low $R$. This is due to the presence of additional, uncoupled components $\vect{n}_q\outprod\vect{p}_q$ in the fMRI factorization, which can absorb some of the variance when the number of coupled components is low, but which lose their relevance at higher ranks.}
	\label{fig:approxerror}
\end{figure}

We show the goodness of fit of the estimated factors for the EEG tensor $\tens{X}$ and the fMRI matrix $\matr{Y}$ in Figure~\ref{fig:approxerror}. 
Due to the normalization steps which have been applied to the data (cfr. \ref{subsec:method_acquisition}), the sCMTF operates in a regime of moderately high relative approximation errors.

\subsection{Spatio-temporo-spectral profiles of interictal discharges}
We analyze for each patient the sources which have been estimated via the sCMTF model. 
Due to space constraints, we discuss the results of two patients in detail in the next subsections, and include complete results for all other patients in the supplementary material.
Every time, we show 1) the thresholded pseudo t-maps of the IED-related source in the fMRI domain, both for significant activation as for significant deactivation; 2) maps highlighting the ROIs of high HRF entropy and extremity; 3) the temporal profile (time-varying power) $\vect{s}_r$, spatial profile (topography) $\vect{m}_r$ and spectral profile $\vect{g}_r$ of each source in the EEG domain; 4) the HRF waveforms in the different ROIs, and the HRF basis functions at convergence of the algorithm.
We plot maximally 800 s of the temporal signatures, to ensure readability.
For ease of comparison, we always overlay the broadband MWF envelope (with an arbitrary vertical offset for visualization only), which is the reference time course $\vect{s}_{\text{ref}}$ for selecting the IED-related component (cfr. \ref{apx:std_iedselection}).
For considerations of space, we generally only show the maps of the fMRI spatial signature $\vect{v}_r$ for the IED-related components, but discuss the maps of other components when relevant.
We show five axial slices of each map: in each case, we show two slices near the highest and lowest voxels of the IOZ or significant regions of the fMRI spatial signature (whichever lies furthest); if applicable, the middle slice is the cross-section with most overlap between IOZ and spatial signature, and the two remaining slices lie halfway between this slice and the extremal slices; otherwise all three bulk slices are chosen with equal spacing between the extremal slices.
We cross-validate the maps against known resting state networks (RSN) of human brain activity from the Stanford atlas \citep{shirer2012decoding}.

We stress again at this point that a \textit{subset} of the results is prone to errors due to imperfect sign normalization (cfr.  \ref{apx:std_ambiguity}).
While it is relatively straightforward to unambiguously determine the `right' sign of the EEG signatures, this is more challenging for fMRI. 
That is, frequently, the polarity of the HRF waveform is ambiguous, and making the `wrong' choice in a voxel $i_v$ (i.e., the HRF which has the opposite effect of the true physical CBF change) immediately leads to the wrong sign of the spatial coefficients in $\vect{r}_{i_v}$ and their pseudo t-values for all sources $r$.
To track the occurrence of this foreseen failure mode, we also investigate the significant deactivations of the sources\footnote{Alternatively, it is possible to use a pseudo F-statistic, e.g. the squared pseudo t-value, to bypass the sign correction altogether. The downside of such an approach is that it is then impossible to distinguish activation and deactivation, which may be meaningful.}.
Note that we designed the HRF variability metrics so that they are \textit{immune} to the polarity of the HRFs.
Hence, any high score of the HRF metrics can be reliably interpreted.
For each case, we separate the twenty waveforms with the highest entropy score, and report how many of those are found in ROIs that overlap with the IOZ, along with the probability (in the form of a p-value) that this would occur by randomly sampling as many ROIs (under a given fraction of brain that is covered by the IOZ).
Hence, this metric is analogous to the positive predictive value (PPV)\footnote{The PPV is the fraction of positive predictions (in a classification or hypothesis testing framework) which are truly positive, which equals one minus the false discovery rate (FDR).}, albeit no rigorous test has been applied in this case.

\subsubsection{Patient 3}
\begin{figure*}
	\centering
	\begin{subfigure}[b]{0.6\textwidth}
		\includegraphics[width=1\linewidth]{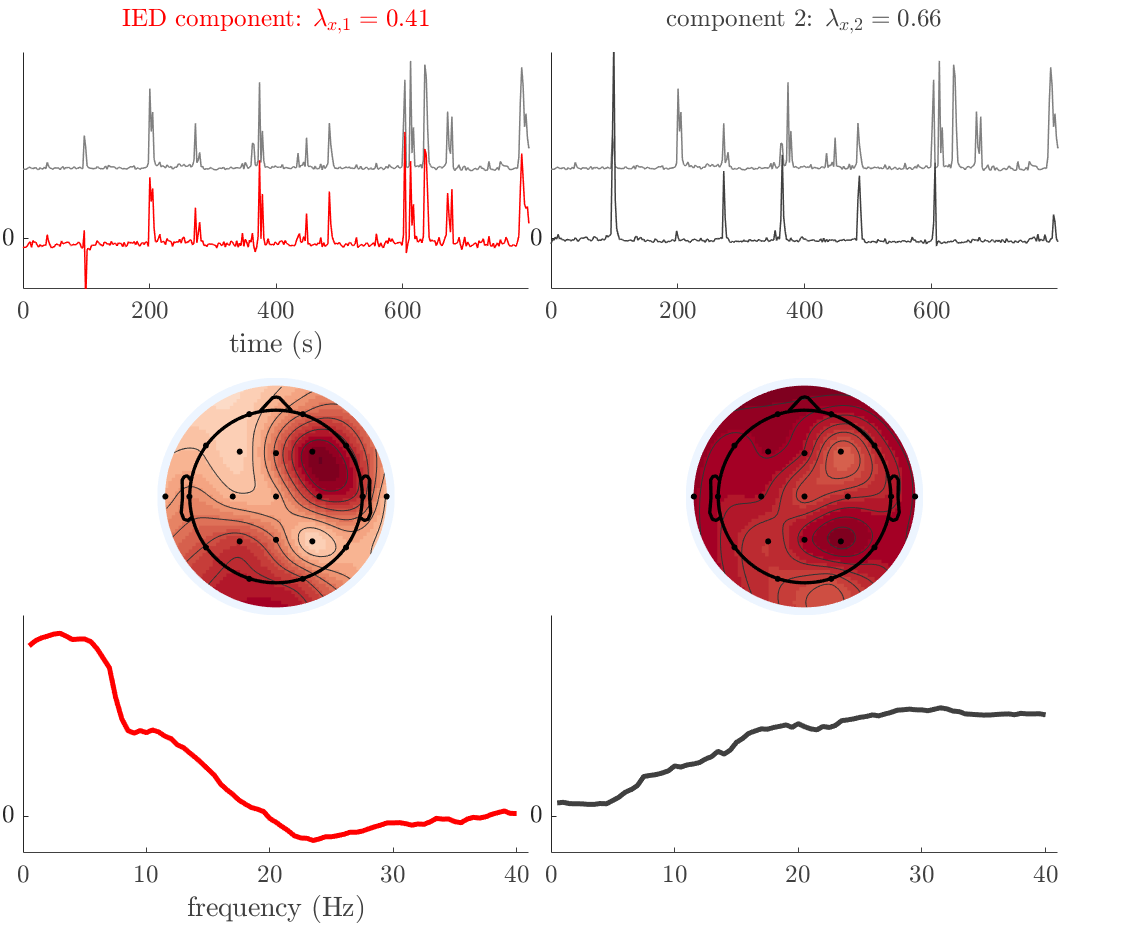}
		\caption{Temporal ($\vect{s}_r$, top), spatial ($\vect{m}_r$, middle), and spectral ($\vect{g}_r$, bottom) profiles of the 2 sources in the EEG domain, and reference IED time course ($\vect{s}_{\text{ref}}$, in grey).}
		\label{fig:p05eeg}
	\end{subfigure}
	
	\begin{subfigure}[b]{0.8\textwidth}
		\includegraphics[width=\textwidth]{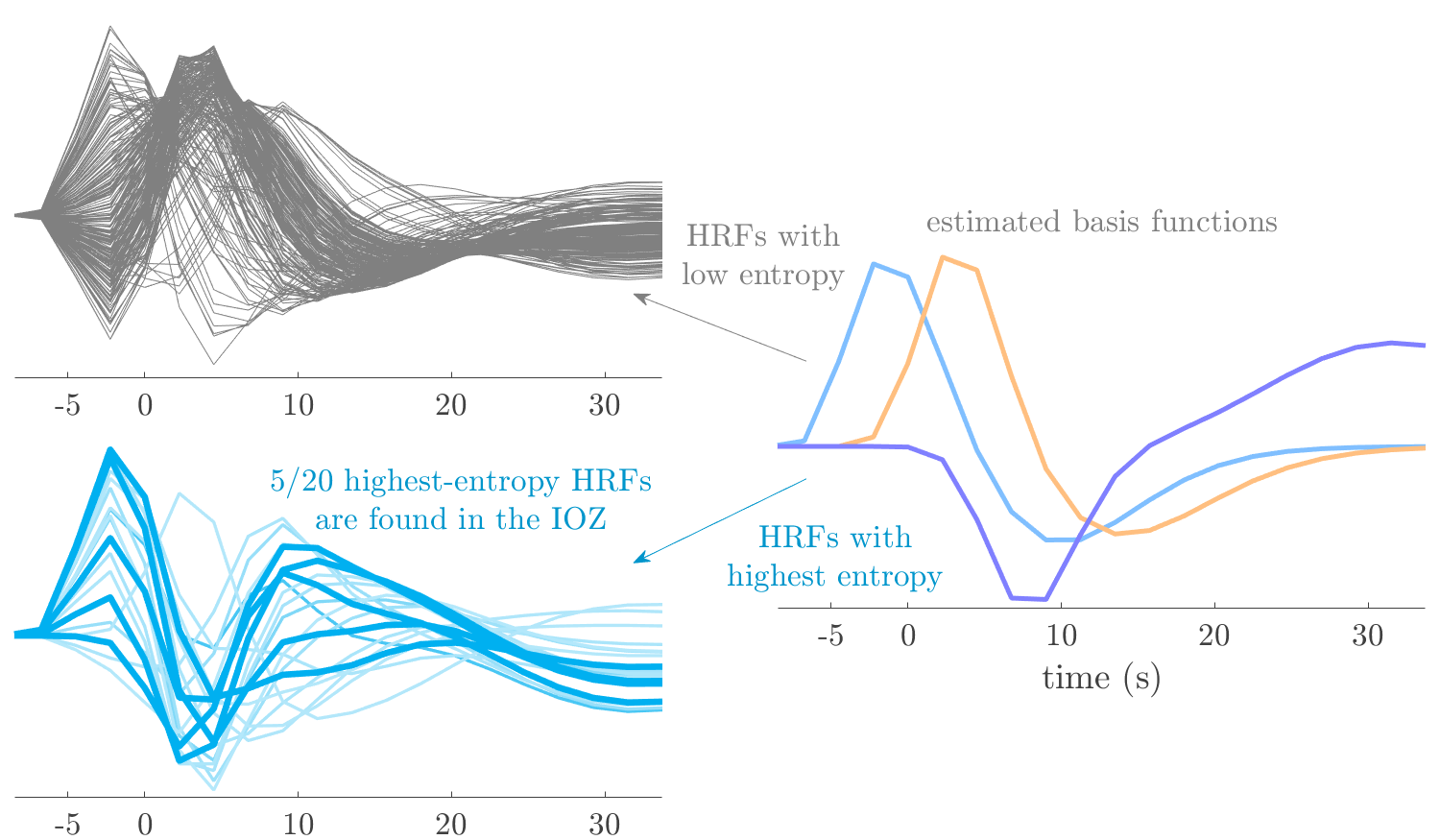}
		\caption{Estimated HRF waveforms in all ROIs, split in HRFs with low (grey) and high (blue) entropy}
		\label{fig:p05hrf}
	\end{subfigure}
	\caption{
		\subfigcap{a} In the selected solution for patient 3 ($R=2$), both sources have a temporal signature that correlated strongly to the reference IED time course. The first source modeled the main onset of IEDs and was low-frequency and topographically focal, while the second source was spatially and spectrally diffuse and captured the propagation of IEDs to remote areas (cfr. Figure~\ref{fig:p05propagation}). \subfigcap{b} Five out of the twenty most deviant HRFs were found inside the ictal onset zone (bold lines, $p<10^{-4}$). These HRFs had main peaks before 0 s, i.e., they led to BOLD changes before the corresponding EEG correlate of the IED was seen.
	}
	\label{fig:p05eeghrf}
\end{figure*}

\begin{figure*}
	\centering
	\includegraphics[width=1\linewidth]{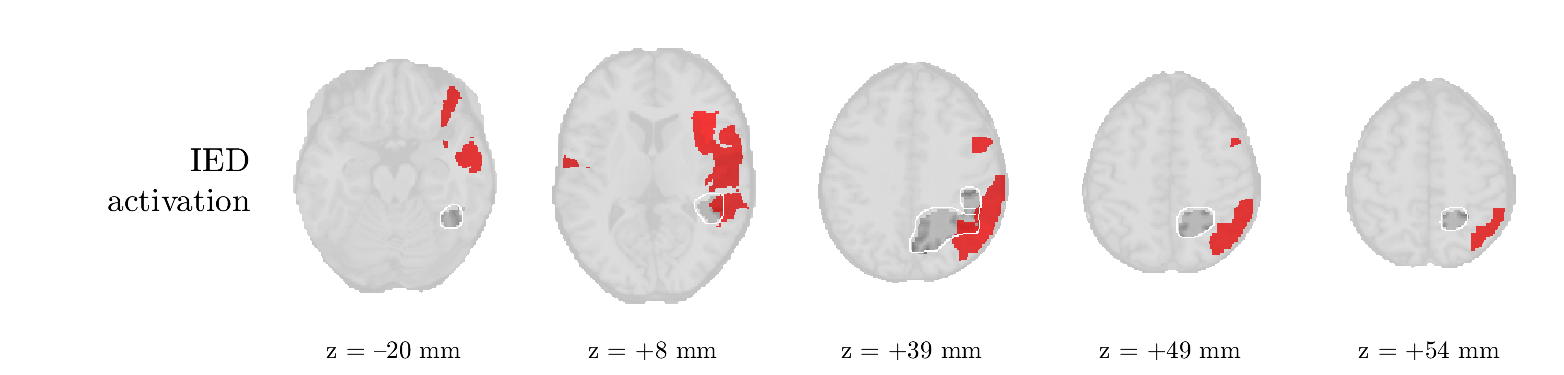}
	\includegraphics[width=1\linewidth]{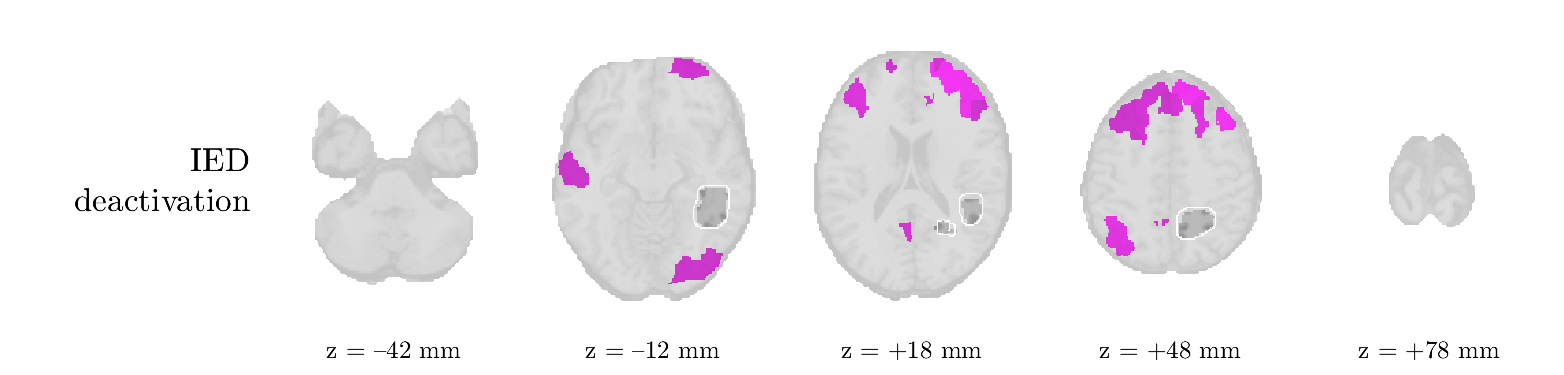}
	\includegraphics[width=1\linewidth]{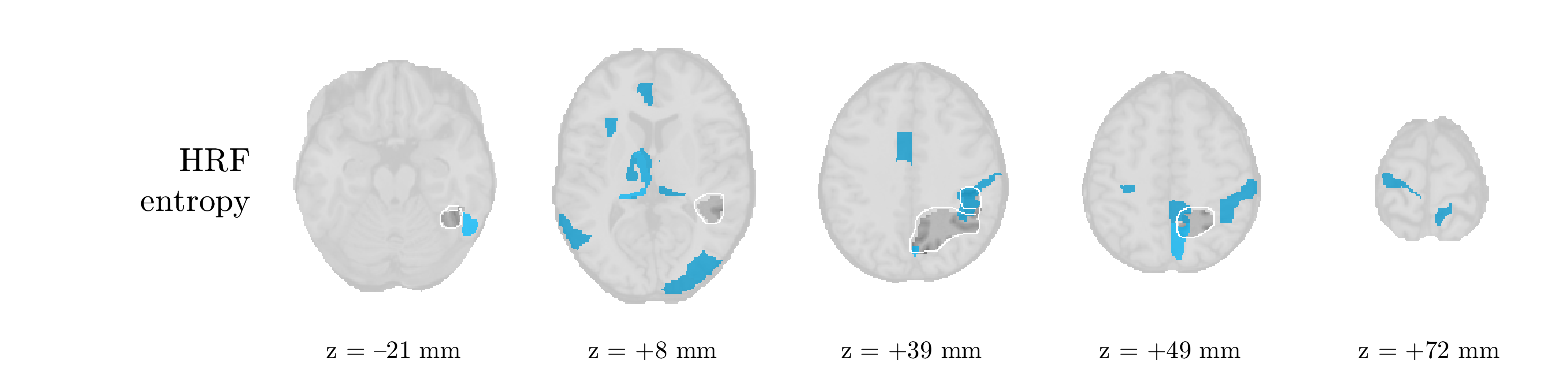}
	\includegraphics[width=1\linewidth]{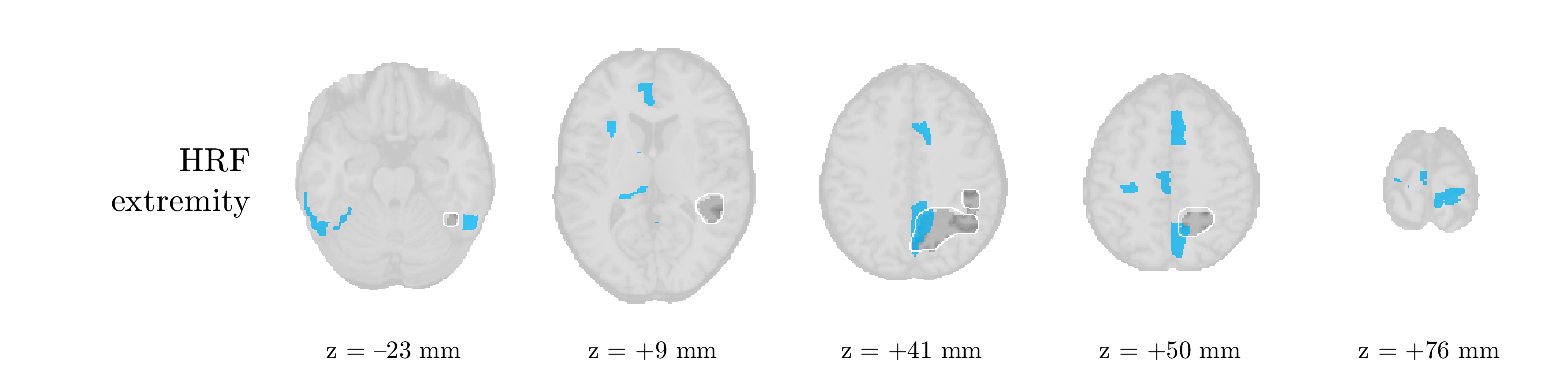}
	\caption{The statistical nonparametric maps of the IED-related component (top two rows) and HRF entropy/extremity maps (bottom two rows) of patient 3 show concordance with the ictal onset zone (IOZ). Especially the regions of significant IED activation were accurate, but also five out of the twenty regions with the most deviant (highest entropy) HRFs were found in the IOZ (cfr. Figure~\ref{fig:p05hrf}). The ground truth ictal onset zone is highlighted in dark gray with a white contour. 	
}
	\label{fig:p05fmri}
\end{figure*}

We analyze the solution with $\hat{R}=2$ sources, and show the results in Figure~\ref{fig:p05eeghrf} and \ref{fig:p05fmri}.
Besides one clear IED-related source, there is one other source that is substantially correlated to the reference time course, but with a homogeneous distribution over the head and an unclear spectrum.
This may signify  that the IEDs do not follow exactly a rank-1 structure in the spectrogram, and that they may be nonstationary in time or space (cfr. the argument made for nonstationary seizures in \citep{hunyadi2014block}).
The second source's pseudo t-map had significantly active areas symmetrically in the left and right parietal lobe, much more focalized than the EEG topography. 
In the EEG time courses, we found indeed IED-like waveforms at the times of the peaks in the temporal signature. 
Hence, we suspect that both sources may reflect the onset and propagation of the IEDs to other areas, respectively.
Five out of the twenty ROIs with high-entropy HRFs overlapped with the IOZ, and a significant finding is that several of them are highly noncausal, i.e., with a positive peak before zero seconds.
Figure~\ref{fig:p05fmri} confirms this, and also shows that the IED-related source is significantly active in different ROIs of the IOZ.

\subsubsection{Patient 10}
\begin{figure*}
	\centering
	\begin{subfigure}[b]{1\textwidth}
		\includegraphics[width=1\linewidth]{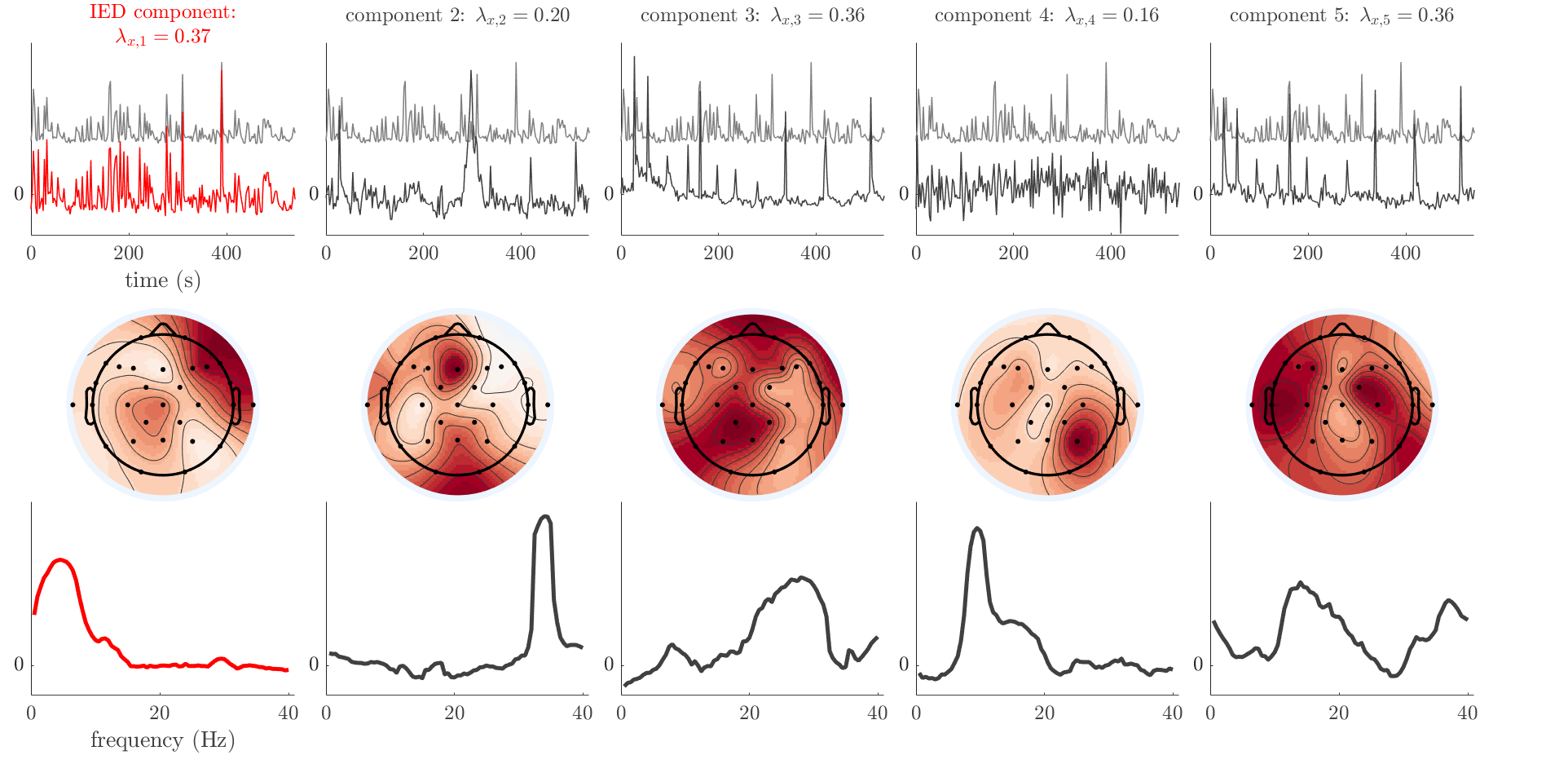}
		\caption{Temporal ($\vect{s}_r$, top), spatial ($\vect{m}_r$, middle), and spectral ($\vect{g}_r$, bottom) profiles of the 5 sources in the EEG domain, and reference IED time course ($\vect{s}_{\text{ref}}$, in grey).}
		\label{fig:p12eeg}
	\end{subfigure}
	
	\begin{subfigure}[b]{0.8\textwidth}
		\includegraphics[width=\textwidth]{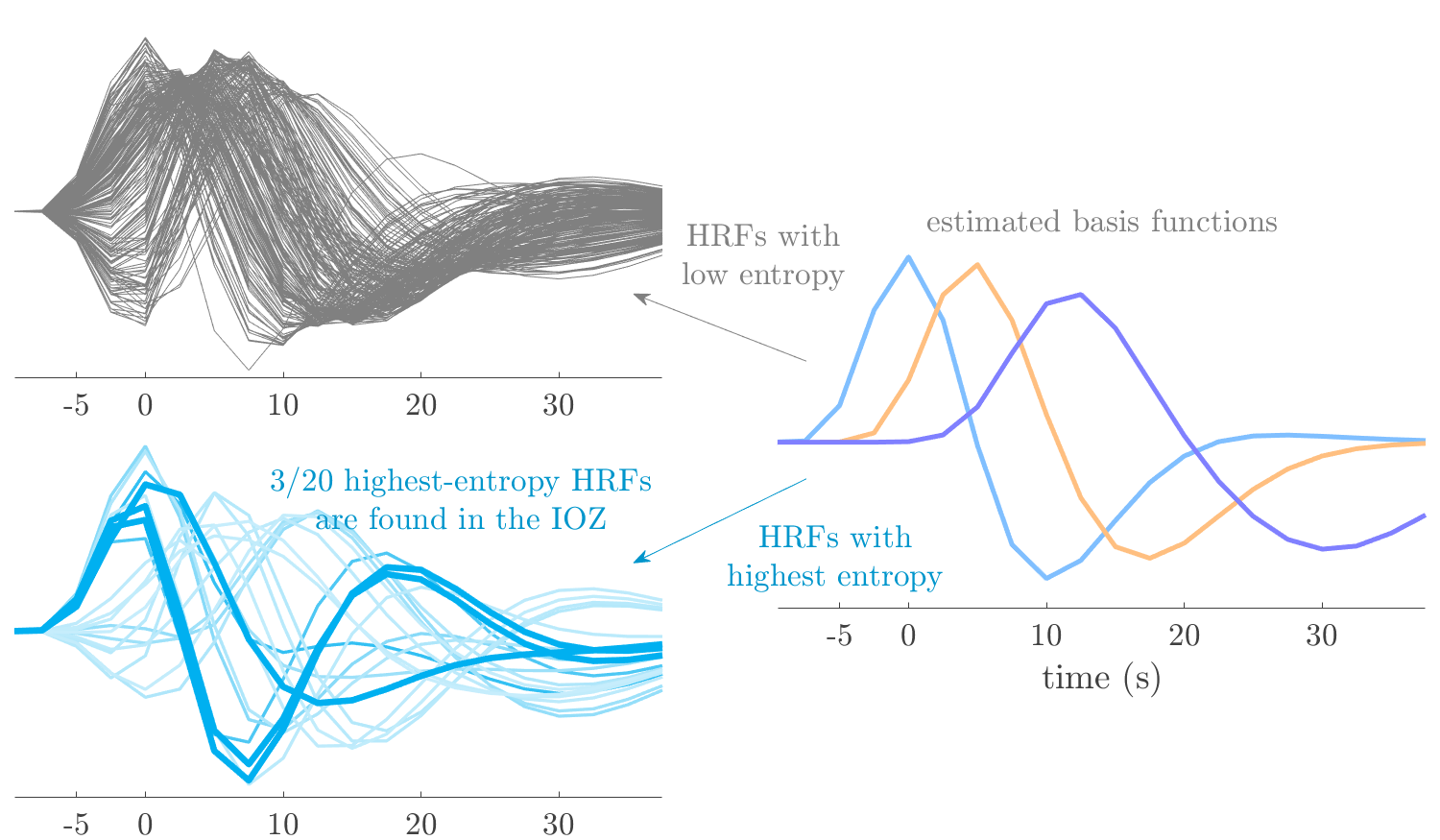}
		\caption{Estimated HRF waveforms in all ROIs, split in HRFs with low (grey) and high (blue) entropy}
		\label{fig:p12hrf}
	\end{subfigure}
	\caption{
		\subfigcap{a} The sCMTF solution with $R=5$ sources was selected for patient 10. 
		One source's temporal signature is highly correlated with the reference IED time course and is identified as the IED-related source, which has a characteristic low-frequency behaviour and with a frontotemporal topography, consistent with the IOZ location.
		The second source, which has very narrowband power around $\pm33$ Hz, likely captured an artifact of the MR acquisition. 
		The fourth source captured $\alpha$ activity in the default mode network (cfr. also Figure~\ref{fig:p12dmm}).
		\subfigcap{b} Three out of the twenty most deviant HRFs were found inside the ictal onset zone (bold lines, $p=0.02$).
	}
	\label{fig:p12eeghrf}
\end{figure*}

\begin{figure*}
	\centering
	\includegraphics[width=1\linewidth]{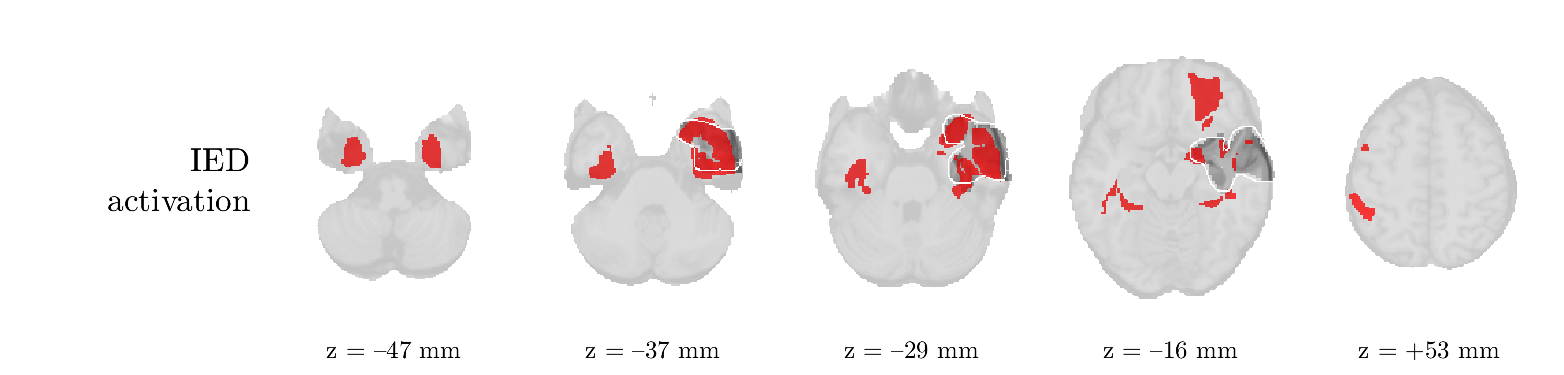}
	\includegraphics[width=1\linewidth]{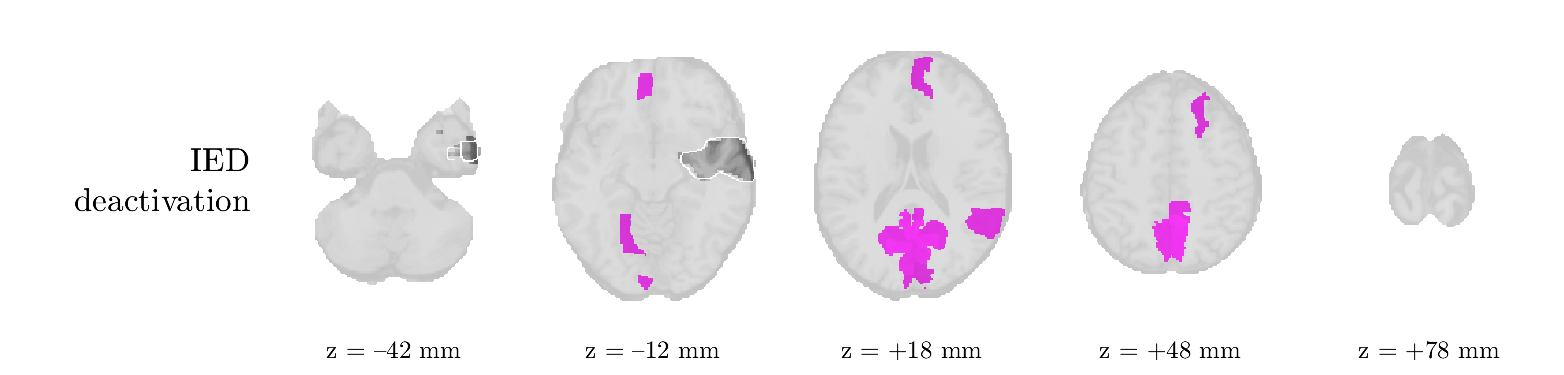}
	\includegraphics[width=1\linewidth]{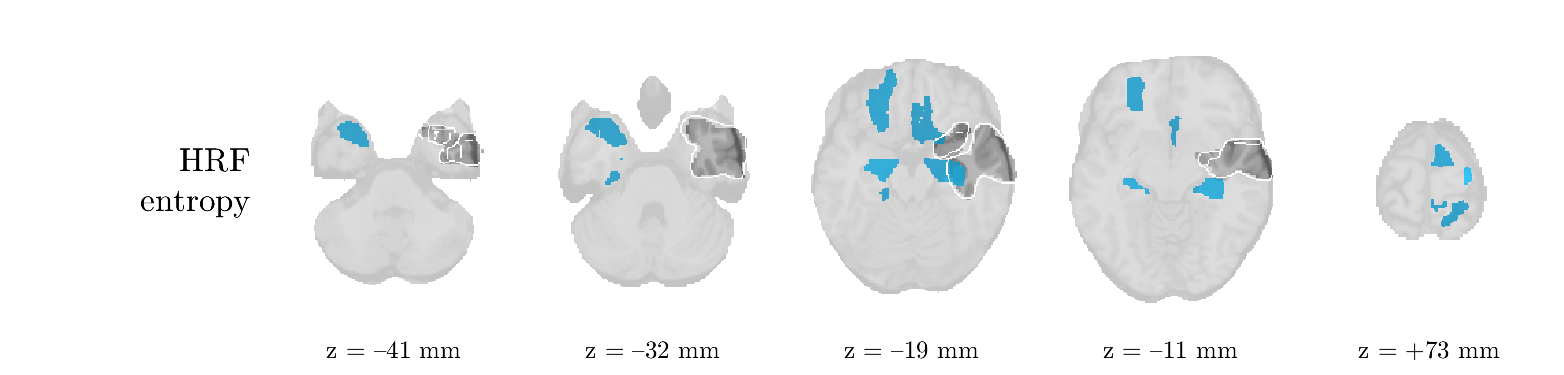}
	\includegraphics[width=1\linewidth]{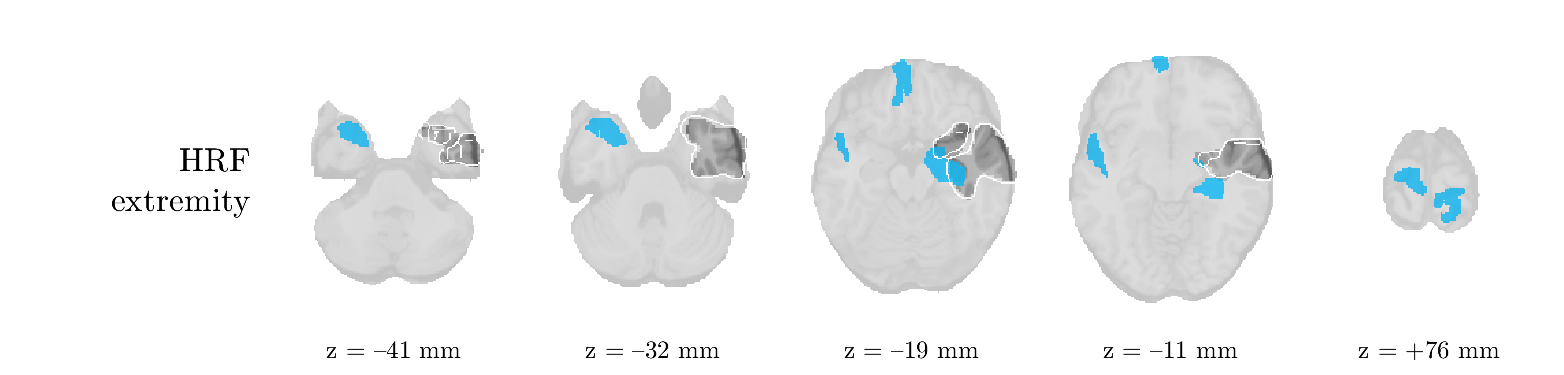}
	\caption{
		The statistical nonparametric maps of the IED-related component (top two rows) and HRF entropy/extremity maps (bottom two rows) of patient 10 show concordance with the ictal onset zone (IOZ). 
		IED occurrences were associated with significantly active regions in and near the IOZ (top row), and at the same time with a deactivation in a part of the default mode network (second row).
		Three out of the twenty regions with the most deviant (highest entropy) HRFs were found in the IOZ (cfr. Figure~\ref{fig:p12hrf}). The ground truth ictal onset zone is highlighted in dark gray with a white contour.
	}
	\label{fig:p12fmri}
\end{figure*}

We analyze the solution with $\hat{R}=5$ sources, and show the results in Figure~\ref{fig:p12eeghrf} and \ref{fig:p12fmri}.
There is a clear IED-related source, and also an artifactual source at $\pm34$ Hz, which is also present in other patients.
Due to its relatively consistent occurrence, we hypothesize that this artifact is due to the MR acquisition.
For example, it may be a remnant of a gradient artifact which is not adequately removed from the data of some channels, cfr. the observation made in \citep{marecek2016can}.
Surprisingly, this source is significantly active in an extended area in the occipital lobe, overlapping with the visual network.
Both HRF metrics correctly reached extreme values at (distinct) ROIs within the IOZ.
The pseudo-t map of the IED-related source shows significantly active ROIs that are concordant with the IOZ, and deactivation of a large part of the default mode network.
Furthermore, the IED-related source's EEG topography is very consistent with the clinical diagnosis.
The fourth source is active in the default mode network, predominantly in the $\alpha$ band (cfr. Figure~\ref{fig:p12dmm}).
The fifth source had an unclear spectrum, but its temporal signature corresponds to the occurrence of high-amplitude IEDs.
Its pseudo t-map shows widespread activations over the brain, which did not include the IOZ. 
We expect that this component captures the propagation of IEDs, after onset near the IOZ, similarly to patient 3.

\subsubsection{Summary of all patient's results}
We provide an overview of the results w.r.t. IOZ detection in Table~\ref{tab:allresults}.
All results taken together, the sCMTF results allow a correct detection of the IOZ based on the significant IED activation (10/12 cases), significant IED deactivation (6/12 cases), HRF entropy (9/12 cases) and HRF \mbox{extremity} (8/12) cases. 
All cases are covered by at least one of the metrics, and all patients besides patient 6 had at least two metrics providing correct and complementary localizing info on the IOZ.
For nearly all cases, the IED-related component's time course was highly correlated to a reference IED time course, and its spectrum was plausible.
In many, but not all cases, this component's EEG topography was also consistent with the location of the IOZ, though this notion is slightly fuzzy because of the very different spatial domains of EEG and (f)MRI---hence we do not use the term `concordant'.
Analysis of the spatial, spectral, and temporal signatures, in combination with inspection of the filtered EEG signals, reveals the identity of RSN oscillations and/or artifacts in the majority of cases.
For several patients, we found sources that are active in a narrow spectral band near 33--34 Hz. 
While this likely reflects a technical artifacts as the result of the MR acquisition, we found no concomitant changes at this frequency in the EEG. 
This may be the result of the normalization procedure which we applied prior to the decomposition: since every frequency bin was given equal importance, even unnoticeable but structured fluctuations at higher frequencies may be captured in a component.

\begin{table*}
	\caption{The sCMTF leads to three types of spatial information, which can be cross-validated against the ground truth IOZ, as defined in Section~\ref{subsec:method_performance} and summarized for all patients in Table~\ref{tab:clindata}: 1) the EEG topography $\vect{v}_{{}_{\text{IED}}}$ of the IED-related component; 2) the significantly activated and deactivated ROIs in the fMRI spatial signature $\vect{v}_{{}_{\text{IED}}}$; 3) the ROIs with strongly deviating HRF waveforms, as measured with entropy and extremity. Since the EEG topography has a very low spatial resolution, and depends on the attenuation properties of the tissue as well as the orientation of the neural sources in the cortex, we only expect partial similarity to the IOZ's spatial focus; hence, we use the term `consistent' rather than `concordant'.}
	\scalebox{0.8}{
			\renewcommand\arraystretch{1}
			\begin{tabular}{
			>{\centering}p{0.06\linewidth} 
	 		c 
	 		>{\centering}p{0.08\linewidth} 
			c 
			>{\centering}p{0.17\linewidth} 
			c 
			*{2}{>{\centering}p{0.10\linewidth}} 
			>{\raggedleft\arraybackslash}p{0.01\linewidth} 
			*{2}{>{\centering}p{0.09\linewidth}} 
			>{\raggedleft\arraybackslash}p{0.01\linewidth} 
			>{\raggedleft}p{0.14\linewidth} 
			>{\raggedright\arraybackslash}p{0.08\linewidth} 
		 	}
			\toprule
			patient  & & \begin{tabular}{@{}c@{}}selected\\ solution\end{tabular} & & 
			\begin{tabular}{@{}c@{}}EEG topography\\ consistent with IOZ? \end{tabular} & & 		
			\multicolumn{2}{c}{\normalsize\begin{tabular}{@{}c@{}}spatial signature $\vect{v}_{{}_{\text{IED}}}$\\ concordant with IOZ?\end{tabular}} & &
			\multicolumn{2}{c}{\normalsize\begin{tabular}{@{}c@{}}HRF variability metrics\\ concordant with IOZ?\end{tabular}} & &
			\multicolumn{2}{c}{20 highest-entropy ROIs}
			\\\cmidrule(lr){1-1}\cmidrule(lr){3-3}\cmidrule(lr){5-5}\cmidrule(lr){7-8}\cmidrule(lr){10-11}\cmidrule(lr){13-14}
			ID & & $\hat{R}$ & & & & activation & deactivation & & entropy & extremity & & \# in IOZ & (p-value)
			\\\cmidrule(lr){1-1}\cmidrule(lr){3-3}\cmidrule(lr){5-5}\cmidrule(lr){7-8}\cmidrule(lr){10-11}\cmidrule(lr){13-14}
			p01 & & 6 & & no & & no & no & & yes & yes & & 1 & (0.34)\\
			p02 & & 3 & & no & & yes & yes & & yes & no & & 1 & (0.59)\\
			p03 & & 2 & & no & & yes & no & & yes & yes & & 5 & ($<10^{-4}$)\\			
			p04 & & 4 & & yes & & yes & no & & yes & yes & & 2 & (0.32)\\			
			p05 & & 5 & & yes & & yes & yes & & yes & yes & & 6 & ($<10^{-3}$)\\			
			p06 & & 2 & & no & & yes & no & & no & no & & 0 & /\\			
			p07 & & 4 & & yes & & yes & no & & yes & no & & 1 & (0.57)\\			
			p08 & & 2 & & no & & yes & yes & & no & no & & 0 & /\\			
			p09 & & 2 & & yes & & yes & yes & & no & yes & & 0 & /\\			
			p10 & & 5 & & yes & & yes & no & & yes & yes & & 3 & (0.02)\\			
			p11 & & 2 & & yes & & yes & yes & & yes & yes & & 4 & (0.01)\\			
			p12 & & 2 & & no & & no & yes & & yes & yes & & 7 & ($<10^{-3}$)		
			\\\bottomrule
			\end{tabular}
	}
	\label{tab:allresults}
\end{table*}

\subsubsection{Sensitivity to model selection}
For many patients, selecting $\hat{R}$ is ambiguous, since more than one solution (with different $R$) score well on some of the criteria (cfr. Table~\ref{tab:Rselection}). 
Therefore, we analyze impact of the choice of $R$ on the sCMTF results.
For each patient, we select the solution with the rank which is next in line, i.e., which would be a second best (or equally good) choice, based on the same criteria.
This is the solution with $R=1$ for patient 12, $R=2$ for patients 1, 2, 5 and 7, $R=3$ for patients 3, 4, 6, 8, 9 and 10, and $R=4$ for patient 11.
For patients 1, 6 and 8, the results deteriorate drastically, as no metric correctly localizes the IOZ.
For patient 11, no ROI within the IOZ is significantly activated due to IEDs anymore, but the HRF metrics are still informative.
The results for patients 9 and 12 improve, since all metrics are now sensitive to the IOZ.
For the other patients, the situation stay more or less the same, i.e., the same metrics are valuable for IOZ localization. 
However, the maximum value under the different metrics is generally attained at different ROIs compared to the initially selected model.

%% file: sections/discussion.tex
\section{Discussion}\label{sec:discussion}

\begin{figure*}
	\centering
	\includegraphics[width=1\linewidth]{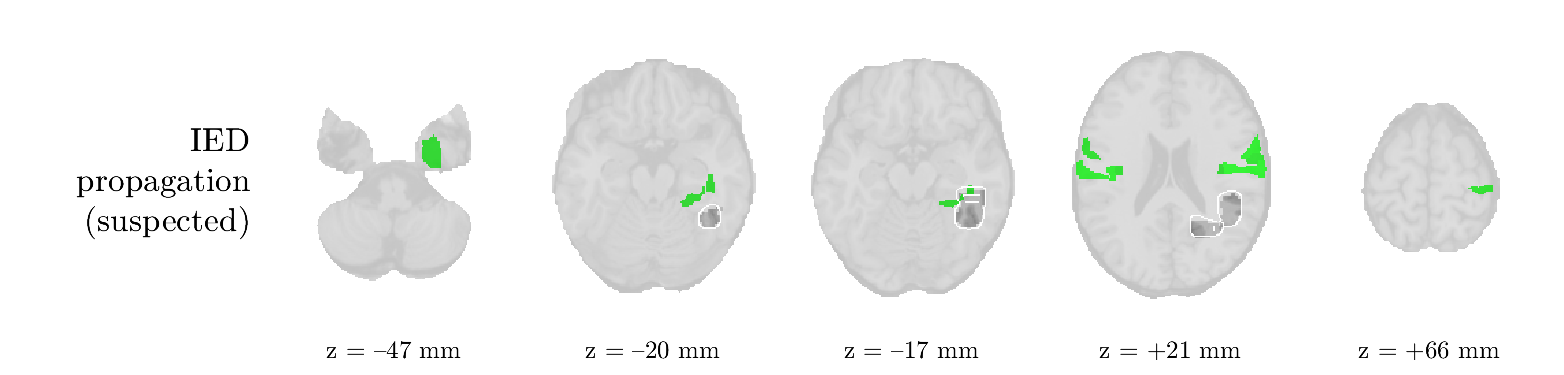}
	\caption{The second component in patient 3 likely captured the  propagation of IEDs from the irritative zone, given its relatively large correlation to the MWF envelope (cfr. Figure~\ref{fig:p05eeg}). The ground truth ictal onset zone is highlighted in dark gray with a white contour.}
	\label{fig:p05propagation}
\end{figure*}

\subsubsection*{A novel EEG-fMRI data fusion framework}
We have proposed an integrated and structured coupled matrix-tensor factorization (sCMTF) framework, which can be used to make inferences on the localization of the ictal onset zone in refractory epilepsy based on simultaneous EEG and fMRI recordings. 
Our approach aims to perform blind source separation of the neural activity related to interictal epileptic discharges (IEDs), and to characterize it in the spatial, temporal, and spectral domain.
To this end, we developed a pipeline consisting of 1) semi-automated EEG enhancement based on annotations of the IEDs; 2) modality-specific preprocessing and tensorization steps, which lead to a third-order EEG spectrogram tensor varying over electrodes, time points, and frequencies, and an fMRI matrix with BOLD time courses for a predefined set of regions of interest or parcels; 3) coupled matrix-tensor factorization of the EEG tensor and fMRI matrix along the shared temporal mode, while accounting for variations in the local neurovascular coupling; 4) automated selection of a robust, and relevant IED-related component, and nonparametric testing to infer its spatial distribution in the brain.

We have stressed the importance of and accounted for the variability of the hemodynamic response function (HRF) over different patients and brain regions, by equipping the CMTF with the required expressive power via a set of adaptive basis functions. 
Moreover, after estimating the EEG and fMRI factor signatures, as well as the HRF parameters, we have computed different summary metrics (entropy and extremity) that measure the local deviance of a ROI's HRF compared to other HRFs in the same brain, and have cross-validated the spatial map of these metrics against the ground truth localization of the ictal onset zone.

The sCMTF pipeline managed to provide correct detection in all twelve patients in this study, with varying degrees of certainty.
The statistical nonparametric map (SnPM) of the spatial signature of the IED-related component, obtained with the sCMTF, is the best biomarker
The ROIs that are significantly activated under influence of IEDs were the best biomarker in the study, which is in line with the traditional EEG-correlated fMRI approach \citep{lemieux2001event}.
In the large majority of patients, several of these regions overlapped with the IOZ.
Also the HRF entropy, as a measure of how unlikely an HRF is within a specific set of other HRFs, is a very good biomarker, which almost always identified regions of the IOZ that were complementary to those already found by tracking significant IED activation.
In roughly half of all cases, we also found regions within the IOZ that significantly deactivated in association to IEDs.
Compared to our earlier work on these data in \citep{vaneyndhoven2019semi}, we achieved an additional correct IOZ detection (for patient 11), which is probably thanks to the increased flexibility of the current model.
In a waterbed effect, the pseudo t-map for patient 1 no longer allowed correct IOZ detection compared to the simpler pipeline.
Luckily, however, this gap is filled here by the new HRF entropy metric.
We inspected the 20 HRFs and ROIs with the highest extremity and entropy. 
Hence, it is inevitable that some or most of these ROIs are not within the IOZ.
Standalone HRF metrics would hence have a high false discovery rate, even though for several patients, the high proportion of IOZ-covering ROIs among the 20 selected ROIs was very unlikely due to chance (as measured with p-values).
However, the ROIs that were highlighted by the HRF metrics were often distinct from the ROIs identified as significantly activated to the IEDs.
Hence, the SnPMs of the IEDs and the entropy metrics provide very complementary information, and when analyzed jointly, they may infer the location of the IOZ with much more certainty, i.e., in the brain area where both IED-related and HRF-related metrics have a high value.

\subsubsection*{HRFs vary strongly over subjects and brain regions}
There were substantial differences in (estimated) neurovascular coupling over patients and brain regions, as expected.
Since we used `regularized' basis functions, which are parametrized as smooth gamma density functions, the resulting HRFs generally had a plausible shape.
However, in some cases we found nonsensical shapes, in which, e.g., the waveform had the same polarity over the whole time course, potentially with a bimodal shape (cfr. patient~4).
This serves as a humble reminder to not blindly trust the outputted HRFs (or other factor signatures, for that matter). 
While we have empirically verified that the optimization algorithm converges properly to the true factor signatures and HRFs for synthetic data under mild conditions, there is no guarantee that this holds true for real-life data, which are orders of magnitude more complex, so that a linear generative model like the sCMTF may not be sufficient to describe the interplay between EEG and fMRI.
Moreover, the proper behavior of the sCMTF estimation depends on careful preprocessing, and on a proper selection of hyperparameters (in casu: a good value for the number of sources $\hat{R}$).
Hence, manual inspection of the data quality and the solution are still required. 
Even if the estimated HRFs or factor signatures may not fully reflect the `correct' underlying physical phenomena, we have demonstrated that they offer actionable information.
Not in the least, via summarizing metrics such as HRF entropy and extremity, our algorithm manages to be reasonably robust to subtle changes in the waveform---which is less of interest here than spatial cues towards the IOZ.

The algorithm used its modeling freedom to fit `noncausal' HRFs, which are ahead of the EEG by as much as 10 s. 
Generally, we indeed found that many of the estimated HRFs had significant positive or negative amplitudes already before the neural correlate visible on the EEG.
This is in line with recurrent findings on BOLD changes that precede the IEDs which were observed in the EEG \citep{hawco2007bold,pittau2011changes,jacobs2009hemodynamic,moeller2008changes}. 
We stress that this noncausality may only be in the observation, and not in the underlying physical chain of events: here, it strictly means that we \textit{observed} BOLD changes in the fMRI data that occur before the corresponding \textit{observed} neural correlate on the EEG.
Despite the fact that many of the HRFs differed substantially from the canonical HRF, which is causal and peaks approximately 6 s \mbox{\textit{after}} its neural input, we obtained good results as well with the latter HRF as a nonadaptive model for neurovascular coupling \citep{vaneyndhoven2019semi}. 
The reason for this agreement between these different models---which differ substantially in terms of flexibility---is likely that the canonical HRF is positively correlated to the true HRFs which are found inside the IOZ, and as such the resulting activation maps may still be sufficiently informative.
In our data and sCMTF results this is indeed the case for many patients.

\subsubsection*{Prior EEG signal enhancement aids analysis}
Importantly, our pipeline heavily relies on a prior enhancement of the interictal spikes in the EEG data, which would otherwise have a too low SNR for the sCMTF algorithm to pick up IED-related sources.
We employ multi-channel Wiener filters, which solely rely on the annotation of a sufficient amount of IEDs in the data itself, or in related data (e.g., data from the same patient, recorded outside the MR scanner).
While this task still frequently relies on the skill of human EEG readers and neurologists, advanced automated solutions for interictal spike detection are available \citep{wilson1999spike,scheuer2017spike}.
Within each solution of a specific rank, we picked the IED component as the one with the highest correlation with a reference time course directly derived from the enhanced EEG. 
Some of the presented results make clear that this reference time course is not completely free from artifacts, hence caution is warranted when many high-amplitude artifacts are still present in the reference. 
In this study, however, we have not encountered any issues that seemed to be the direct results of a noisy reference during IED component selection.

\begin{figure*}
	\centering
	\includegraphics[width=1\linewidth]{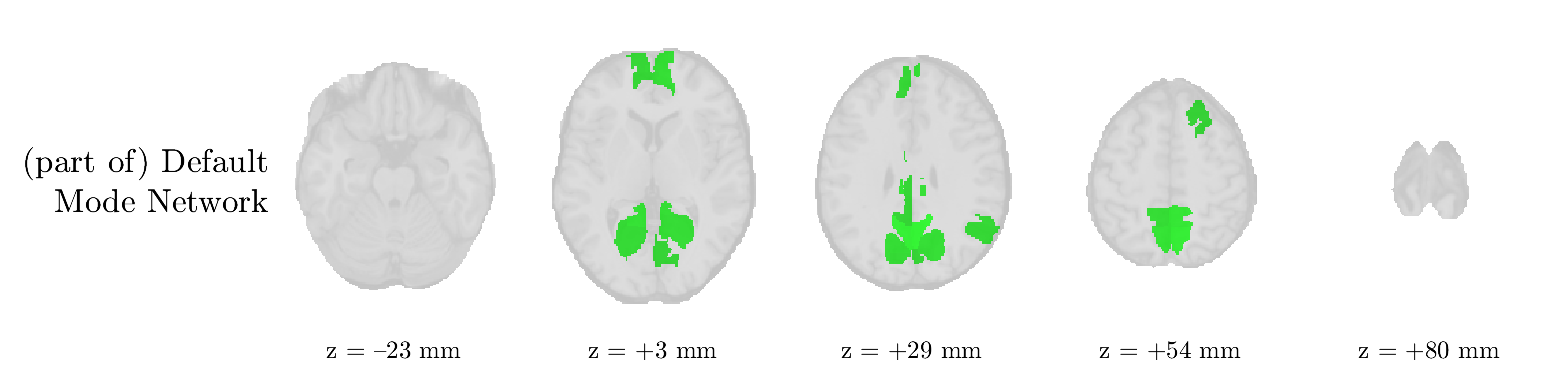}
	\caption{The fourth component in patient 10 seemed to pick up activity in the Default Mode Network (DMN), predominantly in the $\alpha$ band (cfr. Figure~\ref{fig:p12eeg}).}
	\label{fig:p12dmm}
\end{figure*}

\subsubsection*{The interpretation of components}
Overall, the sCMTF pipeline succeeded in extracting meaningful IED-related components, alongside components that modeled resting-state neural fluctuations and physiological and technical artifacts.
The fact that the sCMTF can estimate signatures and statistical maps for multiple components is a powerful advantage over classical EEG-correlated analysis.
As we demonstrated in the experiments, artifactual influences may be isolated in separate components, which could reduce their impact on IED mapping in the brain.
Additionally, we encountered cases where two components were correlated to the IED occurrences: the component with the highest temporal correlation to a reference IED time course then correctly revealed the localization of the IOZ, while the other component presumably modeled the propagation of IEDs to remote brain regions.
This observation is analogous to the finding in \citep{hunyadi2016fusion}, where a different type of CMTF was applied to average EEG waveforms of IEDs and statistical BOLD maps, which revealed a dissociation between the early IED spike and the subsequent wave, which were related to the onset and spread of the IEDs, respectively. 
Since we transformed the data with a time-frequency transform that used windows of length TR, our algorithm is unable to unravel different phases within one IED, since they occur in a shorter time frame. 
However, we identified these different IED-related sources by their significantly correlated temporal signatures, and their distinct spatial and spectral profiles. 
While we did not impose nonnegativity constraints, many estimated EEG spectral and spatial signatures were approximately nonnegative.
This need not be the case, however, since the EEG data are normalized in a way such that the resulting signatures would reveal relative increase/decrease, rather than absolute time-varying spectral power \citep{marecek2016can}.
For example, if a certain component is associated with a power increase in one spectral band, and a simultaneous power decrease in another band, this would be reflected in a spectrum with both a positive and a negative peak.

\subsubsection*{Practical considerations}
The end-to-end sCMTF pipeline can provide a richer set of results compared to classical EEG-correlated fMRI analysis.
In this respect, it is a more powerful data exploration tool.
The tradeoff to be made is that \mbox{significant} computation time goes into the sCMTF and subsequent inference---if one wants to apply it as rigorously as we have done in the current experiments.
We seem to be doing a lot of unnecessary work, by computing the sCMTF factors for several numbers of sources, and by repeating the optimization several times for a fixed number of sources.
Unfortunately, both ways of repetition seem required to obtain robust results, as we have argued in \ref{apx:optimization} and \ref{apx:standardization}.
However, the EEG-only CP decomposition, which lies at the heart of our initialization strategy, seemed very robust: we found highly similar EEG signatures for almost all random initializations.
Probably, this is thanks to the use of the powerful Gauss--Newton-type of optimization.
Hence, fewer repetitions of the sCMTF may be already sufficient to arrive at the same robust results. 
Despite the very reproducible EEG signatures in the initial CP decompositions, we still performed 50 repetitions of the sCTMF, each time slightly varying the initial HRF parameters.
As such, we believe our findings are reasonably robust to poor initialization of the HRFs.
Performing the sCMTF for many choices of $R$ may still be required, as the quality of the result depends on the extraction of an appropriate number of sources.
Luckily, however, most `optimal' ranks were quite low in our study, under the heuristic selection procedure.
This is reassuring, as it signifies that a typical rank encountered in this context is not problematically high, and no prohibitive computations are needed.
Furthermore, we have demonstrated in our experiments that the summary metrics (sensitivity for localizing the IOZ based on different statistical scores) are fairly robust to the choice of $R$, although the estimated signatures themselves differ.

For many patients, the available data was split across multiple runs (i.e., with a few minutes break in between), and we opted to temporally concatenate data over runs, as explained in Section~\ref{subsec:method_acquisition}.
While this violates the coupling model based on HRFs for time samples near the boundaries, we consider the effect minimal, given that the number of those `affected' time samples represents a very tiny fraction of the whole time series.
However, a more rigorous approach would be to `inverse-impute' these samples and consider them as missing values: as such, they are ignored during the sCMTF optimization and will not affect the results.

\subsubsection*{Strategies to alleviate the computational demand}
Due to the repeated decomposition and the nature of the nonparametric inference, the computations are highly parallellizable.
For a typical dataset with available IED annotations, and with the parameters we have used for this study, the end-to-end computation for one patient took a few hours on a machine with twelve cores. 
To alleviate the computational burden, we have parcellated the fMRI data into 246 regions, based on the Brainnetome atlas \citep{fan2016human}.
This is clearly suboptimal, as the atlas is not patient-specific, and is mostly designed to study healthy brains.
There is a serious risk for partial volume effects, in which the IOZ is scattered over several ROIs.
As such, the IED-related BOLD changes in the part of the IOZ that falls within a certain ROI may get swamped by the remaining BOLD fluctuations within the ROI delineation.
Hence, we hope to be able overcome this problem, either by algorithmic improvements, including a speed up of the optimization, or by the use of a patient-specific parcellation or PCA-like compression of the fMRI data. 
As of yet, it is hard to say whether the fixed atlas had an adverse effect on the results, and it is not so straightforward to compare the statistical maps from this study to maps which are voxel-based.
We are currently pursuing experiments in which we employ a hierarchical parcellation: in a first step, the BOLD time series are grouped (but not yet averaged) according to the Brainnetome atlas; subsequently, we use spectral clustering to further refine each Brainnetome parcel based on the correlation matrix of its BOLD time series. As such, this hybrid approach combines a fixed, coarse-grained atlas with a further data-driven subdivision, which can mitigate partial volume effects, while still providing a significant data compression.
Alternatively, it is possible to achieve a data reduction while still preserving voxelwise BOLD signals, by limiting the scope of the sCMTF to an a priori defined ROI (e.g., based on a clinical hypothesis stemming from other modalities).

\subsubsection*{Summary}
In summary, we have developed and empirically validated a fully integrated framework for EEG-fMRI data fusion, which yielded a rich characterization of the interictal activity in time, space, and frequency, and which accounts for and exploits neurovascular coupling variation over the brain.
The ability to separate local (de)activation of IEDs from local deviations in the HRF makes the sCMTF a powerful tool for exploratory analysis of interictal EEG and fMRI data.
We envision that this approach, with some minor modifications, may also be used to analyze resting-state\footnote{Since in such a case, no IEDs are present, EEG-enhancement like we have done in this study would no longer take place. However, an MWF may still be used to clean up the EEG, e.g. by annotating artifactual periods, which can be removed from the data by the MWF in its dual form (or another tool) \citep{somers2018generic}.} EEG-fMRI activity.

Our complete MATLAB code to execute the pipeline is available at \url{https://github.com/svaneynd/structured-cmtf}.

%% file: sections/acknowledgment.tex
\section*{Acknowledgment}\label{sec:ack}
The research leading to these results has received funding from the European Research Council under the European Union's Seventh Framework Programme (FP7/2007-2013)/ERC Advanced Grant: BIOTENSORS (no. 339804). This paper reflects only the authors' views and the Union is not liable for any use that may be made of the contained information.
This research furthermore received funding from the Flemish Government under the ``Onderzoeksprogramma Artificiële Intelligentie (AI) Vlaanderen'' programme; from the Bijzonder Onderzoeksfonds KU Leuven (BOF) under the project numbers C24/15/036 and C24/18/097; from the Agentschap Innoveren en Ondernemen (VLAIO) under the project number 150466; from the EU for Horizon 2020 projects 766456, 813120 and 813483; and from EIT for the project SeizeIT (no. 19263).

%% file: sections/declaration.tex
\section*{Disclosure of competing interests}\label{sec:declaration}
Declarations of interest: none

%% file: sections/appendix-optimization.tex
\appendix
\section{Nonlinear fitting of the sCMTF model}\label{apx:optimization}

\subsection{Accommodating noncausal HRFs}\label{apx:opt_noncausal}
In section \ref{subsec:methods_coupling}, we derived the implementation of the convolution with an HRF as a left multiplication of the temporal signatures $\matr{S}$ with a Toeplitz matrix, whose diagonals hold the HRF samples. 
For a causal convolution, in which the BOLD signal strictly lags its neural correlate, $\matr{H}_k\left(i,j\right) = h_k\left(i-j\right)$ if $i-j \geqslant 0$, $\matr{H}_k\left(i,j\right) = 0$ otherwise, hence the matrix is lower triangular. This is the situation depicted also in Figure~\ref{fig:sCMTF}.

However, a recurring observation is that BOLD changes can be observed that precede the IEDs themselves \citep{hawco2007bold,pittau2011changes,jacobs2009hemodynamic,moeller2008changes}. 
Hence, we allowed noncausal HRFs that start at most 4 samples before the EEG, which allows for BOLD responses preceding the IEDs by up to 10 s at a typical TR of 2.5 s.

\subsection{Initialization and optimization}\label{apx:opt_init}
Since the cost function $J$ in \eqref{eq:costfunction} is nonconvex, any optimization procedure can only guarantee to converge to a local optimum, hence selecting a good starting point is crucial to obtain a reliable solution.	

Firstly, we decomposed the EEG data $\tens{X}$ individually according to the CP or PARAFAC model \citep{harshman1970foundations,kruskal1977three,bro1997parafac}, to obtain a good initialization for the factors $\left\llbracket\matr{S},\matr{G},\matr{M}\right\rrbracket$ in the sCMTF model.
To this end, we used a Gauss--Newton algorithm (\texttt{cpd\_nls} with 2000 iterations, 400 conjugate gradient iterations for the step computation, and tolerance on the relative cost function update of $10^{-8}$, in Tensorlab 3.0 \citep{vervliet2016tensorlab}), which we ran 50 times, from randomly drawn initial factors.
We observed that the resulting factors lied very often close together over runs, indicating the algorithm had found a robust solution.

\begin{figure}
	\centering
	\includegraphics[width=1\linewidth,trim=00mm 00mm 00mm 00mm,clip = true]{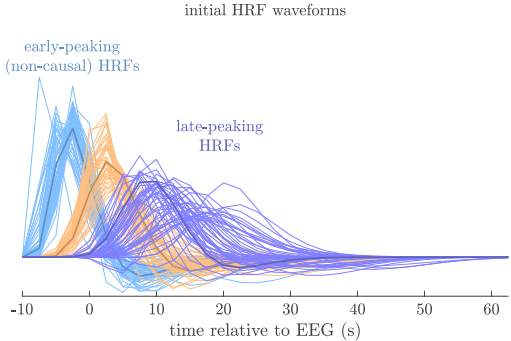} 	
	\caption{Different initializations of the HRF basis functions are used in every repetition of the sCMTF fitting procedure. In each repetition, one early-peaking, one mid-peaking and one late-peaking HRF were sampled from probability distributions of the HRF's generating parameters ($\boldsymbol{\theta}_1$, $\boldsymbol{\theta}_2$ and $\boldsymbol{\theta}_3$, respectively) that was created by applying some multiplicative noise to baseline  parameters (shown as the bold waveforms).
		The support extends to four negative samples, which gives the model the freedom to fit noncausal HRFs (relatively to the synchronized EEG).}
	\label{fig:hrf_init}
\end{figure}

We always employed $K=3$ HRFs, which we manually initialized.
To assess whether the eventual sCMTF solution was also robust to the initialization of the HRF basis functions, we used a slightly different set of HRF-generating parameters $\boldsymbol{\theta}_k$ in each repetition of the optimization.
Figure \ref{fig:hrf_init} shows some typical HRF waveforms, which are used to generate the Toeplitz blocks in Figure~\ref{fig:Ydecomp}.

From there, we initialized also the fMRI factors in the sCMTF model in \eqref{eq:cmtf_fmrirank1}--\eqref{eq:cmtf_fmriflattened}. 
We constructed a flattened `design matrix' $\matr{D}=\begin{bmatrix}
\matr{H}_1\matr{S}~\ldots~\matr{H}_K\matr{S} 
\end{bmatrix}$ in \eqref{eq:cmtf_fmriflattened} and obtained a rough estimate for $\matr{B}\transp \kr \matr{V}\transp$ as $\matr{U}\transp=\matr{D}^\dagger\matr{Y}$ via regression\textemdash albeit this does not yet disentangle $\matr{B}$ and $\matr{V}$. 
To obtain initializations for the individual spatial factors, we exploit the fact that in every ROI $i_v$, the Khatri--Rao product of the $i_v$-th columns of $\matr{B}\transp$ and $\matr{V}\transp$ corresponds to a rank-1 constraint when folded into a $K\times R$ matrix \citep{bousse2018linear,beckmann2005tensorial}; hence, a rank-1 truncated singular value decomposition of the folded $i_v$-th column of $\matr{U}\transp$ leads to the desired vectors \citep{beckmann2005tensorial}, which are further refined via a constrained Gauss--Newton algorithm \citep{bousse2018linear}. 
We approximated the residual of the fMRI data under the initialized (coupled) factors using a rank-$Q$ truncated SVD to capture fMRI nuisances. 
The parameter $Q$ was chosen as twice the number of acquisition runs that had been done for a subject.

After each of the 50 runs of the initialization procedure, we iteratively optimized \eqref{eq:costfunction} with a quasi-Newton algorithm (\texttt{sdf\_minf} with 1000 iterations, and tolerance on the relative cost function update of $10^{-8}$, in Tensorlab 3.0 \citep{vervliet2016tensorlab}).
Both $\tens{X}$ and $\matr{Y}$ were divided by their Frobenius norm, such that afterwards $\pnorm{}{F}{\tens{X}} = \pnorm{}{F}{\tens{Y}} = 1$, and we chose $\beta_x = \beta_y = 1$ so their fit had equal contribution to the cost function.
For the regularization penalties, we pick $\gamma_x = \gamma_y = 10^{-3}$, as in \citep{acar2014structure}

%% file: sections/appendix-standardization.tex
\section{Model selection}\label{apx:standardization}

We fitted the sCMTF model to each patient's data for varying number of sources (rank), i.e., $R=1\ldots 6$.
For each choice of $R$, we ran the optimization procedure 50 times, as explained in \ref{apx:opt_init}. 
Afterwards, the results need to be aggregated, such that clear conclusions on the sources of interest can be drawn.
This involves several steps, which we explain below in chronological order.

\subsection{Sign and scale standardization of factor estimates}\label{apx:std_ambiguity}
Some of the factors that are estimated via minimization of \eqref{eq:costfunction} are subject to sign and scale ambiguities, which are inevitable in many BSS contexts \citep{sidiropoulos2017tensor,kolda2009tensor}.
In the EEG factor model in \eqref{eq:cmtf_eegrank1}, the factor vectors $\vect{s}_r$, $\vect{g}_r$ and $\vect{m}_r$ belonging to the same component $r$ may be multiplied with arbitrary scaling factors whose product is one, without altering the goodness of fit. 
Similarly, in the fMRI factor model in \eqref{eq:cmtf_fmrirank1}, sign and scale are exchangeable between corresponding columns of $\matr{S}$ and $\matr{V}$, and between rows of $\matr{V}$ and rows of $\matr{B}$.
However, to conduct proper statistical inference on each source $r$'s spatial amplitudes, the elements in $\vect{v}_r$ must be calibrated, in the sense that it must be possible to compare them across ROIs, and determining the sign is crucial to distinguish local activation from deactivation.
Hence, we sequentially fix these ambiguities as follows: 
\begin{enumerate}
	\item For every component $r$:\begin{enumerate}
		\item $\vect{s}_r$ and $\vect{g}_r$ are rescaled to unit $\ell_2$-norm, and $\vect{m}_r$ is counterscaled
		\item for both $\vect{s}_r$ and $\vect{g}_r$, the sign is flipped if the sum of squares of the negative elements exceeds that of the positive elements; the sign of $\vect{m}_r$ is adjusted to preserve the global sign of the EEG rank-1 term
		\item if the sign of $\vect{s}_r$ was flipped, also the sign of $\vect{v}_r$ is flipped to preserve the global sign of the $r$-th fMRI block term
	\end{enumerate}
	\item For every ROI $i_v$:\begin{enumerate}
		\item the local HRF $\vect{h}_{i_v}$ is reconstructed using \eqref{eq:convolution_sumofKoperator2}
		\item the $i_v$-th row of $\matr{B}$ is rescaled and sign-corrected as to make $\vect{h}_k$ unit $\ell_1$-norm and as to ensure that the HRF's largest overshoot precedes the largest undershoot; the $i_v$-th row of $\matr{V}$ is counterscaled
	\end{enumerate}
\end{enumerate}

\subsection{Stability analysis}\label{apx:std_stability}
To assess the reproducibility of the factors, we use the graph-structured clustering algorithm that we proposed in \citep{vaneyndhoven2019identifying} and briefly summarize here.
We represent the factor sets for all 50 repetitions of the fitting procedure as $\left\llbracket\matr{S},\matr{G},\matr{M},\matr{V}\right\rrbracket$, and use a threshold of 0.85 to construct a binary link matrix that encodes similarities between components from different runs of the optimization (empirically, we found that this threshold led to acceptable cluster definitions in this context).
Via low-rank matrix approximation of this link matrix, we then obtained clusters of components that were encountered in varying numbers of repetitions.
High cardinality of a cluster is then a sign that the involved component is very reproducible or `stable', since it is part of the factor set upon convergence in many repetitions.
We suggest to assign higher trust in such components, as opposed to components in small clusters, which are likely specific to one (potentially poor) local minimum.
For the further steps, we condensed each cluster to one of its components, i.e., its centroid.
In each cluster, the centroid component is defined as the component which has the largest accumulated similarity with all other components in the cluster\footnote{By extension, the centroid repetition is defined as the repetition to which the centroid component belongs.}. 

\subsection{IED component selection}\label{apx:std_iedselection}
Out of the centroids of the clustered components, we identified the (most) IED-related component as the one whose temporal signature $\vect{s}_r$ was most correlated to a reference time course, which is constructed as the average over channels of the MWF's output signal's time-varying (broadband) power.
This reference is the BOLD predictor we have proposed for EEG-correlated fMRI analysis in \citep{vaneyndhoven2019semi}, and which provides a good baseline for identifying temporal dynamics that are timelocked to the IEDs.

\subsection{Choice of the rank $R$ of the factorization}\label{apx:std_Rselection}
After the previous steps have been carried out for each setting of $R$, we are left to select the rank \mbox{$R\in\{1,2,3,4,5,6\}$} whose set of results we proceed with.
We heuristically determined an appropriate value $\hat{R}$ by selecting the model which fulfills several criteria:
\begin{enumerate}
	\item \textit{high core consistency of the EEG decomposition}\\
	We compute the core consistency diagnostic (CorConDia) \citep{bro2003new} for the EEG tensor in combination with its estimated factor set $\left\llbracket\matr{S},\matr{G},\matr{M}\right\rrbracket$ from the centroid repetition. The consistency describes how suitable a rank-$R$ CPD is for the given tensorial data and given factors, and is expressed as a percentage (100\% being a very adequate model, and percentages below 70--80\% indicating that the model is not appropriate)\footnote{CorConDia is a popular and robust model selection tool for tensor decompositions \citep{morup2009automatic,acar2007multiway,miwakeichi2004decomposing,liu2016detection,papalexakis2016automatic}. To compute it, first the core tensor which is most appropriate (in minimum mean squared error sense) for the given data and CPD-derived factors is estimated. Subsequently, CorConDia is computed as the fraction of the core tensor's sum of squares which is due to off-superdiagonal elements. When for a given set of factor matrices the CP structure is indeed ideal, the core tensor is superdiagonal and CorConDia equals 100\%. Note that for a rank-1 model, this notion is meaningless, since the core tensor is a scalar, and CorConDia would trivially be 100\% always.}.
	\item \textit{reproducible IED-related component}\\
	We count the number of repetitions in the cluster that was most related to the IED (cfr. \ref{apx:std_stability} and \ref{apx:std_iedselection}), and used this as a measure of reproducibility. We rejected clusters whose cardinality was lower than 10.
	\item \textit{similarity to a reference IED time course}\\
	We track the correlation between the IED-related component's temporal signature $\vect{s}_{{}_{\text{IED}}}$ and the reference temporal signature $\vect{s}_{\text{ref}}$, as explained in \ref{apx:std_iedselection}. We expect higher correlations to signify a more suitable model, since $\vect{s}_{\text{ref}}$ generally led to good results in our previous study \citep{vaneyndhoven2019semi}.
	\item \textit{high significance in the IED-related spatial map}\\
	We track the highest pseudo t-value in the SnPM that was created based on the IED-related component's spatial signature $\vect{v}_{{}_{\text{IED}}}$ (cfr. Section \ref{subsec:method_inference}). A high statistical score indicates a good model fit for the IED-related component \citep{abreu2018eeg}.
\end{enumerate}

\begin{landscape}
	\begin{table}
		\caption{For every patient (ID 1--12), the sCMTF model can be fitted for a varying number of sources or rank $R$. We select a `good' value for $R$ post hoc, based on four criteria which are checked intra-patient: 1) the core consistency of the EEG tensor decomposition should be high ($> 70\%$); 2) the IED-related source should be found in sufficiently many ($\geqslant 10$) of the 50 repetitions of the estimation procedure; 3) the correlation of the IED-related source's temporal signature with the reference time course, namely the MWF's broadband envelope, is preferably high; 4) the maximal pseudo t-statistic for the IED-related source's spatial signature is preferably high.}
		\scalebox{0.65}{
			\begin{small}		
				\settowidth{\mycellwidth}{100}
				\renewcommand\arraystretch{2}
				\begin{tabular}{>{\raggedright}p{0.03\linewidth} >{\raggedleft}p{0.03\linewidth} *{5}{>{\raggedleft}p{0.035\linewidth}}>{\raggedleft\arraybackslash}p{0.035\linewidth}>{\centering}p{0.005\linewidth}*{5}{>{\raggedleft\arraybackslash}p{0.015\linewidth}}>{\raggedleft\arraybackslash}p{0.015\linewidth}>{\centering}p{0.005\linewidth} *{5}{>{\raggedleft}p{0.03\linewidth}}>{\raggedleft\arraybackslash}p{0.03\linewidth}>{\centering}p{0.005\linewidth} *{5}{>{\raggedleft}p{0.03\linewidth}}>{\raggedleft\arraybackslash}p{0.03\linewidth}>{\centering}p{0.005\linewidth} >{\raggedleft\arraybackslash}p{0.06\linewidth}}
					\toprule
					ID & & 
					\multicolumn{6}{c}{\normalsize core consistency diagnostic (\%)} & &
					\multicolumn{6}{c}{\normalsize\begin{tabular}{@{}c@{}}reproducibility\\ (\# repetitions in IED cluster)\end{tabular}} & & 
					\multicolumn{6}{c}{\normalsize\begin{tabular}{@{}c@{}}correlation of $\vect{s}_{{}_{\text{IED}}}$ with\\reference MWF envelope $\vect{s}_{\text{ref}}$\end{tabular}} & & 
					\multicolumn{6}{c}{\normalsize maximal t-statistic of $\vect{v}_{{}_{\text{IED}}}$} & & 
					{\normalsize\begin{tabular}{@{}c@{}}selected\\ rank \end{tabular}} 
					\\\cmidrule(lr){1-1}\cmidrule(lr){3-8}\cmidrule(lr){10-15}\cmidrule(lr){17-22}\cmidrule(lr){24-29}\cmidrule(lr){31-31} 
					& $R=$ & 1 & 2 & 3 & 4 & 5 & 6 & & 1 & 2 & 3 & 4 & 5 & 6 & & 1 & 2 & 3 & 4 & 5 & 6 & & 1 & 2 & 3 & 4 & 5 & 6 & & \\ \cmidrule(lr){3-8}\cmidrule(lr){10-15}\cmidrule(lr){17-22}\cmidrule(lr){24-29} 
					p01
					& & $-$ & 100.0 & 99.6 & 97.1 & 90.0 & 76.8 & & 25 & 19 & 14 & 7 & 16 & 20 & & 0.35 & 0.91 & 0.38 & 0.94 & 0.97 & 0.97 & & 14.2 & 20.0 & 21.5 & 16.0 & 7.7 & 20.1 & & $\hat{R}=6$\\
					p02
					& & $-$ & 100.0 & 94.9 & 58.0 & 19.6 & 7.2 & & 14 & 23 & 27 & 26 & 17 & 18 & & 0.10 & 0.93 & 0.95 & 0.83 & 0.96 & 0.88 & & 8.7 & 17.0 & 23.4 & 22.6 & 17.7 & 16.9 & & $\hat{R}=3$\\
					p03
					& & $-$ & 100.0 & 95.6 & 70.7 & 29.0 & 28.4 & & 29 & 23 & 24 & 21 & 23 & 19 & & 0.29 & 0.93 & 0.92 & 0.93 & 0.93 & 0.93 & & 15.2 & 16.5 & 14.9 & 14.3 & 13.9 & 15.0 & & $\hat{R}=2$\\
					p04
					& & $-$ & 100.0 & 87.0 & 74.4 & 37.3 & 12.3 & & 25 & 22 & 15 & 27 & 15 & 20 & & 0.09 & 0.20 & 0.61 & 0.71 & 0.68 & 0.81 & & 9.7 & 6.7 & 15.4 & 14.3 & 11.7 & 23.4 & & $\hat{R}=4$\\
					p05
					& & $-$ & 100.0 & 98.7 & 94.2 & 93.2 & 89.7 & & 3 & 14 & 14 & 21 & 13 & 17 & & 0.28 & 0.85 & 0.00 & 0.01 & 0.92 & 0.01 & & 19.6 & 5.5 & 6.5 & 4.4 & 9.0 & 4.6 & & $\hat{R}=5$\\
					p06
					& & $-$ & 100.0 & 94.6 & 80.9 & 33.8 & $-409$ & & 15 & 21 & 14 & 14 & 11 & 13 & & 0.55 & 0.92 & 0.87 & 0.80 & 0.80 & 0.22 & & 9.9 & 17.1 & 9.9 & 9.6 & 16.4 & 36.9 & & $\hat{R}=2$\\
					p07
					& & $-$ & 100.0 & 97.7 & 96.7 & $-76.0$ & $-0.6$ & & 18 & 17 & 12 & 15 & 11 & 12 & & 0.09 & 0.74 & 0.80 & 0.80 & 0.80 & 0.22 & & 7.4 & 10.0 & 9.4 & 8.6 & 11.2 & 9.9 & & $\hat{R}=4$\\
					p08
					& & $-$ & 100.0 & 99.7 & 30.6 & $-67.2$ & $-178$ & & 12 & 21 & 21 & 11 & 14 & 23 & & 0.61 & 0.95 & 0.44 & 0.95 & 0.94 & 0.94 & & 13.8 & 16.0 & 10.1 & 20.3 & 15.5 & 18.0 & & $\hat{R}=2$\\	
					p09
					& & $-$ & 100.0 & 98.8 & 95.9 & 90.1 & 23.6 & & 21 & 27 & 21 & 15 & 19 & 19 & & 0.19 & 0.66 & 0.67 & 0.13 & 0.48 & 0.49 & & 14.7 & 22.5 & 21.9 & 10.1 & 8.0 & 19.2 & & $\hat{R}=2$\\
					p10
					& & $-$ & 100.0 & 98.0 & 95.0 & 91.7 & 80.3 & & 23 & 34 & 13 & 24 & 25 & 14 & & 0.47 & 0.29 & 0.96 & 0.96 & 0.96 & 0.19 & & 11.9 & 12.9 & 10.0 & 8.6 & 9.9 & 12.8 & & $\hat{R}=5$\\
					p11
					& & $-$ & 100.0 & 97.9 & 78.3 & $-307$ & 49.1 & & 21 & 18 & 12 & 22 & 13 & 12 & & 0.65 & 0.91 & 0.50 & 0.74 & 0.66 & 0.67 & & 18.8 & 12.8 & 12.0 & 28.9 & 18.6 & 12.6 & & $\hat{R}=2$\\ 
					p12
					& & $-$ & 100.0 & 97.3 & 89.0 & 59.3 & 69.4 & & 23 & 14 & 14 & 12 & 15 & 12 & & 0.78 & 0.79 & 0.60 & 0.07 & 0.47 & 0.50 & & 15.8 & 15.0 & 10.9 & 7.9 & 10.2 & 5.5 & & $\hat{R}=2$\\\bottomrule
				\end{tabular}
			\end{small}
		}
		\label{tab:Rselection}
	\end{table}
\end{landscape}

%% file: sections/appendix-entropy.tex
\section{Computing HRF deviation metrics}\label{apx:hrfentropy}

\subsection{HRF extremity}\label{apx:ent_extremity}

The extremity of a specific ROI's HRF is computed as one minus the average of the absolute values of the Pearson correlation between the HRF waveform and all other ROIs' HRFs waveforms.
I.e., for the $j_v$-th ROI, the extremity is computed as
\begin{equation}
\text{extremity}(j_v) = 1 - \frac{1}{I_v-1}\sum_{i_v\neq j_v}^{}\lvert\text{corr}(h_{i_v},h_{j_v})\rvert
\end{equation}
Only the first twenty samples ($\sim$50 s) are considered. 
Note that the extremity does not change if the (global, not samplewise) sign (polarity) of one or more HRFs changes.

\subsection{HRF entropy}\label{apx:ent_entropy}
The entropy of a specific ROI's HRF is computed as the negative logarithm of the probability of this HRF, conditional on all other ROIs' HRFs.
For example, we first estimated a probability density in HRF space based on all other ROIs' HRFs, and then evaluated this density at the HRF of the ROI under inspection.
From every HRF, we considered the first twenty samples, and then estimated a nonparametric multivariate kernel density in 20-dimensional space, by placing a multivariate Gaussian probability kernel at the location of each HRF except one.
We made this entropy metric insensitive to the signs of the HRFs, by extending the set of HRF waveforms by their flipped counterparts, and computing the nonparametric density using the resulting $2(I_v-1)$ HRFs in a leave-one-ROI-out fashion.
\begin{equation}
\begin{split}
&\text{entropy}(j_v) = \\ &-\text{log}\left(
\frac{1}{2(I_v-1)}\sum_{i_v\neq j_v}^{}\left( K(h_{i_v},h_{j_v};\matr{\Sigma}) +  K(h_{i_v},-h_{j_v};\matr{\Sigma}) \right)\right)\;,
\end{split}
\end{equation}
in which $K(h_{i_v},h_{j_v};\matr{\Sigma})$ is a Gaussian kernel distance, which is proportional to
\begin{equation}
\text{exp}\left(-\frac{1}{2}(\vect{h}_{i_v}-\vect{h}_{j_v})\transp\matr{\Sigma}^{-1}(\vect{h}_{i_v}-\vect{h}_{j_v})\right)\;,
\end{equation}
in which $\vect{h}_{i_v}$ and $\vect{h}_{j_v}$ are column vectors that store the twenty first samples of the HRFs $h_{i_v}(t)$ and $h_{j_v}(t)$, and $\matr{\Sigma}$ is a diagonal covariance or bandwidth matrix.
We used Silverman's heuristic to set the kernel bandwidths for each individual dimension, corresponding to one HRF time sample \citep{silverman1986density}.
I.e., the $n$-th bandwidth $\sigma^2_{nn}$, which corresponds to the HRF amplitudes at sample $n$, is given by 
\begin{equation}
\sigma^2_{nn} = \left(\frac{4}{20+2}\right)^\frac{2}{20+4}\left(2(I_v-1)\right)^\frac{-2}{20+4}s^2_n\;,
\end{equation}
in which $s^2_n$ is the observed variance (over ROIs) of the HRFs' amplitudes at the $n$-th sample.

%% file: sections/suppfigures.tex
\section*{Supplement: sCMTF signatures for all patients}\label{sec:results}

We show the EEG spatial, temporal and spectral signatures, HRF waveforms, and fMRI spatial maps of significant IED (de)activation and HRF variability for all patients (except patients 3 and 10, whose results have been analyzed in the main text).

\subsection*{Patient 1}
\begin{figure*}
	\centering
	\begin{subfigure}[b]{1\textwidth}
		\includegraphics[width=1\linewidth]{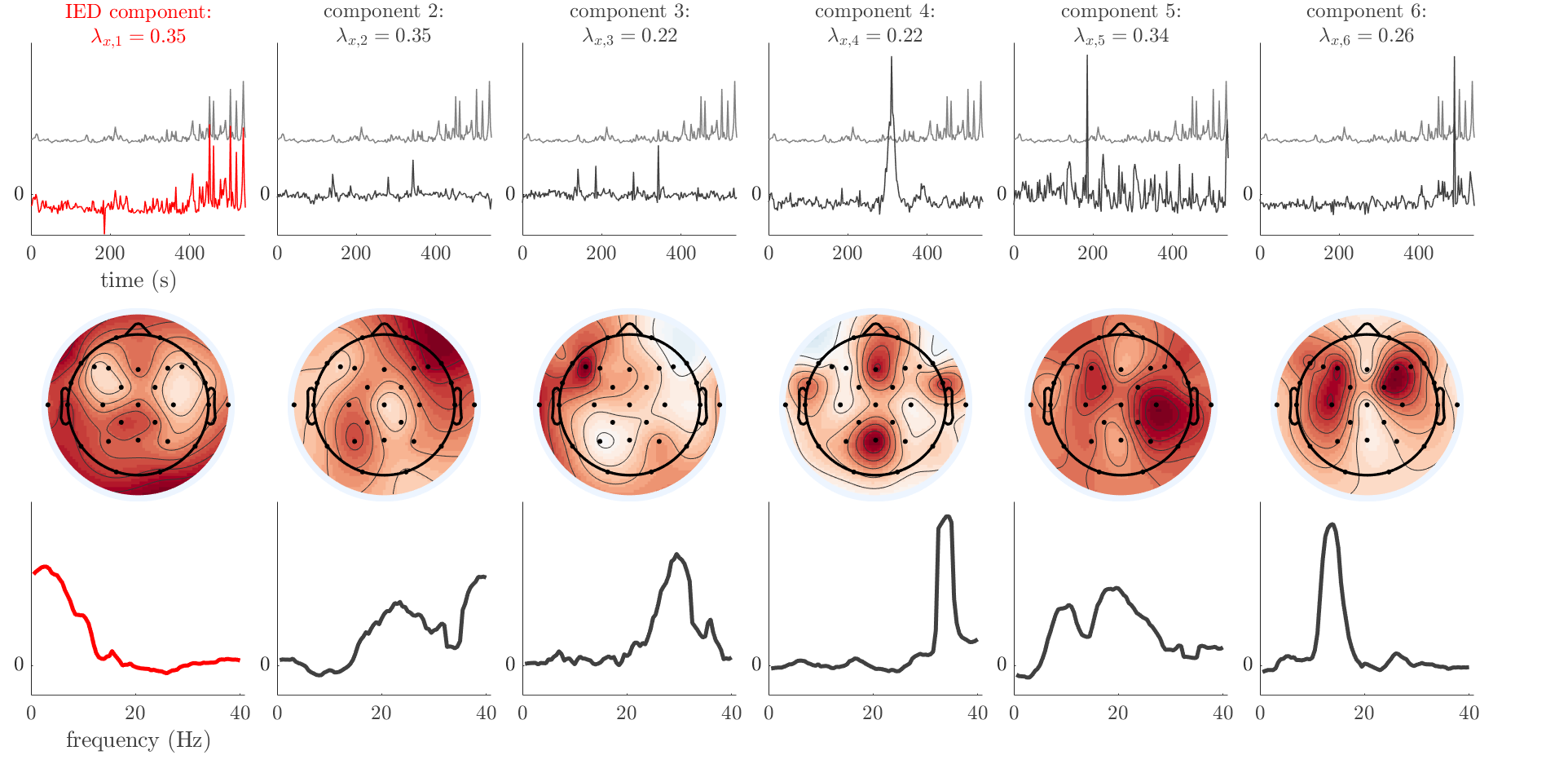}
		\caption{Temporal ($\vect{s}_r$, top), spatial ($\vect{m}_r$, middle), and spectral ($\vect{g}_r$, bottom) profiles of the 6 sources in the EEG domain, and reference IED time course ($\vect{s}_{\text{ref}}$, in grey).}
		\label{fig:p02eeg}
	\end{subfigure}
	
	\begin{subfigure}[b]{0.8\textwidth}
		\includegraphics[width=\textwidth]{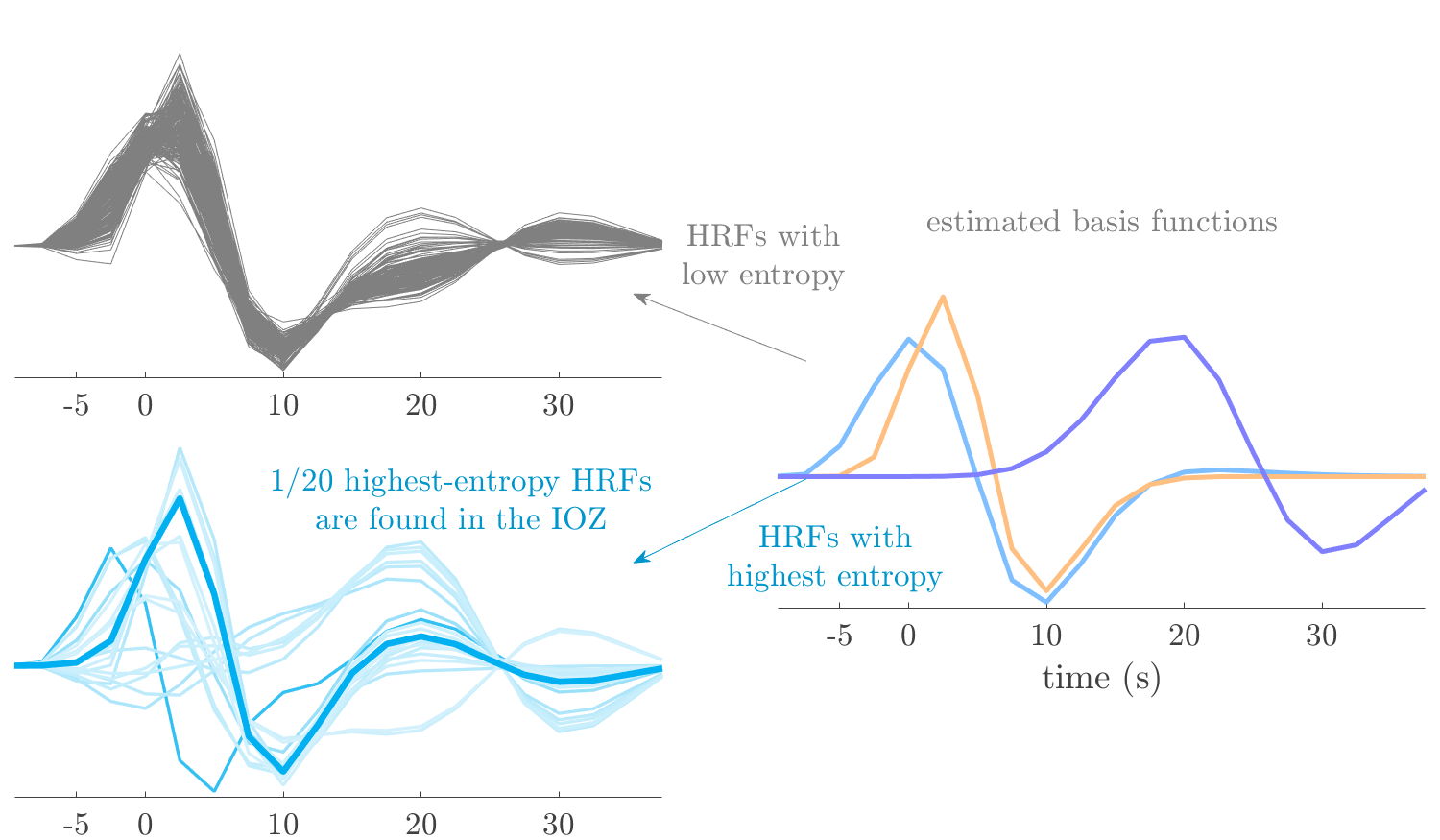}
		\caption{Estimated HRFs with low and high entropy}
		\label{fig:p02hrf}
	\end{subfigure}
	\caption{Patient 1's estimated sources and neurovascular coupling parameters. \subfigcap{a} The IED-related component correlates well with the reference time course, and is mostly a low-frequency phenomenon. The fourth source is most likely an artifact (of unknown origin), picked up mostly by two central channels.  \subfigcap{b} One of the ROIs with the highest-entropy HRFs belongs to the ictal onset zone (bold line, $p=0.34$).}
	\label{fig:p02eeghrf}
\end{figure*}

\begin{figure*}
	\centering
	\includegraphics[width=1\linewidth]{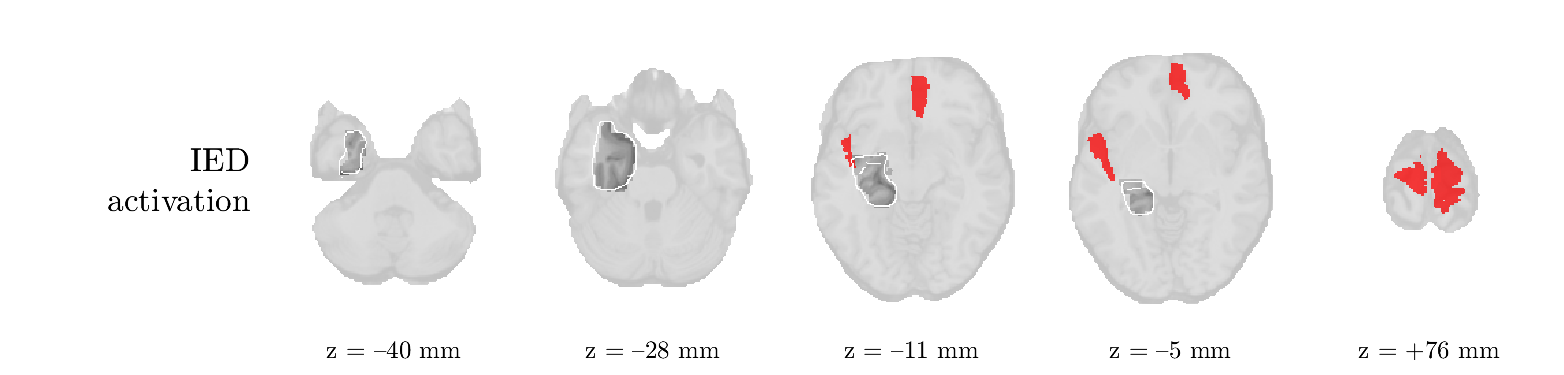}
	\includegraphics[width=1\linewidth]{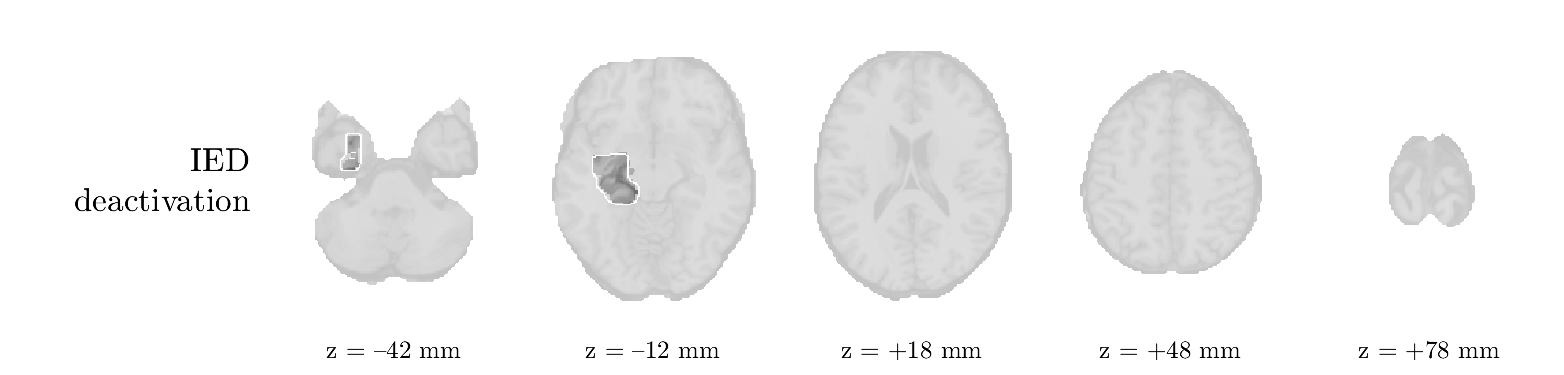}
	\includegraphics[width=1\linewidth]{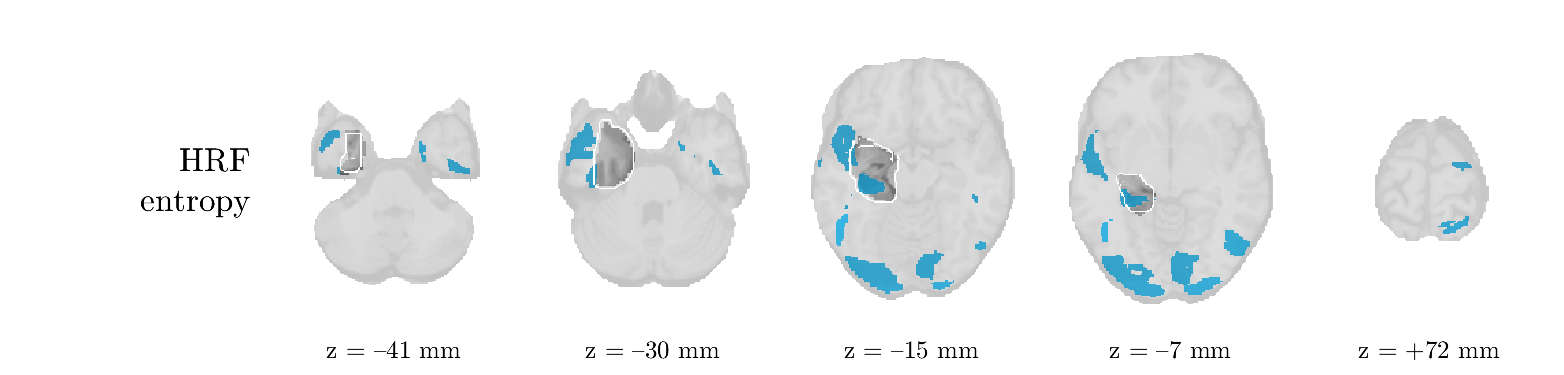}
	\includegraphics[width=1\linewidth]{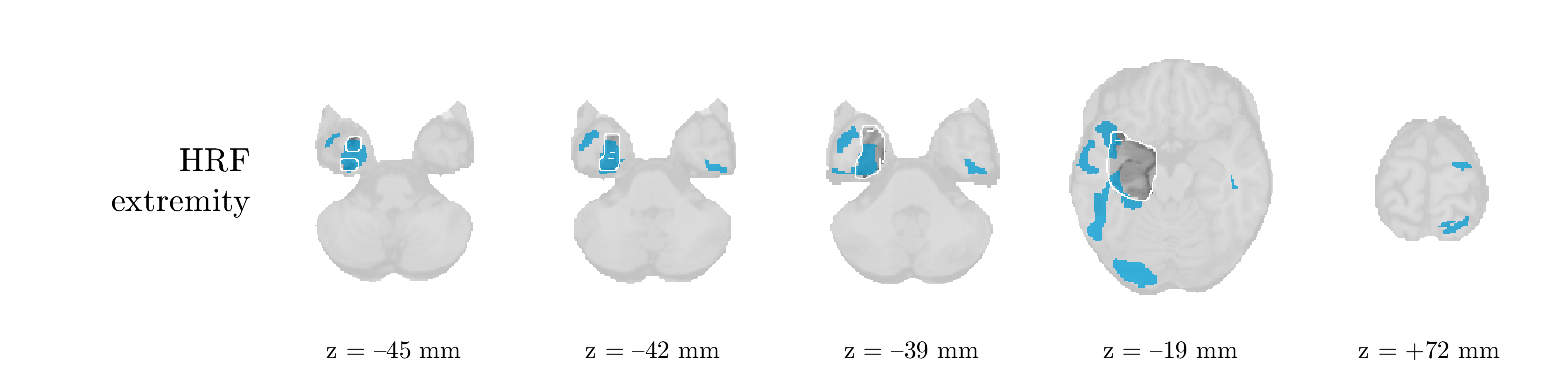}
	\caption{Patient 1's statistical nonparametric maps and HRF entropy/extremity maps. The ground truth ictal onset zone is highlighted in dark gray with a white contour.}
	\label{fig:p02fmri}
\end{figure*}

We analyze the solution with $\hat{R}=6$ sources.
Figure~\ref{fig:p02eeghrf} shows the EEG signatures and HRF waveforms.
One of the sources is highly correlated to the MWF reference (in grey), which was already known from Table~B.3. 
This IED-related source had a typical low-frequency spectrum, which is expected for the typical spike-and-wave interictal discharges. 
The topography is relatively diffuse, although the highest amplitudes are mostly in the left hemisphere. 
This is in accordance with the lateralization of ictal onset zone (left temporal lobe, cfr. Table~1).
There are some noteworthy observations to be made about some of the other components.
The fourth has an unusually sharp spectrum, is mainly localized on two nonadjacent center electrodes, and is sustained for a single period of many seconds 
Hence, this component likely captured an artifact (of yet unknown origin), although we spotted no large-amplitude changes in the EEG itself.
Similarly, the third source is only present at one frontal electrode, and exists in a frequency range above 20 Hz. 
It might represent a muscle artifact, e.g., due to frowning or twitching of some muscles in the forehead.
The HRFs of all ROIs are shown in Figure~\ref{fig:p02hrf}.
Two of the basis functions seem to have converged to a very similar waveform, which is an unfortunate possibility if two initial HRFs are too close to the same local optimum in their respective parameters.
This reduces the expressive power of the basis set, which is clearly visible, since many ROIs have a nearly identical HRF.
One of the twenty ROIs with the highest-entropy HRF overlapped the IOZ, although clearly this HRF (bold line) is not among the most dissimilar waveforms for this patient.
This is also visible in Figure~\ref{fig:p02fmri}: both the HRF entropy and extremity maps show a small overlap with the delineated IOZ.
Despite the good correspondence in the EEG domain, no significant (de)activation of the IED-component is found inside the IOZ.

\subsection*{Patient 2}
\begin{figure*}
	\centering
	\begin{subfigure}[b]{0.8\textwidth}
		\includegraphics[width=1\linewidth]{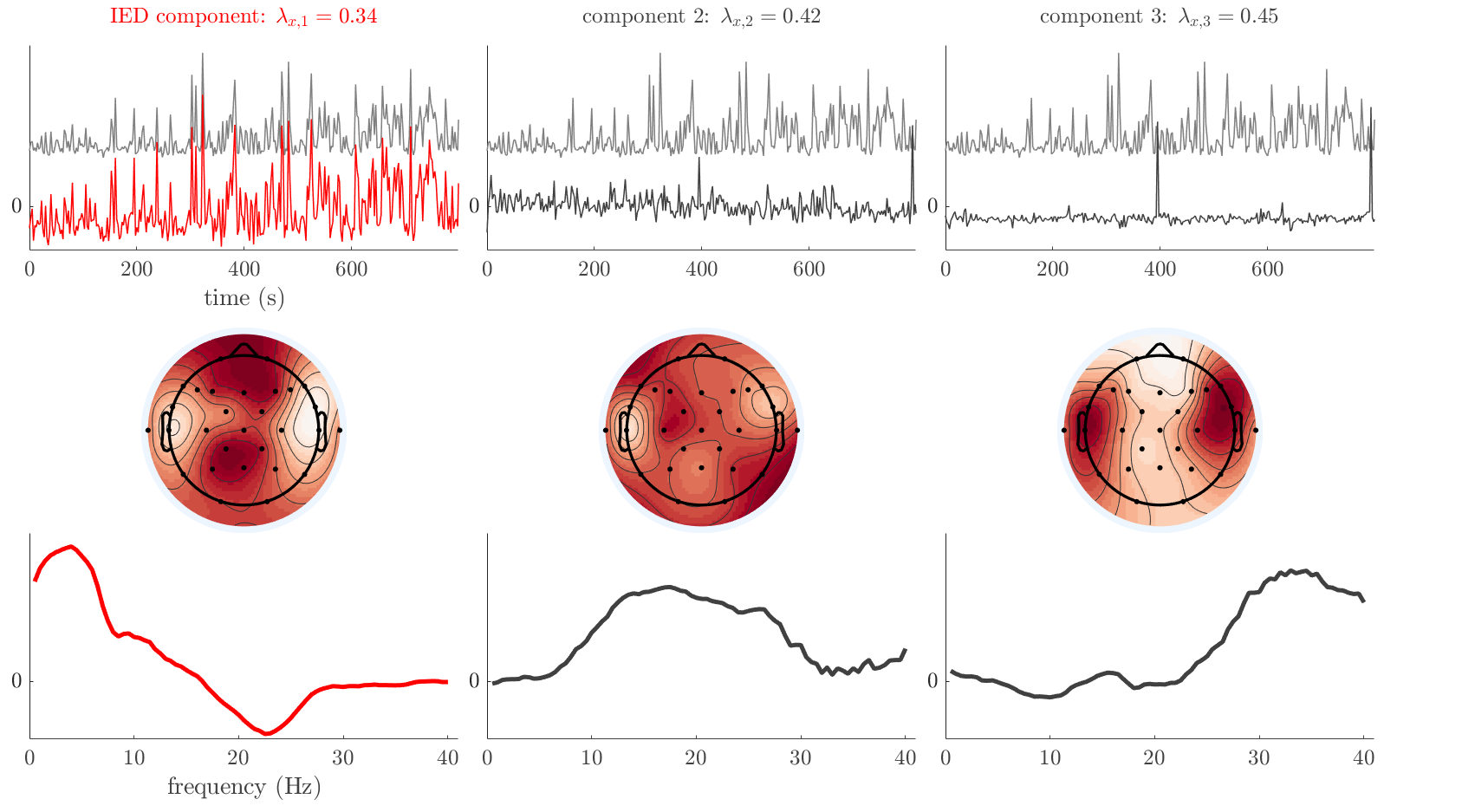}
		\caption{Temporal ($\vect{s}_r$, top), spatial ($\vect{m}_r$, middle), and spectral ($\vect{g}_r$, bottom) profiles of the 3 sources in the EEG domain, and reference IED time course ($\vect{s}_{\text{ref}}$, in grey).}
		\label{fig:p04eeg}
	\end{subfigure}
	
	\begin{subfigure}[b]{0.8\textwidth}
		\includegraphics[width=\textwidth]{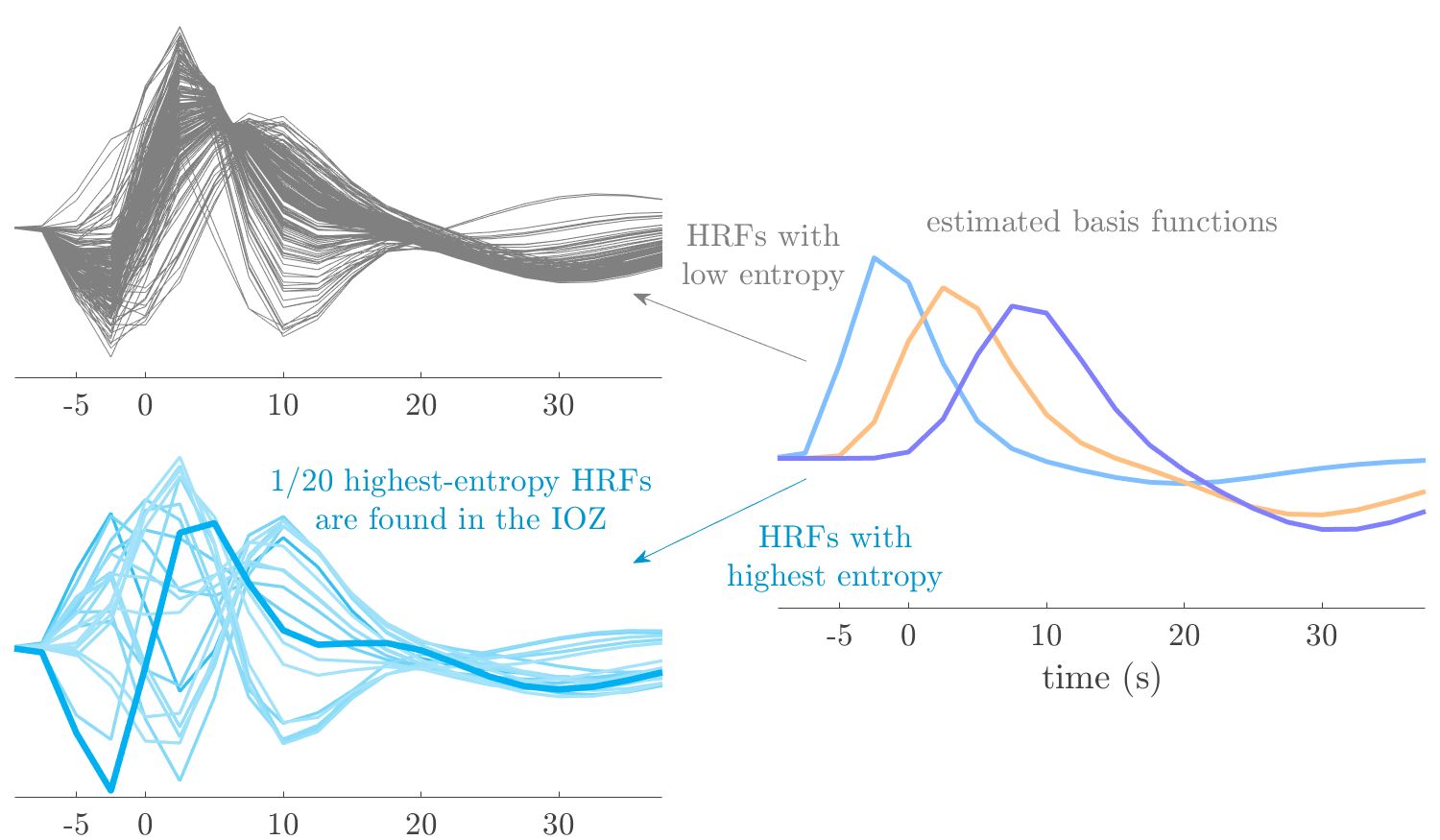}
		\caption{Estimated HRFs with low and high entropy}
		\label{fig:p04hrf}
	\end{subfigure}
	\caption{Patient 2's estimated sources and neurovascular coupling parameters. \subfigcap{a} The IED-related component correlates well with the reference time course, and is mostly a low-frequency phenomenon. \subfigcap{b} One of the ROIs with the highest-entropy HRFs belongs to the ictal onset zone (bold line, $p=0.59$).}
	\label{fig:p04eeghrf}
\end{figure*}

\begin{figure*}
	\centering
	\includegraphics[width=1\linewidth]{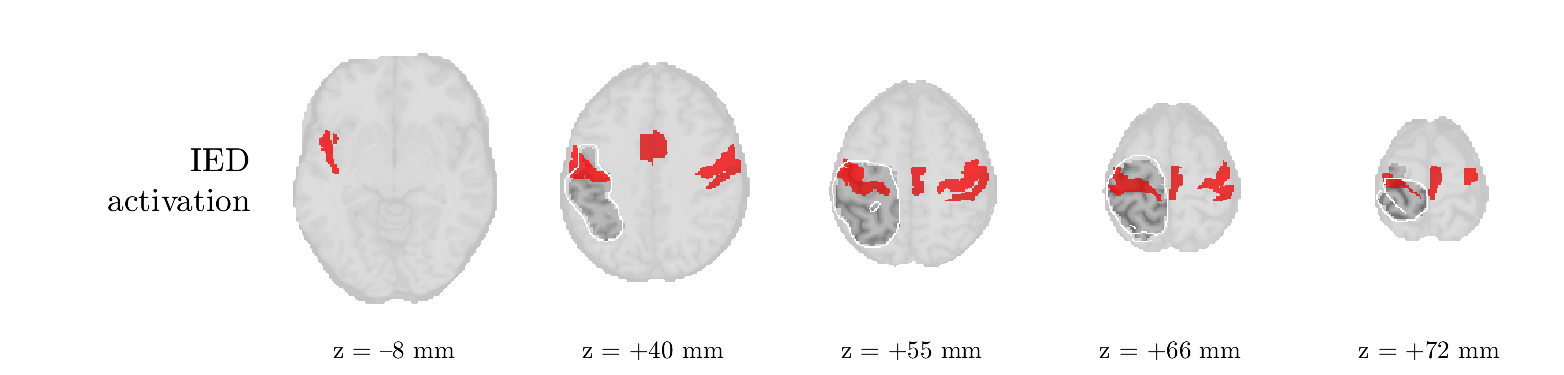}
	\includegraphics[width=1\linewidth]{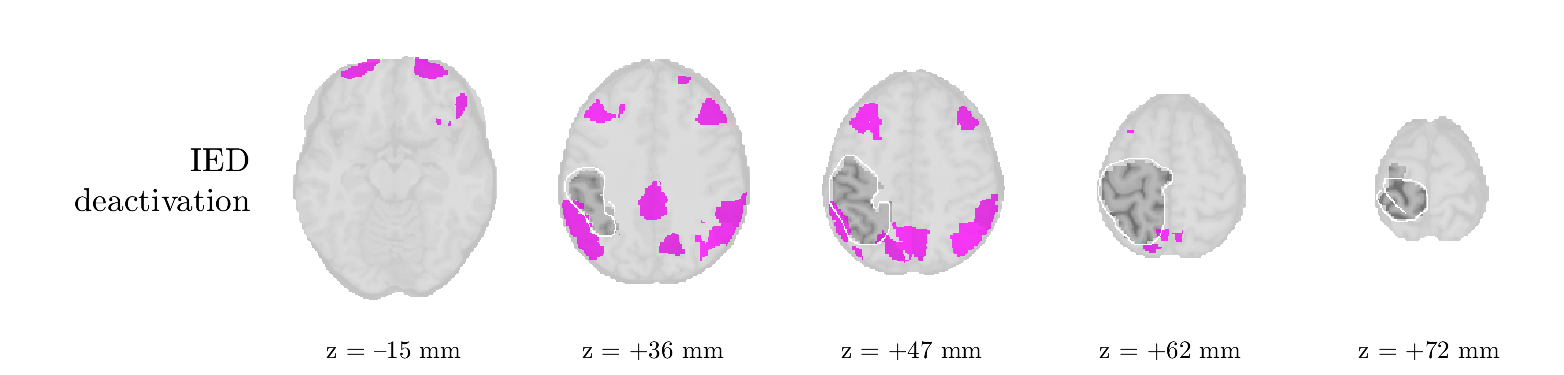}
	\includegraphics[width=1\linewidth]{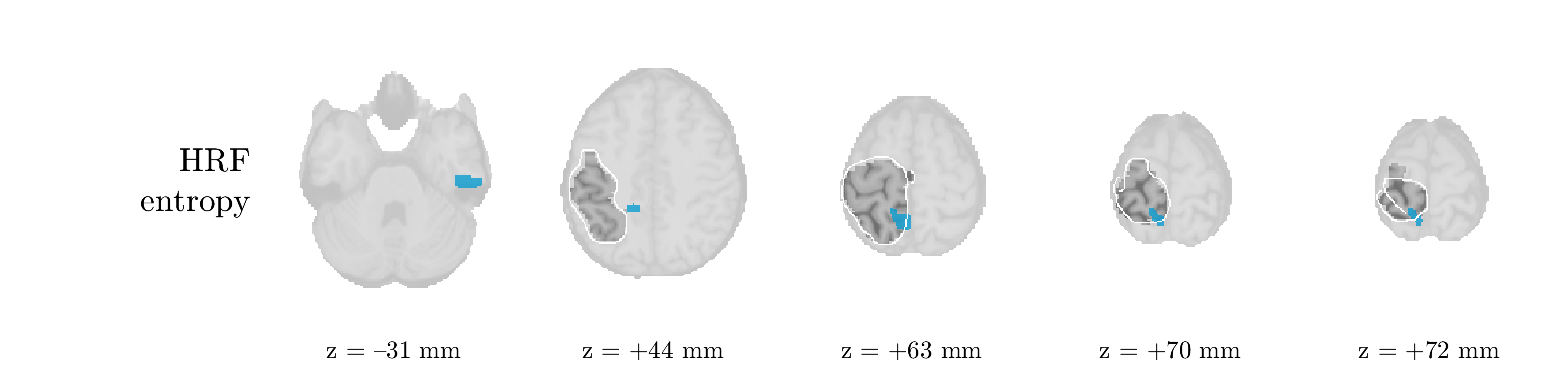}
	\includegraphics[width=1\linewidth]{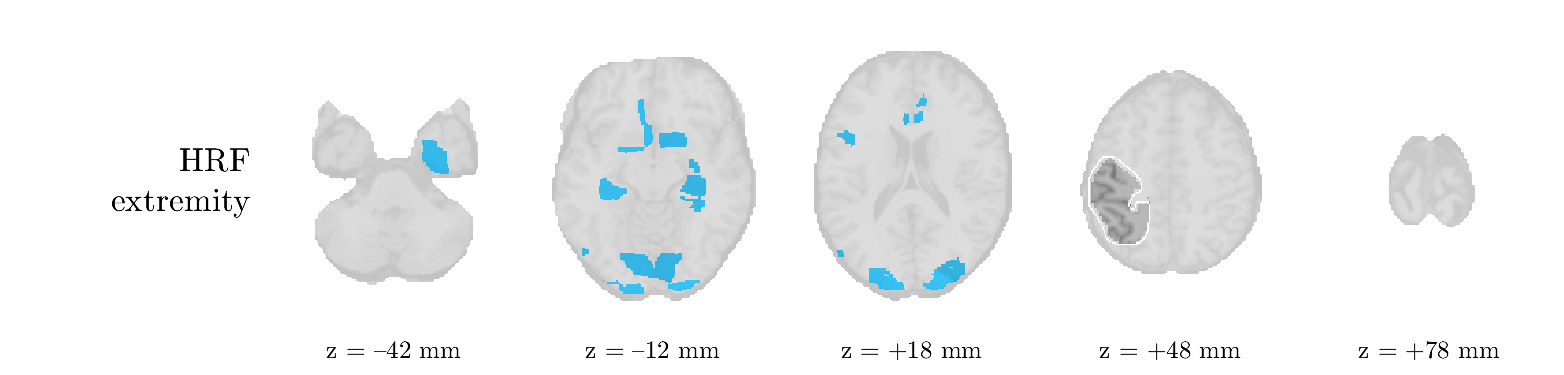}
	\caption{Patient 2's statistical nonparametric maps and HRF entropy/extremity maps. The ground truth ictal onset zone is highlighted in dark gray with a white contour.
	}
	\label{fig:p04fmri}
\end{figure*}

We analyze the solution with $\hat{R}=3$ sources, and show the results in Figure~\ref{fig:p04eeghrf} and \ref{fig:p04fmri}.
As for patient 1, we found a source which is strongly correlated to the MWF envelope, and which had a mostly low-frequency behavior characteristic for spikes.
The topography is mostly uninformative, and does not clearly correspond to the patient's clinical data.
The third source is mostly present at both sides of the head, is very sparsely active in time, and has a high-frequency content: this is most likely an artifact due to the neck muscles.
Again, there is one of the highest-entropy HRFs which belongs to a ROI in the IOZ. 
Now, the waveform is clearly resolved from the other HRFs, through the strong initial dip (before 0 seconds). 
Such a dip is sometimes observed in HRFs, but its underlying physiological mechanism is not yet fully understood.
It is possible that this dip reflects altered vascular autoregulation near the IOZ (cfr. the explanation in the Section~1 of the main text), or a rapid depletion in oxygen due to IED generation (before the IED becomes visible on the EEG).
Figure~\ref{fig:p04fmri} furthermore shows that the IED-related component is significantly active in parts of the IOZ, and deactive in others.
As mentioned earlier, this deactivation may or may not be due to errors in sign correction.
Interestingly, the ROI with the high alteration in neurovascular coupling is distinct from both the activated and deactivated ROIs.

\subsection*{Patient 4}
\begin{figure*}
	\centering
	\begin{subfigure}[b]{1\textwidth}
		\includegraphics[width=1\linewidth]{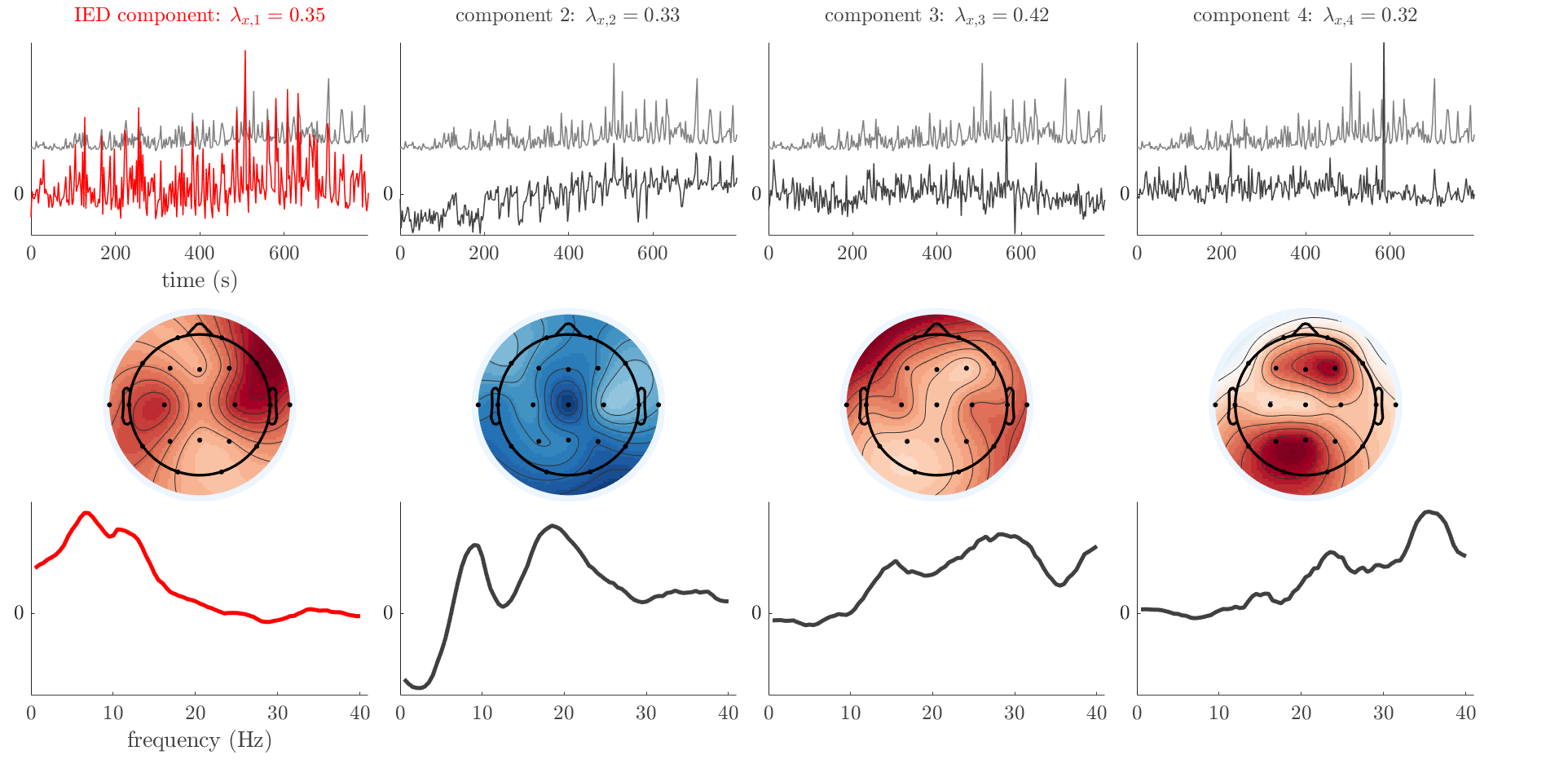}
		\caption{Temporal ($\vect{s}_r$, top), spatial ($\vect{m}_r$, middle), and spectral ($\vect{g}_r$, bottom) profiles of the 4 sources in the EEG domain, and reference IED time course ($\vect{s}_{\text{ref}}$, in grey).}
		\label{fig:p06eeg}
	\end{subfigure}
	
	\begin{subfigure}[b]{0.8\textwidth}
		\includegraphics[width=\textwidth]{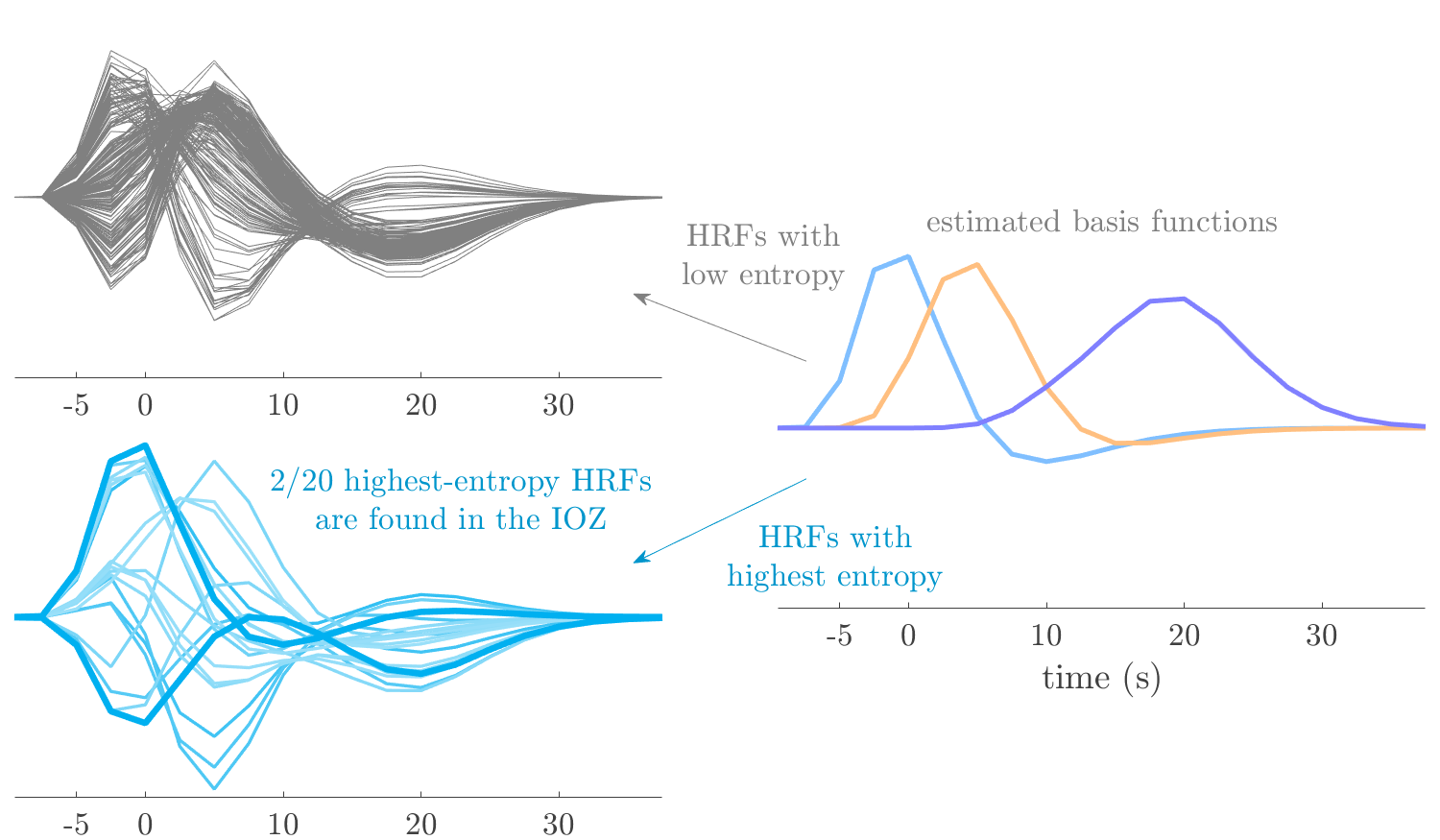}
		\caption{Estimated HRFs with low and high entropy}
		\label{fig:p06hrf}
	\end{subfigure}
	\caption{Patient 4's estimated sources and neurovascular coupling parameters. \subfigcap{b} Two of the ROIs with the highest-entropy HRFs belong to the ictal onset zone (bold line, $p=0.32$).}
	\label{fig:p06eeghrf}
\end{figure*}

\begin{figure*}
	\centering
	\includegraphics[width=1\linewidth]{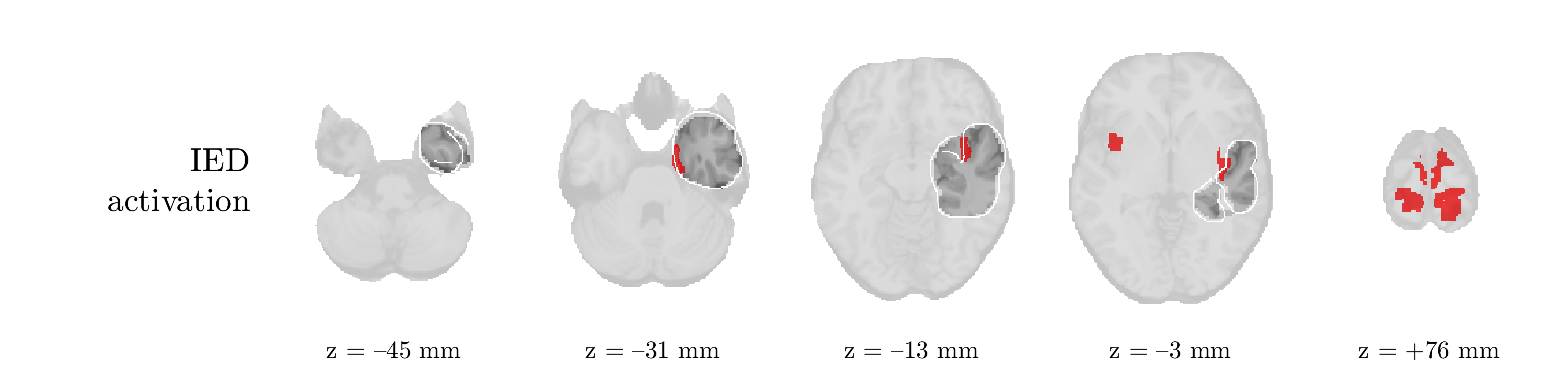}
	\includegraphics[width=1\linewidth]{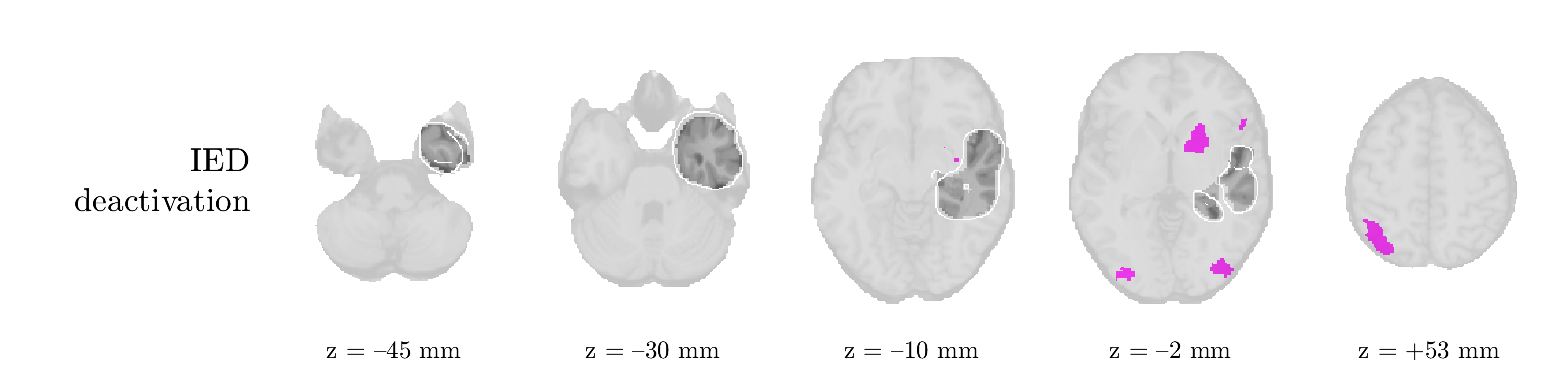}
	\includegraphics[width=1\linewidth]{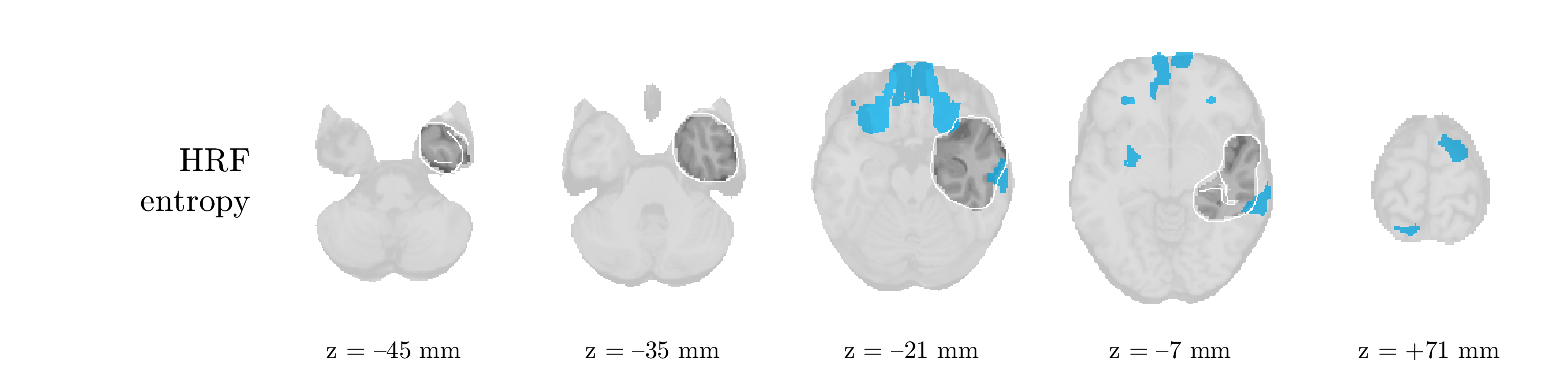}
	\includegraphics[width=1\linewidth]{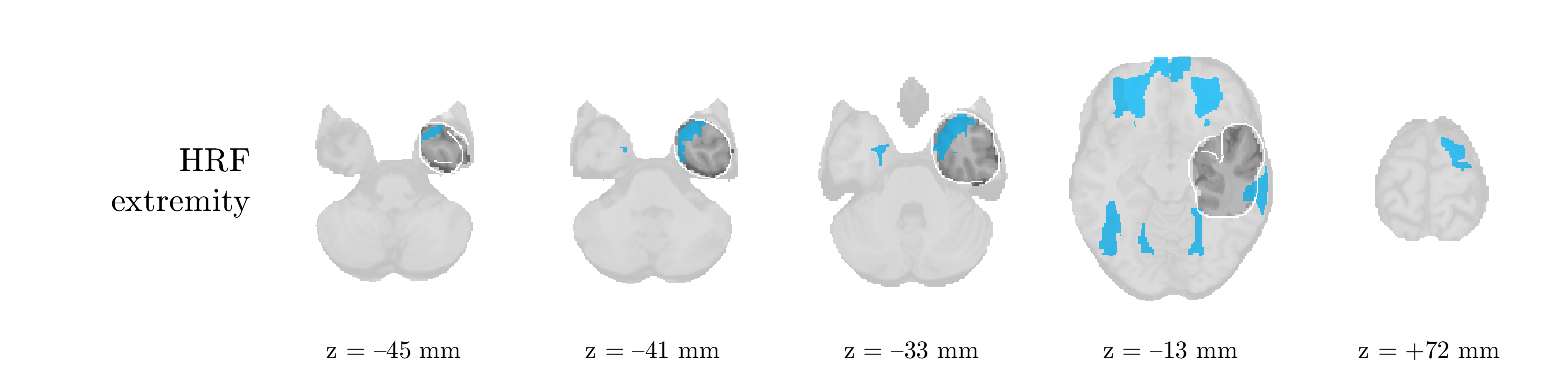}
	\caption{Patient 4's statistical nonparametric maps and HRF entropy/extremity maps. The ground truth ictal onset zone is highlighted in dark gray with a white contour.
	}
	\label{fig:p06fmri}
\end{figure*}

We analyzed the solution with $\hat{R}=4$ sources, and show the results in Figure~\ref{fig:p06eeghrf} and \ref{fig:p06fmri}.
There is one source which is mostly correlated to the reference (but not extremely, see also Table~B.3).
This source had a right-temporal focus, conform the diagnosis in Table~1.
The second source illustrates the phenomenon of an erroneous sign exchange between the spatial and spectral profiles.
Also one of the HRFs has a negative polarity, which is a failure of the sign correction procedure (in this case, because there is exceptionally no positive overshoot).
However, the HRF variability metrics are still interpretable, and indeed two ROIs among the ones with the highest-entropy HRFs overlap with the IOZ. 
The IED component is significantly active in a tiny portion of the IOZ (cfr. Figure~\ref{fig:p06fmri}). 
The second source is significantly active in symmetrical parts of the parietal lobe.
Given its ongoing fluctuation over time, we hypothesize that this source captures a resting state network (RSN).

\subsection*{Patient 5}
\begin{figure*}
	\centering
	\begin{subfigure}[b]{1\textwidth}
		\includegraphics[width=1\linewidth]{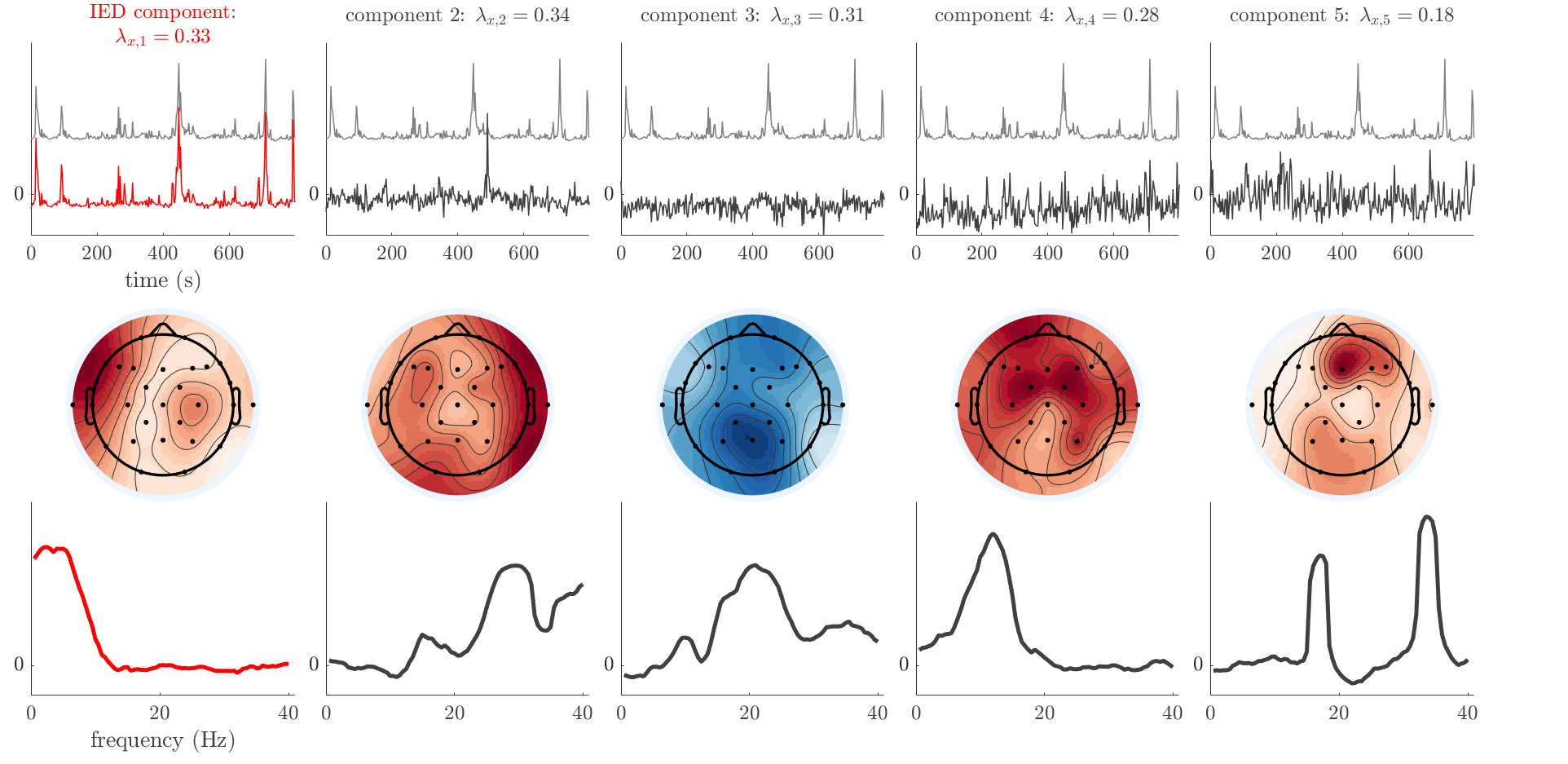}
		\caption{Temporal ($\vect{s}_r$, top), spatial ($\vect{m}_r$, middle), and spectral ($\vect{g}_r$, bottom) profiles of the 5 sources in the EEG domain, and reference IED time course ($\vect{s}_{\text{ref}}$, in grey).}
		\label{fig:p07eeg}
	\end{subfigure}
	
	\begin{subfigure}[b]{0.8\textwidth}
		\includegraphics[width=\textwidth]{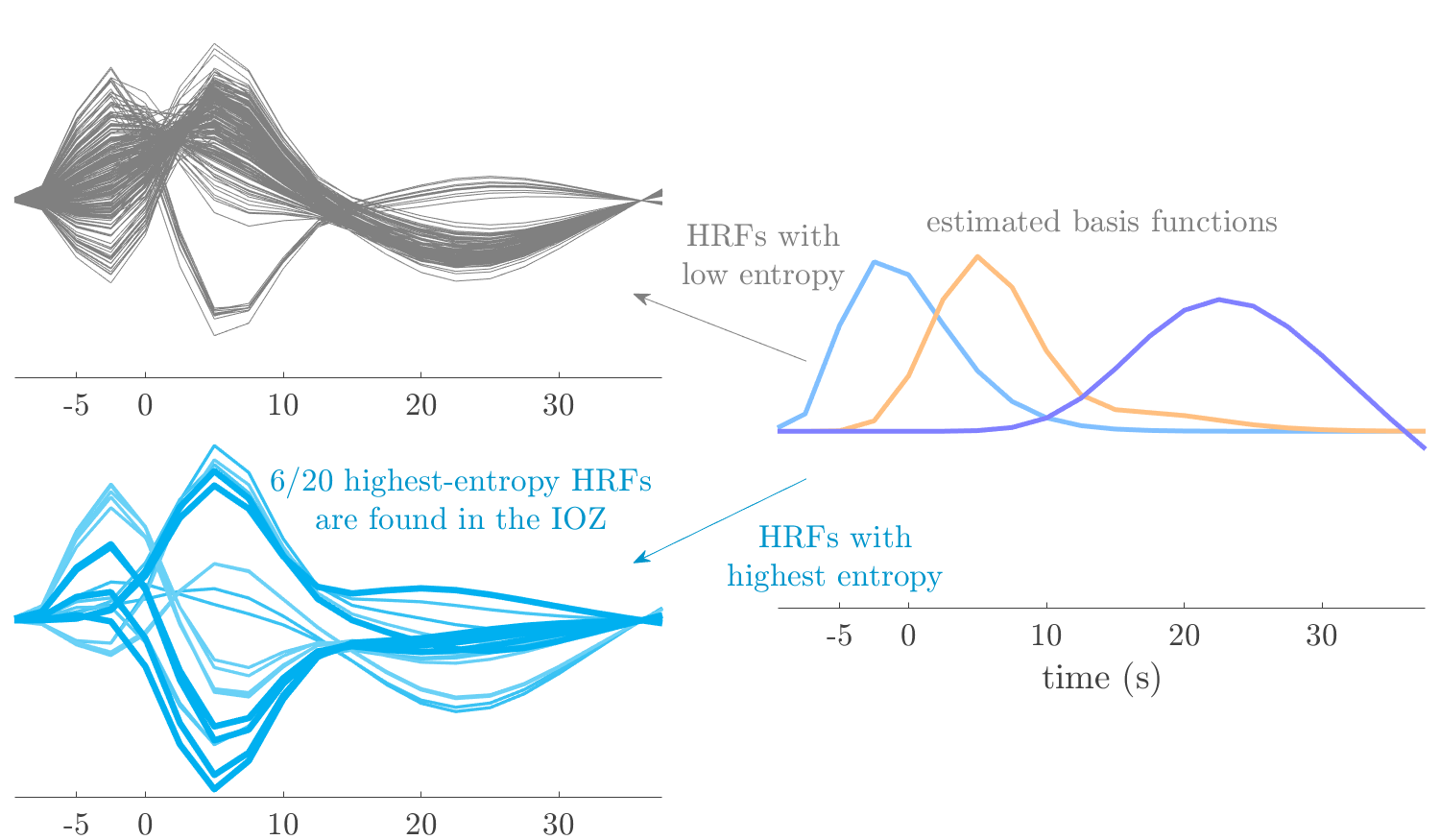}
		\caption{Estimated HRFs with low and high entropy}
		\label{fig:p07hrf}
	\end{subfigure}
	\caption{Patient 5's estimated sources and neurovascular coupling parameters. \subfigcap{b} Six of the ROIs with the highest-entropy HRFs belong to the ictal onset zone (bold line, $p<10^{-3}$).}
	\label{fig:p07eeghrf}
\end{figure*}

\begin{figure*}
	\centering
	\includegraphics[width=1\linewidth]{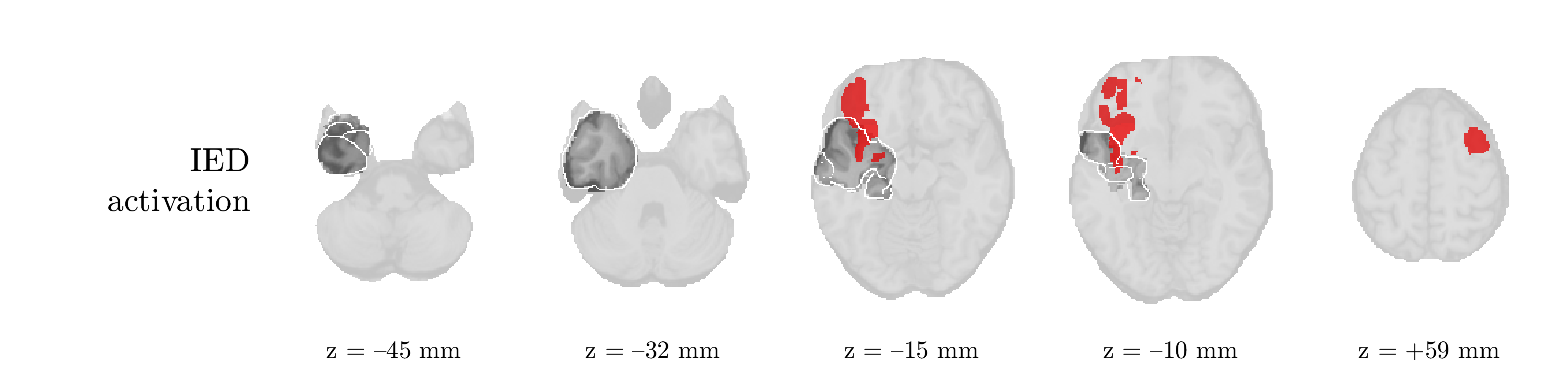}
	\includegraphics[width=1\linewidth]{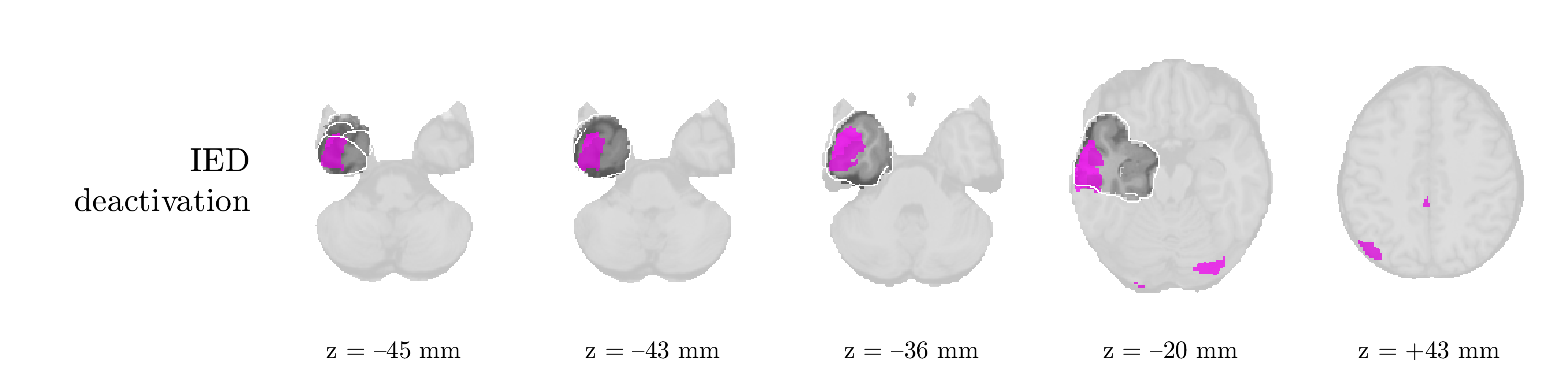}
	\includegraphics[width=1\linewidth]{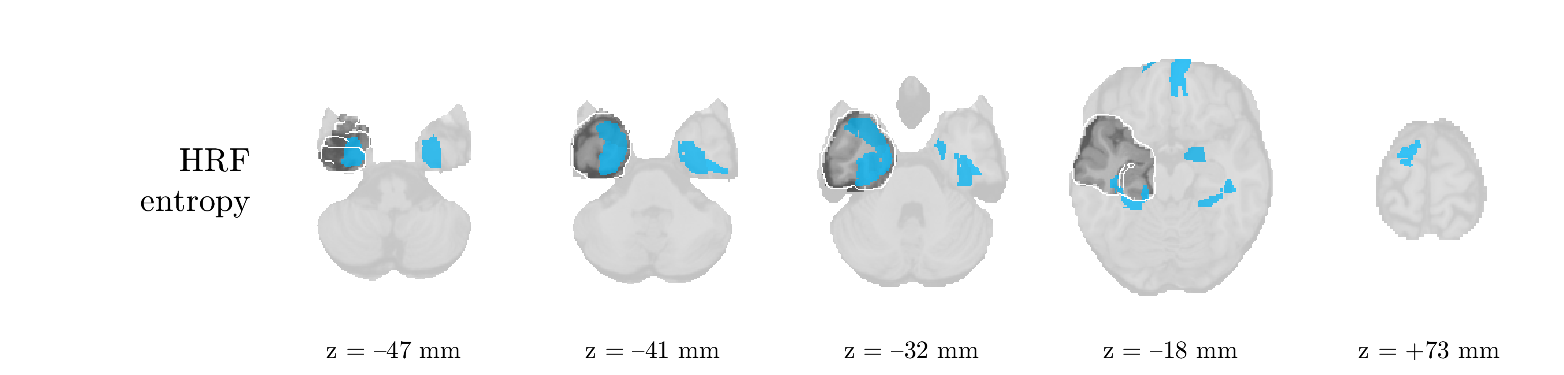}
	\includegraphics[width=1\linewidth]{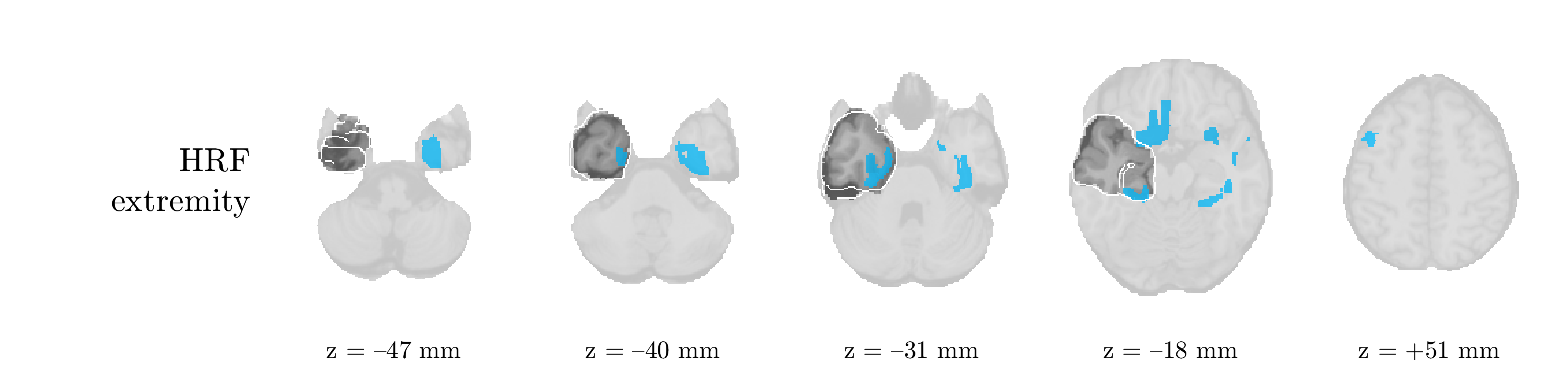}
	\caption{Patient 5's statistical nonparametric maps and HRF entropy/extremity maps. The ground truth ictal onset zone is highlighted in dark gray with a white contour.
	}
	\label{fig:p07fmri}
\end{figure*}

We analyze the solution with $\hat{R}=5$ sources, and show the results in Figure~\ref{fig:p07eeghrf} and \ref{fig:p07fmri}.
There is a clear IED-related component, with a very high correlation to the MWF reference, a typical spectrum, and an anterior-temporal focus, which corresponds very well to the patient's diagnosis (cfr. Table~1).
The fifth source seems present at only one channel, and has spectral harmonic at $\pm17$ Hz and $\pm34$ Hz.
One of these peaks is reminiscent of the fourth component in patient 1.
As Figure~\ref{fig:p07fmri} shows, the HRF entropy and extremity prove to be strong biomarkers for the IOZ in this case, and also the significant IED activation and deactivation allow correct localization. 
In Figure~\ref{fig:p07eeghrf}, it is clear that some HRFs may still have the wrong sign, which means that the interpretation of `active' and `deactivated' is flipped in those ROIs. 
Hence, regions of significant deactivation are in fact significantly activated.
The fourth source had a significant overlap with the auditory RSN, and its spectrum reveals activity in the $\beta$ band.

\subsection*{Patient 6}
\begin{figure*}
	\centering
	\begin{subfigure}[b]{0.6\textwidth}
		\includegraphics[width=1\linewidth]{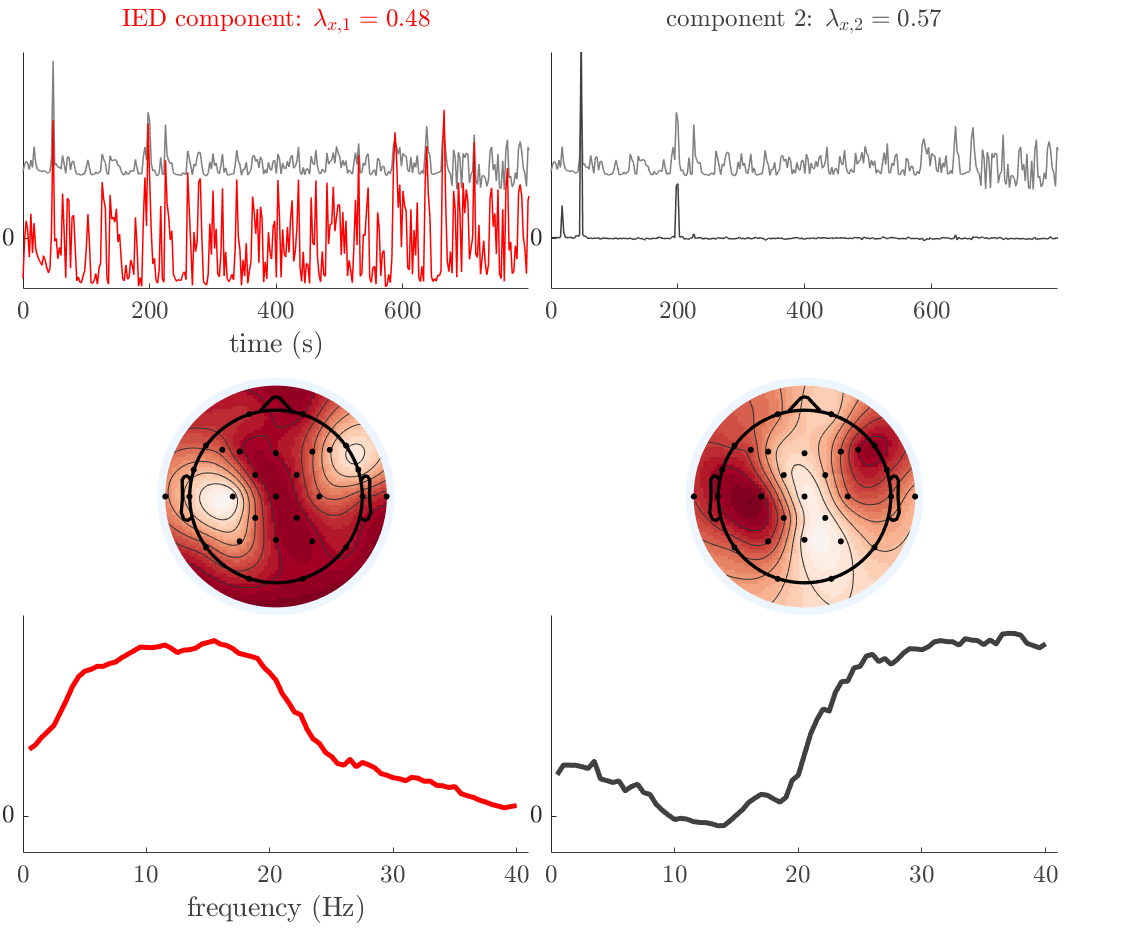}
		\caption{Temporal ($\vect{s}_r$, top), spatial ($\vect{m}_r$, middle), and spectral ($\vect{g}_r$, bottom) profiles of the 2 sources in the EEG domain, and reference IED time course ($\vect{s}_{\text{ref}}$, in grey).}
		\label{fig:p08eeg}
	\end{subfigure}
	
	\begin{subfigure}[b]{0.8\textwidth}
		\includegraphics[width=\textwidth]{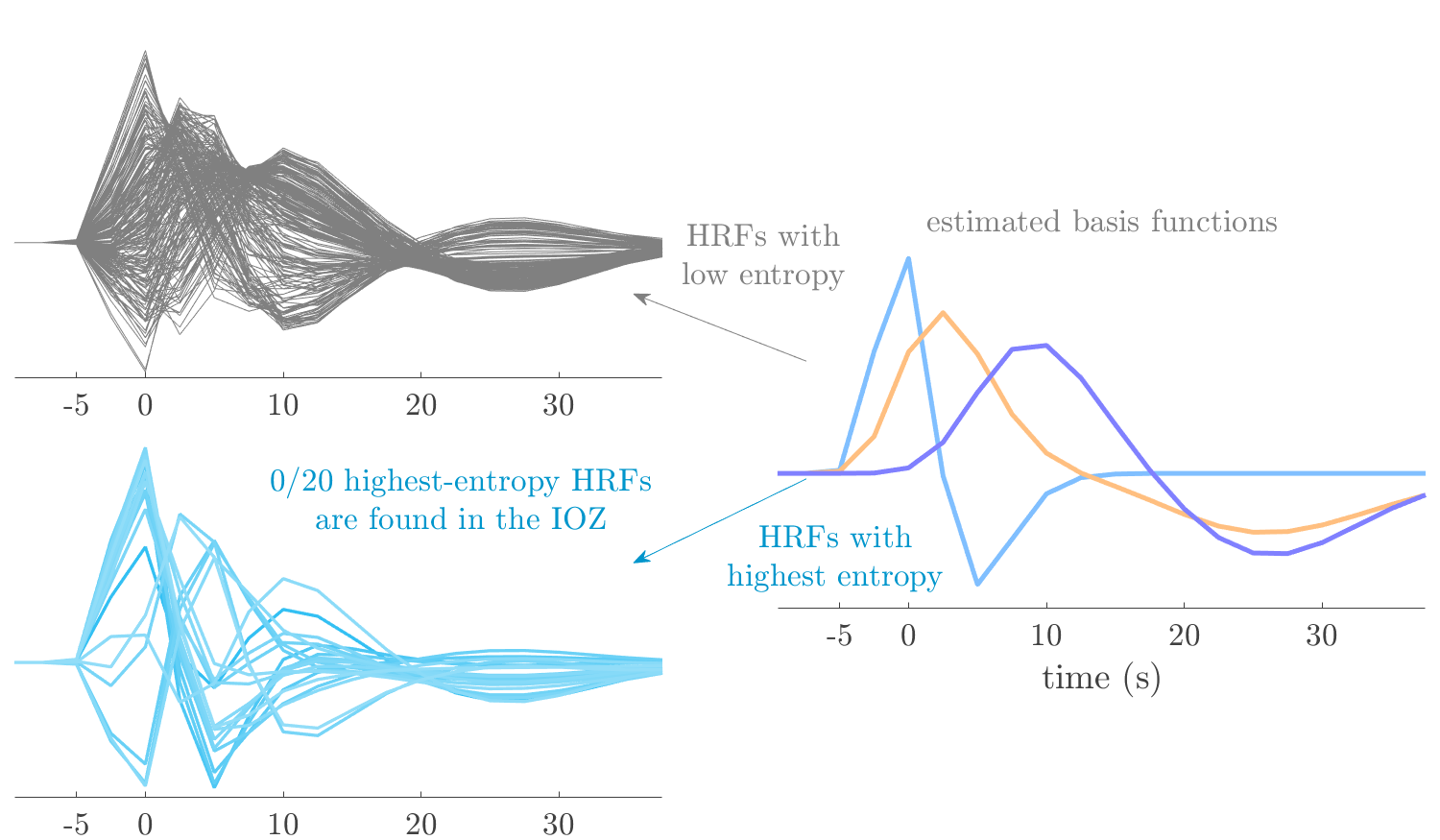}
		\caption{Estimated HRFs with low and high entropy}
		\label{fig:p08hrf}
	\end{subfigure}
	\caption{Patient 6's estimated sources and neurovascular coupling parameters. \subfigcap{b} None of the ROIs with the highest-entropy HRFs belong to the ictal onset zone.}
	\label{fig:p08eeghrf}
\end{figure*}

\begin{figure*}
	\centering
	\includegraphics[width=1\linewidth]{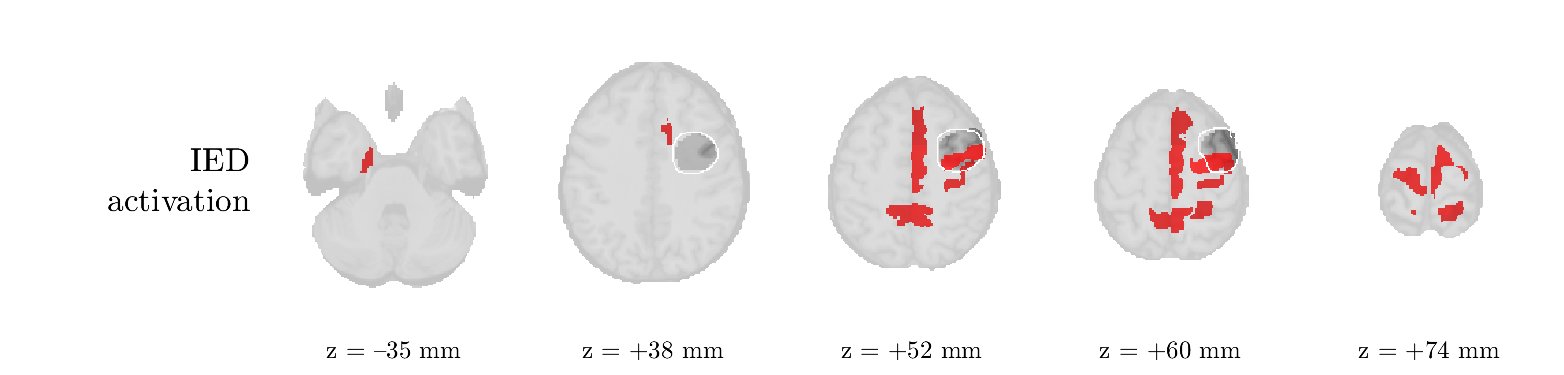}
	\includegraphics[width=1\linewidth]{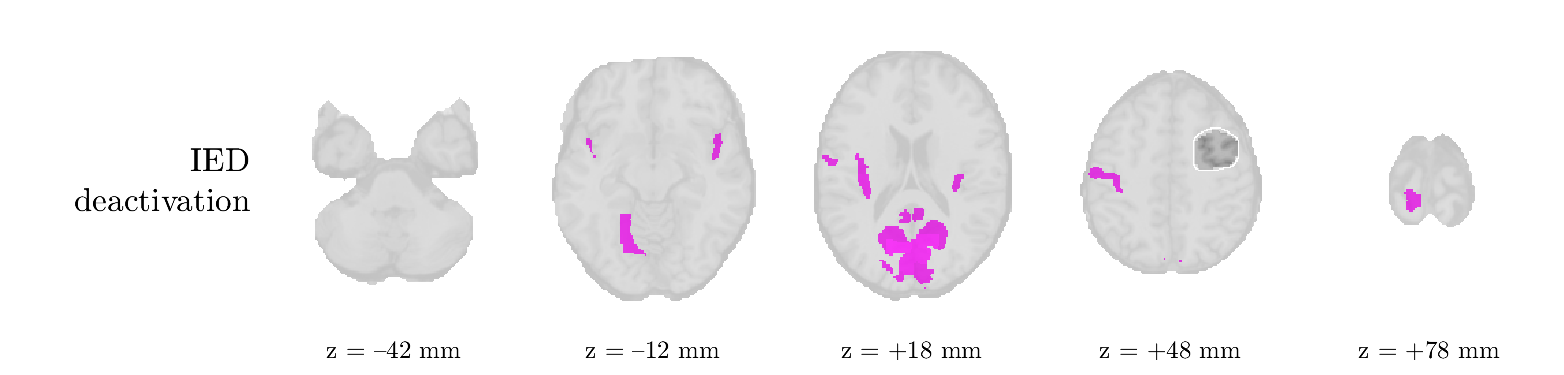}
	\includegraphics[width=1\linewidth]{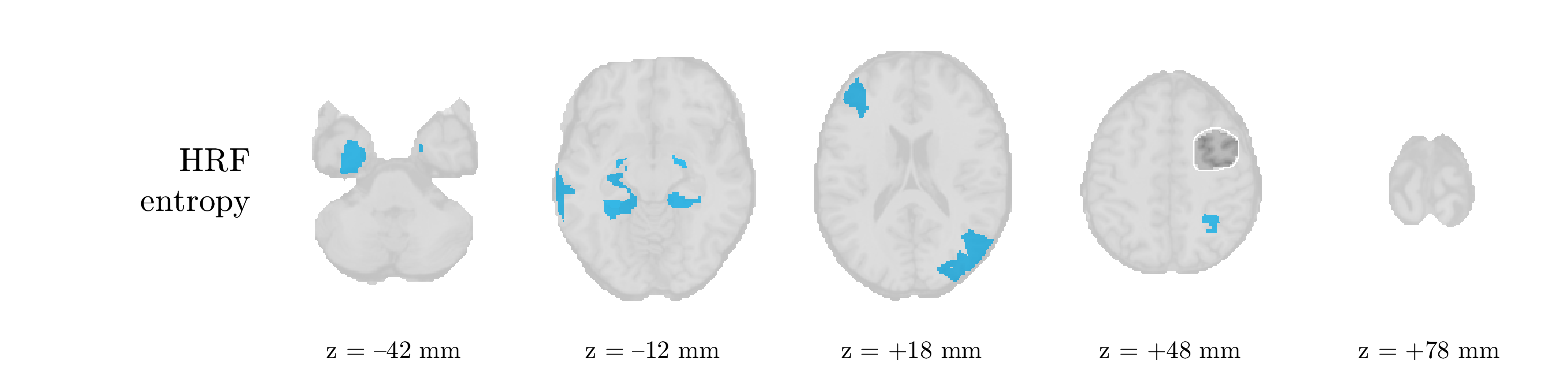}
	\includegraphics[width=1\linewidth]{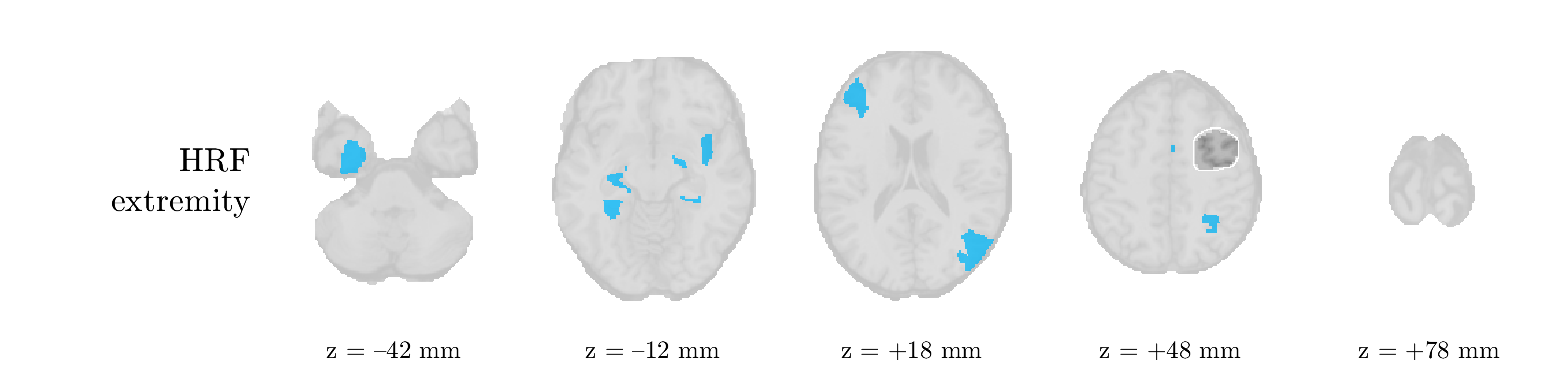}
	\caption{Patient 6's statistical nonparametric maps and HRF entropy/extremity maps. The ground truth ictal onset zone is highlighted in dark gray with a white contour.
	}
	\label{fig:p08fmri}
\end{figure*}

We analyze the solution with $\hat{R}=2$ sources, and show the results in Figure~\ref{fig:p08eeghrf} and \ref{fig:p08fmri}.
One source is strongly correlated to the MWF, while the other source is likely an artifact, given its very sparse temporal profile.
Both sources coincide at one high-amplitude peak, by which we infer that this is probably an artifactual period in the signal.
Indeed, when inspecting the original EEG signals, we found high-frequency muscle artifacts at these times.
This source also had no significant activation in its spatial map, which corroborates its non-neuronal origin.
The IED-related source had a broader spectrum than most other cases, and an uninformative topography.
None of the ROIs with high-entropy HRFs is located in the IOZ.
The pseudo t-map provides correct localization of the IOZ, however.

\subsection*{Patient 7}
\begin{figure*}
	\centering
	\begin{subfigure}[b]{1\textwidth}
		\includegraphics[width=1\linewidth]{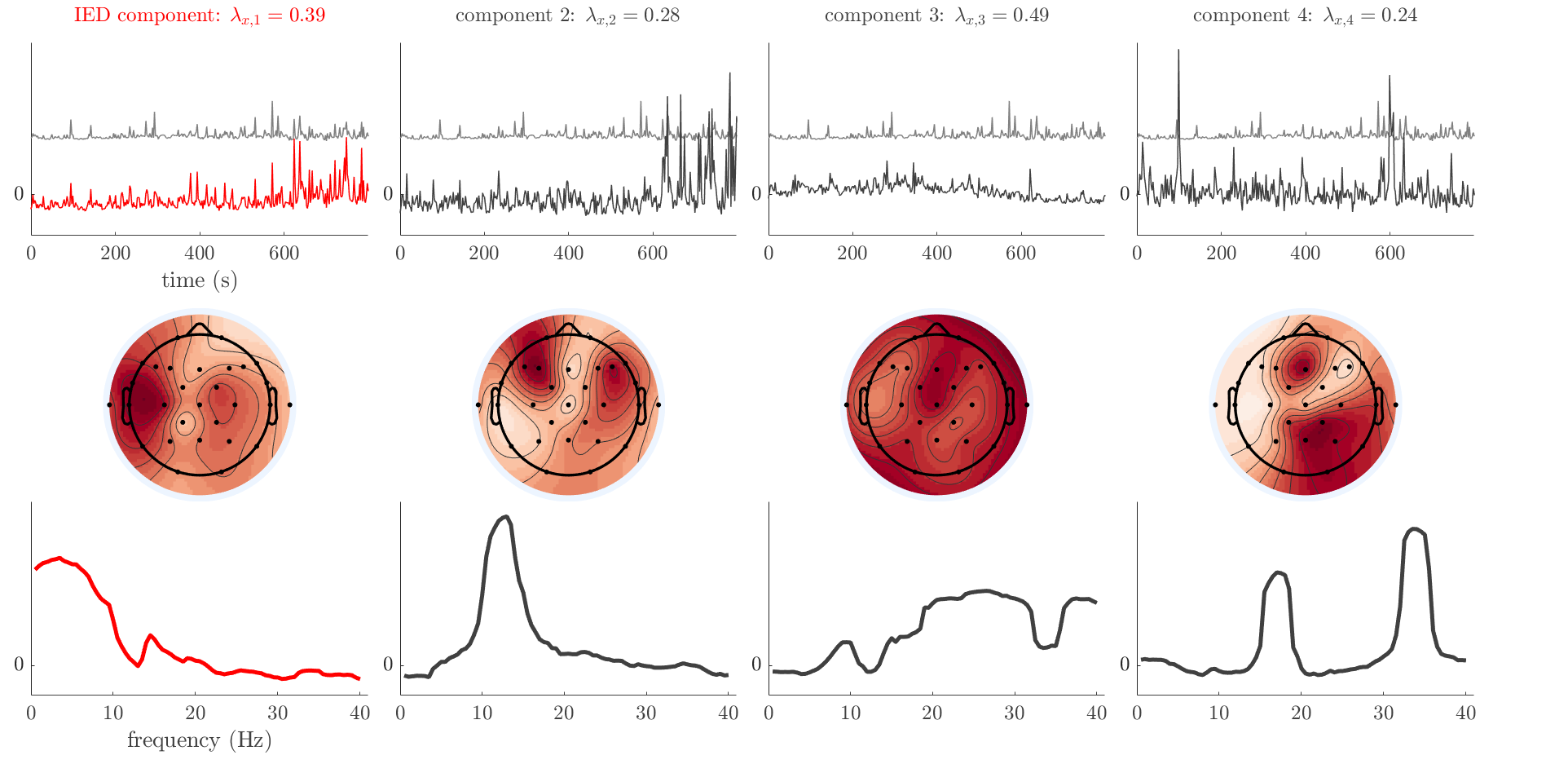}
		\caption{Temporal ($\vect{s}_r$, top), spatial ($\vect{m}_r$, middle), and spectral ($\vect{g}_r$, bottom) profiles of the 4 sources in the EEG domain, and reference IED time course ($\vect{s}_{\text{ref}}$, in grey).}
		\label{fig:p09eeg}
	\end{subfigure}
	
	\begin{subfigure}[b]{0.8\textwidth}
		\includegraphics[width=\textwidth]{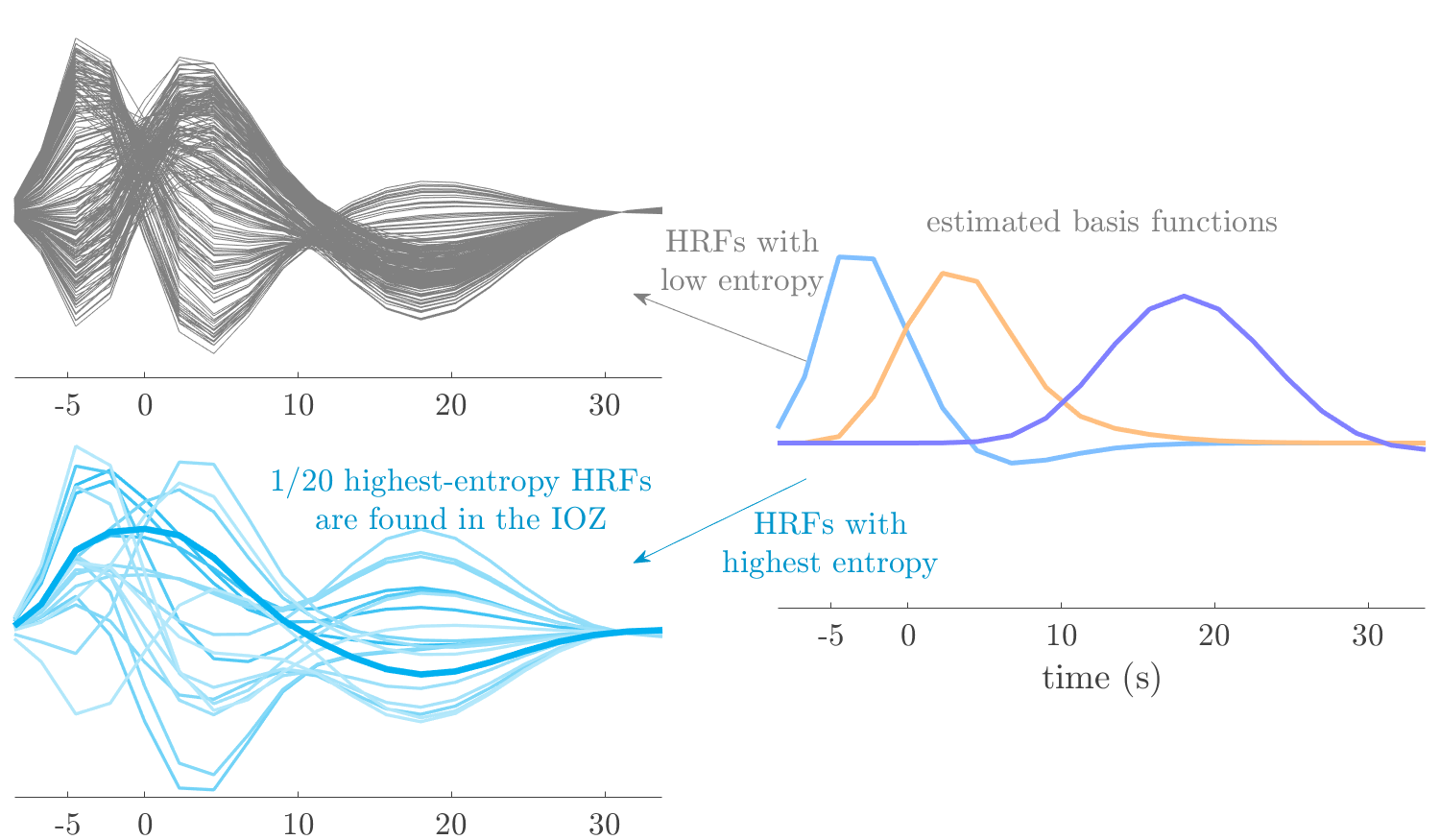}
		\caption{Estimated HRFs with low and high entropy}
		\label{fig:p09hrf}
	\end{subfigure}
	\caption{Patient 7's estimated sources and neurovascular coupling parameters. \subfigcap{b} One of the ROIs with the highest-entropy HRFs belongs to the ictal onset zone (bold line, $p=0.57$).}
	\label{fig:p09eeghrf}
\end{figure*}

\begin{figure*}
	\centering
	\includegraphics[width=1\linewidth]{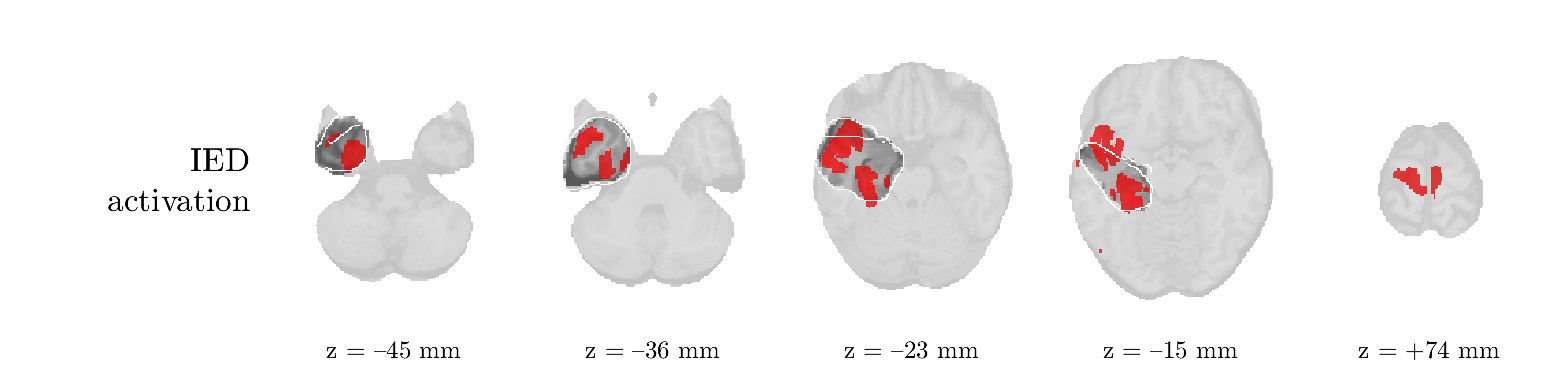}
	\includegraphics[width=1\linewidth]{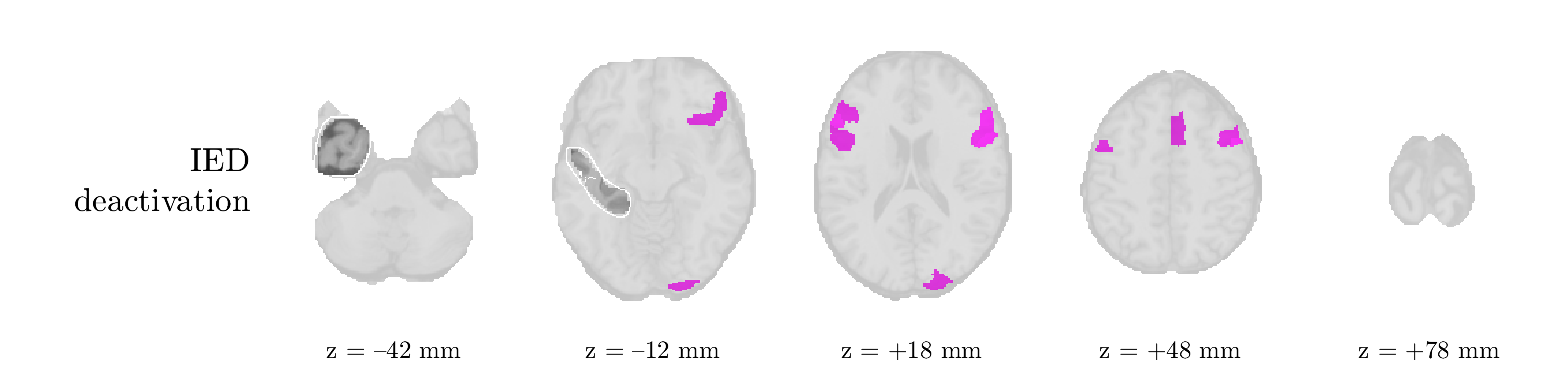}
	\includegraphics[width=1\linewidth]{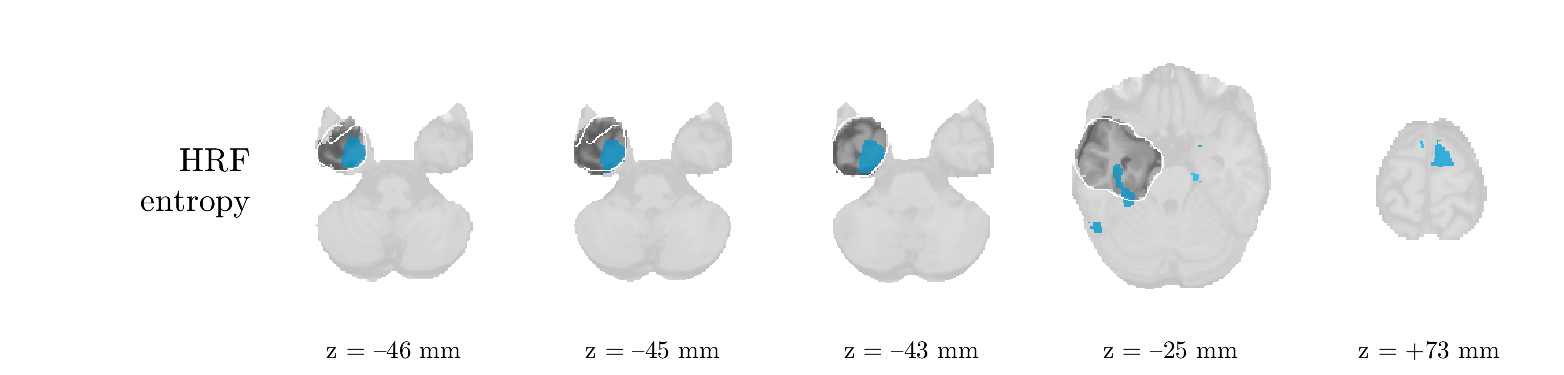}
	\includegraphics[width=1\linewidth]{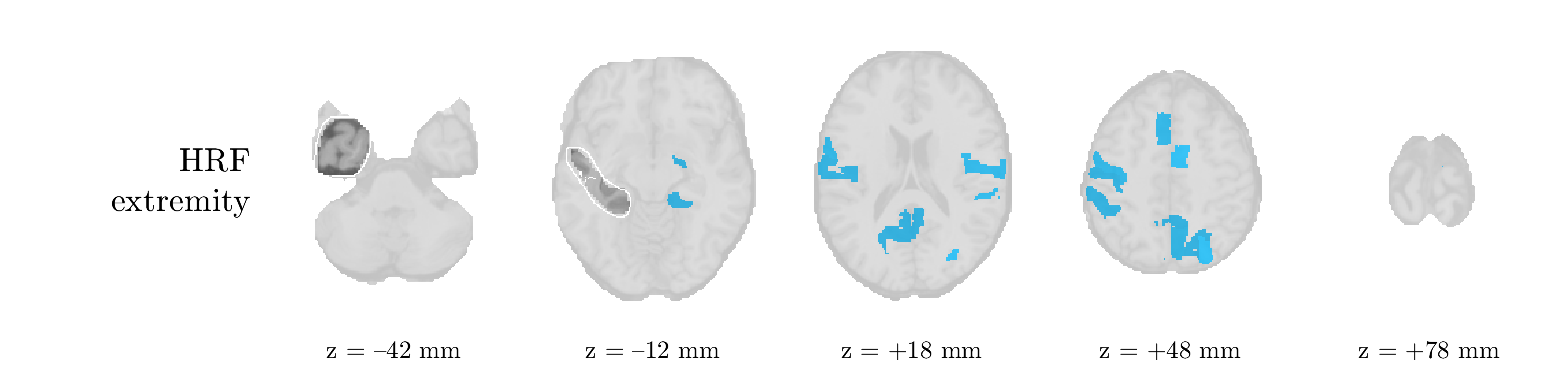}
	\caption{Patient 7's statistical nonparametric maps and HRF entropy/extremity maps. The ground truth ictal onset zone is highlighted in dark gray with a white contour.
	}
	\label{fig:p09fmri}
\end{figure*}

We analyze the solution with $\hat{R}=4$ sources, and show the results in Figure~\ref{fig:p09eeghrf} and \ref{fig:p09fmri}.
We found a clear IED-related component, with a characteristic spectrum and a topography which is backed up by the patient's diagnosis (left anterior-temporal IOZ).
The fourth source has a very similar topography and spectrum to the fifth source in patient 5.
One HRF inside the IOZ had a high-entropy, and is distinguishable from the others by its very sluggish waveform, i.e., it is smeared out in time, with no sharp over- or undershoot.
Also the pseudo t-map provided an accurate localization of the IOZ.
Notably, in this patient, the extremity metric misses the deviating HRF in the IOZ (while the entropy metric picks it up).
The second source overlapped with the frontal part of the default mode network (DMN), and is active in the $\alpha$ and low $\beta$ bands.

\subsection*{Patient 8}
\begin{figure*}
	\centering
	\begin{subfigure}[b]{0.6\textwidth}
		\includegraphics[width=1\linewidth]{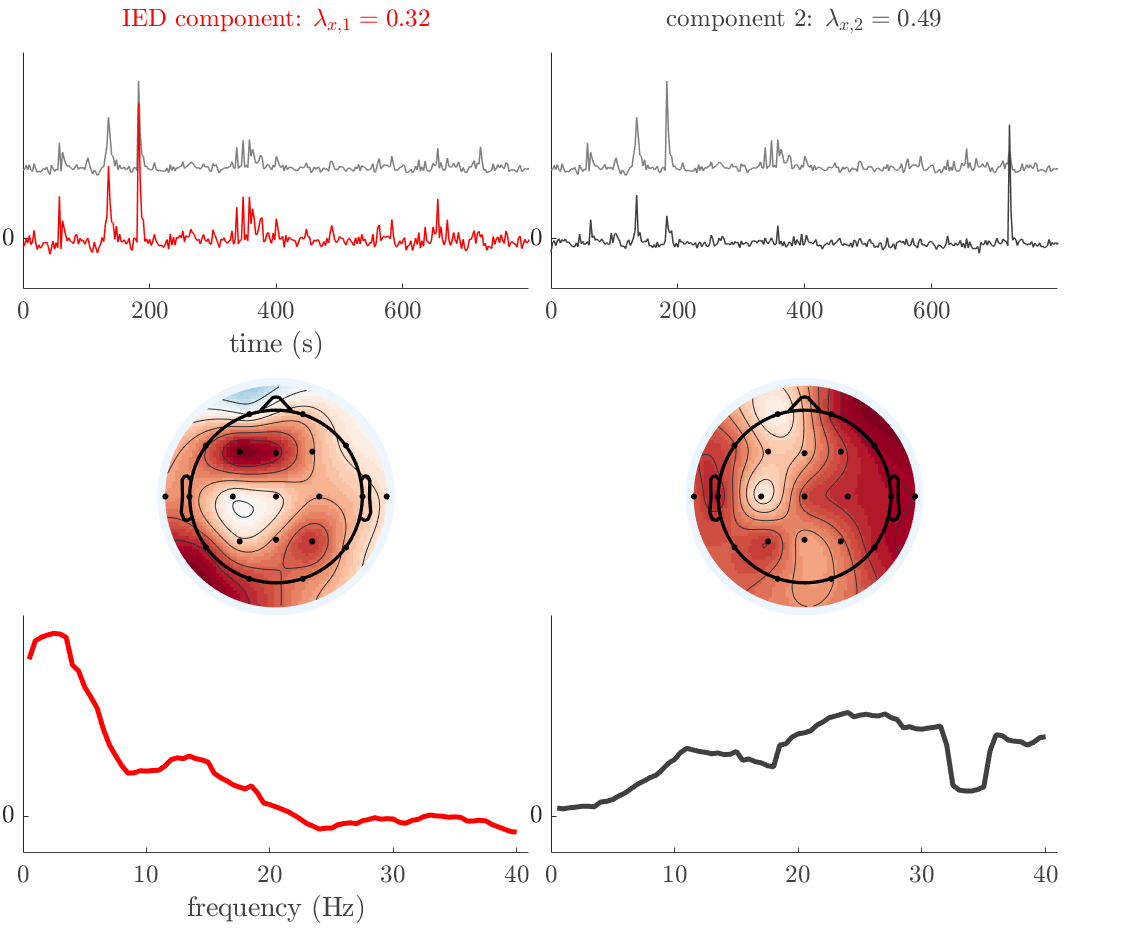}
		\caption{Temporal ($\vect{s}_r$, top), spatial ($\vect{m}_r$, middle), and spectral ($\vect{g}_r$, bottom) profiles of the 2 sources in the EEG domain, and reference IED time course ($\vect{s}_{\text{ref}}$, in grey).}
		\label{fig:p10eeg}
	\end{subfigure}
	
	\begin{subfigure}[b]{0.8\textwidth}
		\includegraphics[width=\textwidth]{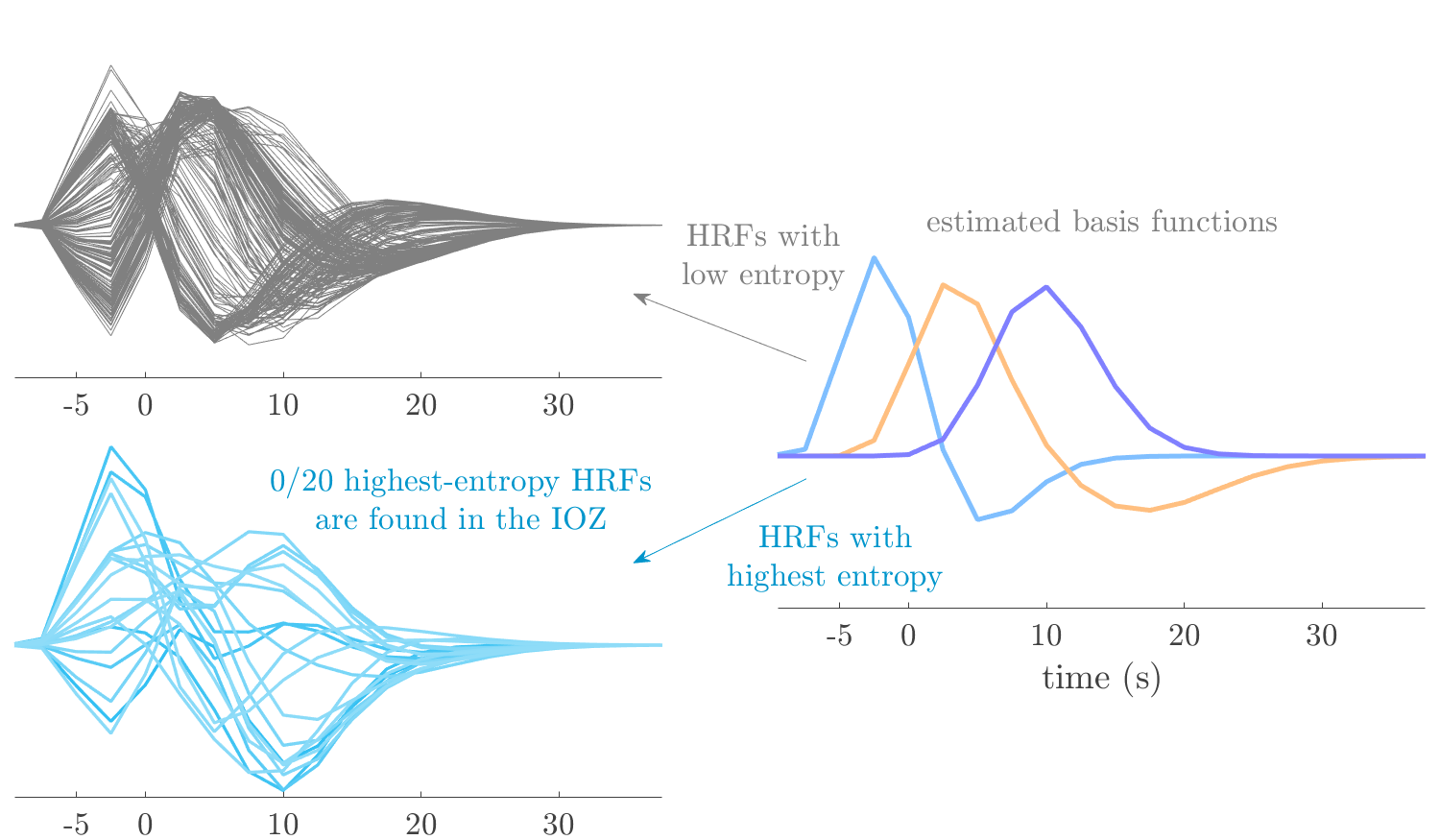}
		\caption{Estimated HRFs with low and high entropy}
		\label{fig:p10hrf}
	\end{subfigure}
	\caption{Patient 8's estimated sources and neurovascular coupling parameters. \subfigcap{b} None of the ROIs with the highest-entropy HRFs belong to the ictal onset zone.}
	\label{fig:p10eeghrf}
\end{figure*}

\begin{figure*}
	\centering
	\includegraphics[width=1\linewidth]{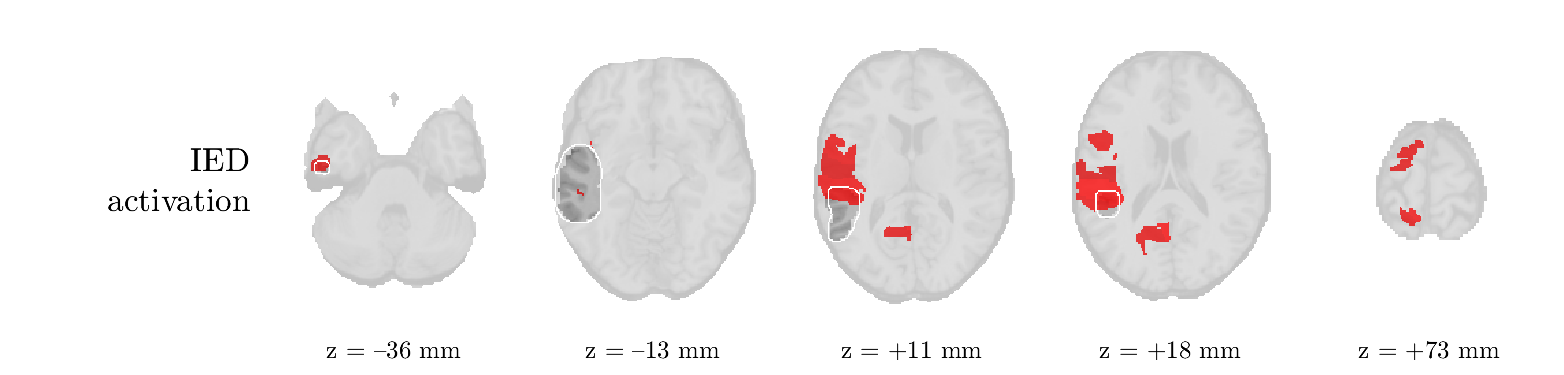}
	\includegraphics[width=1\linewidth]{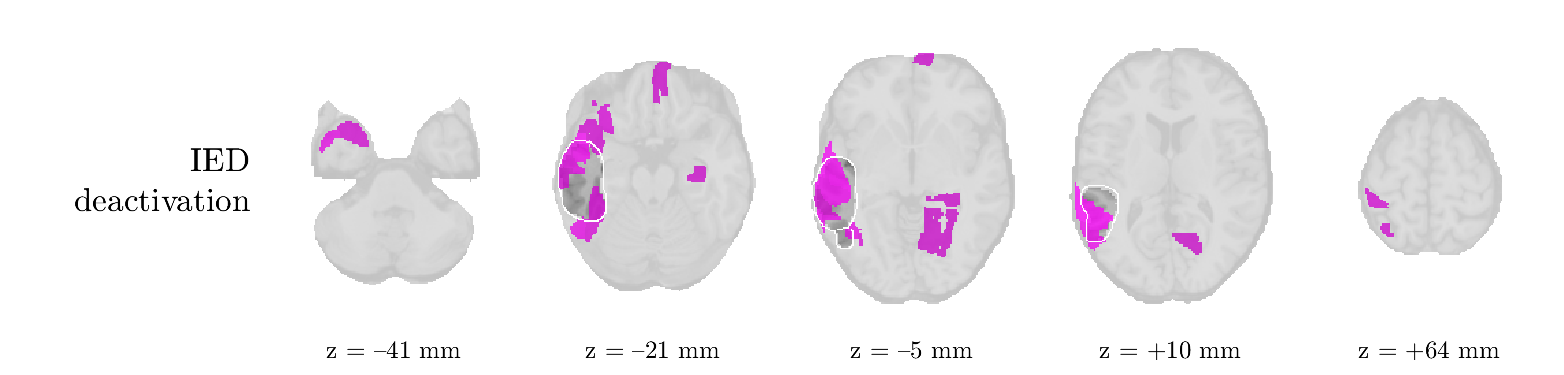}
	\includegraphics[width=1\linewidth]{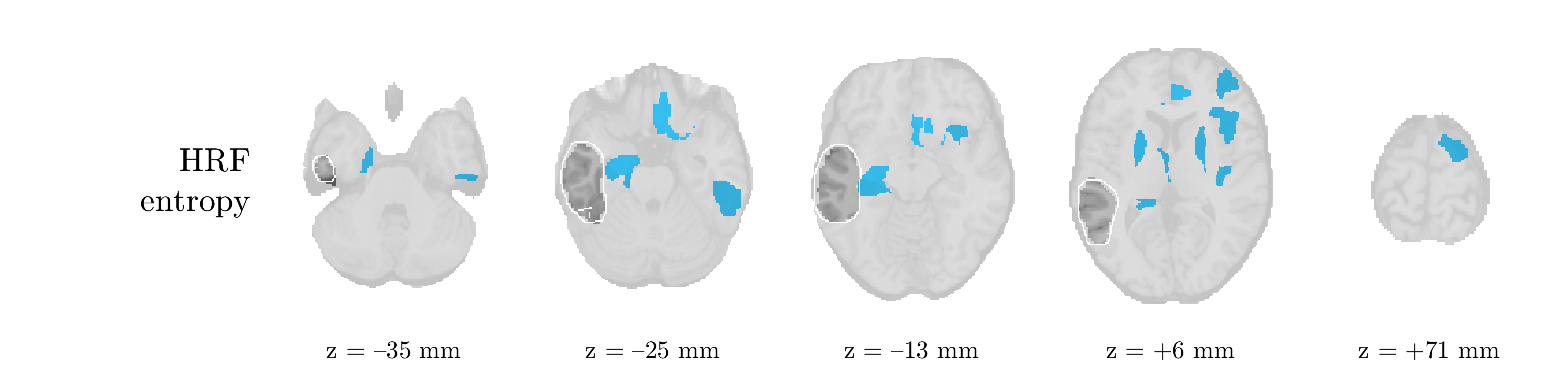}
	\includegraphics[width=1\linewidth]{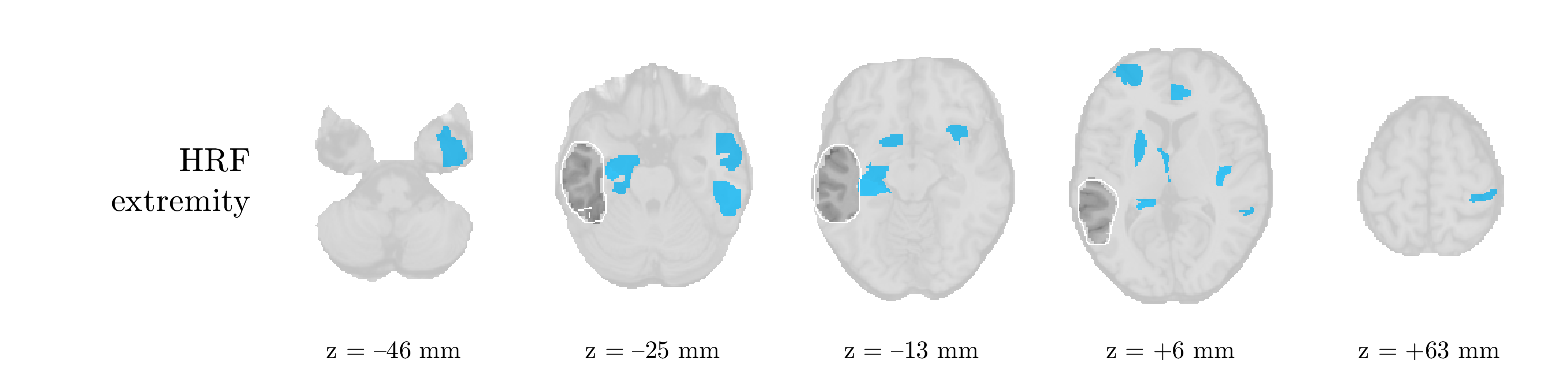}
	\caption{Patient 8's statistical nonparametric maps and HRF entropy/extremity maps. The ground truth ictal onset zone is highlighted in dark gray with a white contour.}
	\label{fig:p10fmri}
\end{figure*}

We analyze the solution with $\hat{R}=2$ sources, and show the results in Figure~\ref{fig:p10eeghrf} and \ref{fig:p10fmri}.
We found two components which had correlated time courses.
At the time of the peaks, we found higher-amplitude events in the EEG with dubious origin, hence they may or may not be artifacts.
One of both components is more strongly correlated to the MWF, and its activation is concordant with the IOZ.
The second component shows high overlap with the sensorimotor network.
For this patient, none of the IOZ's ROIs had extreme values of either HRF metric.

\subsection*{Patient 9}
\begin{figure*}
	\centering
	\begin{subfigure}[b]{0.6\textwidth}
		\includegraphics[width=1\linewidth]{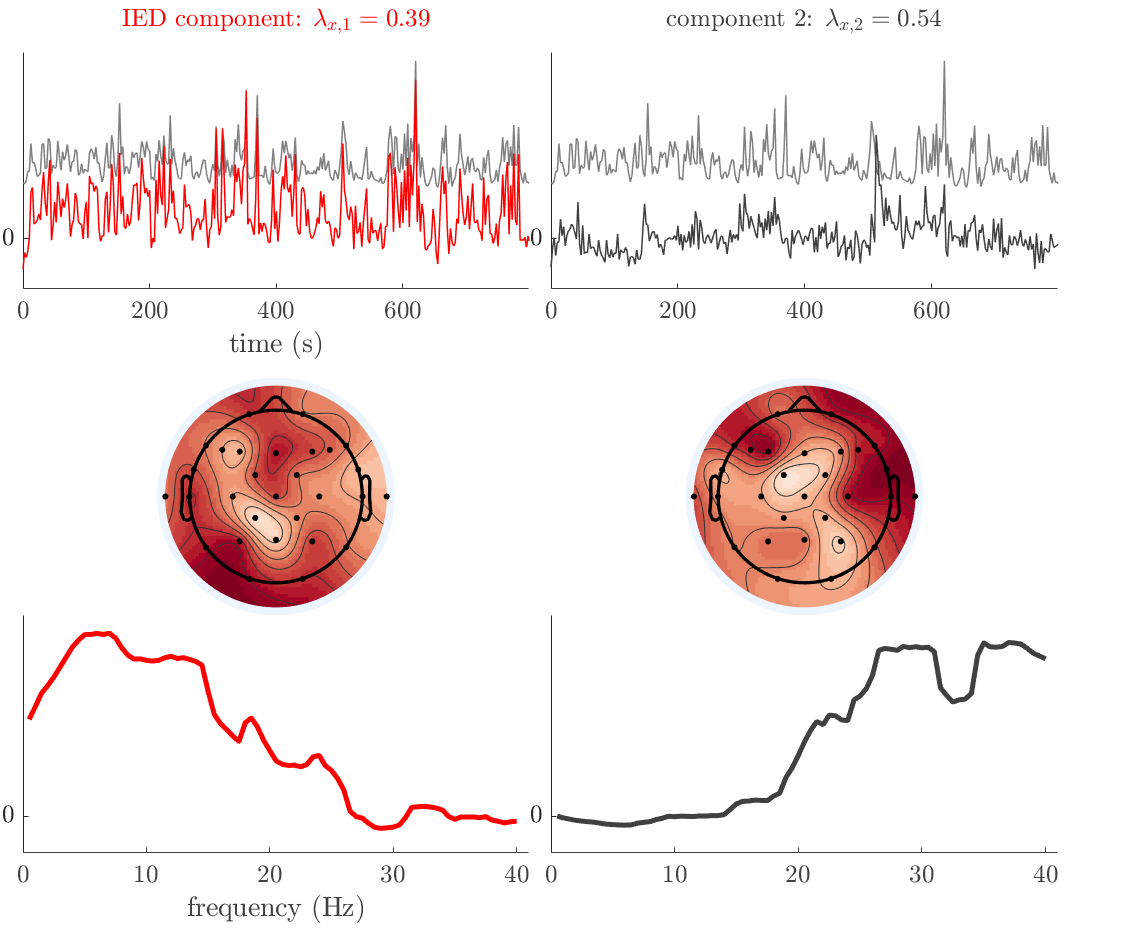}
		\caption{Temporal ($\vect{s}_r$, top), spatial ($\vect{m}_r$, middle), and spectral ($\vect{g}_r$, bottom) profiles of the 2 sources in the EEG domain, and reference IED time course ($\vect{s}_{\text{ref}}$, in grey).}
		\label{fig:p11eeg}
	\end{subfigure}
	
	\begin{subfigure}[b]{0.8\textwidth}
		\includegraphics[width=\textwidth]{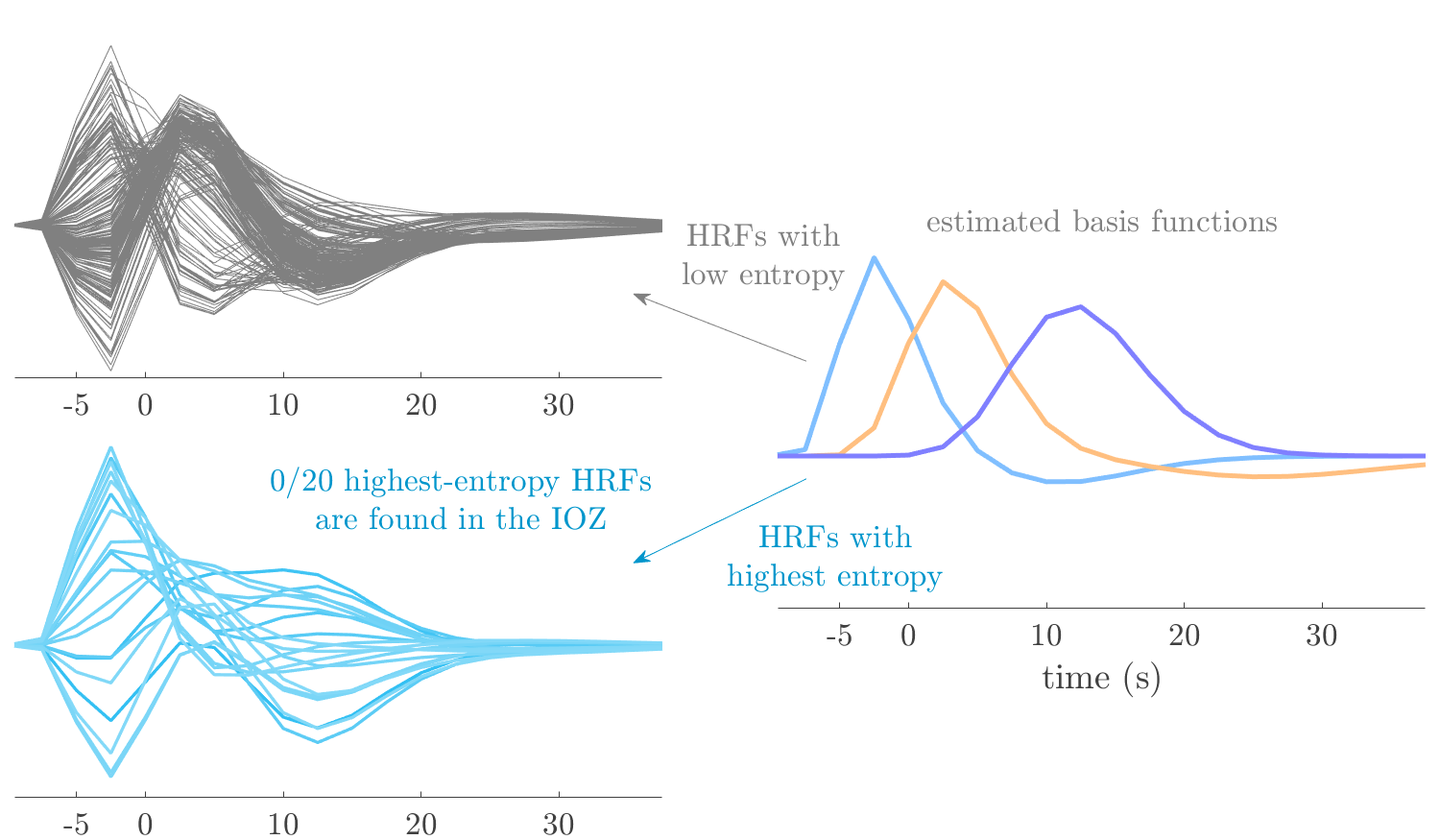}
		\caption{Estimated HRFs with low and high entropy}
		\label{fig:p11hrf}
	\end{subfigure}
	\caption{Patient 9's estimated sources and neurovascular coupling parameters. \subfigcap{b} None of the ROIs with the highest-entropy HRFs belong to the ictal onset zone.}
	\label{fig:p11eeghrf}
\end{figure*}

\begin{figure*}
	\centering
	\includegraphics[width=1\linewidth]{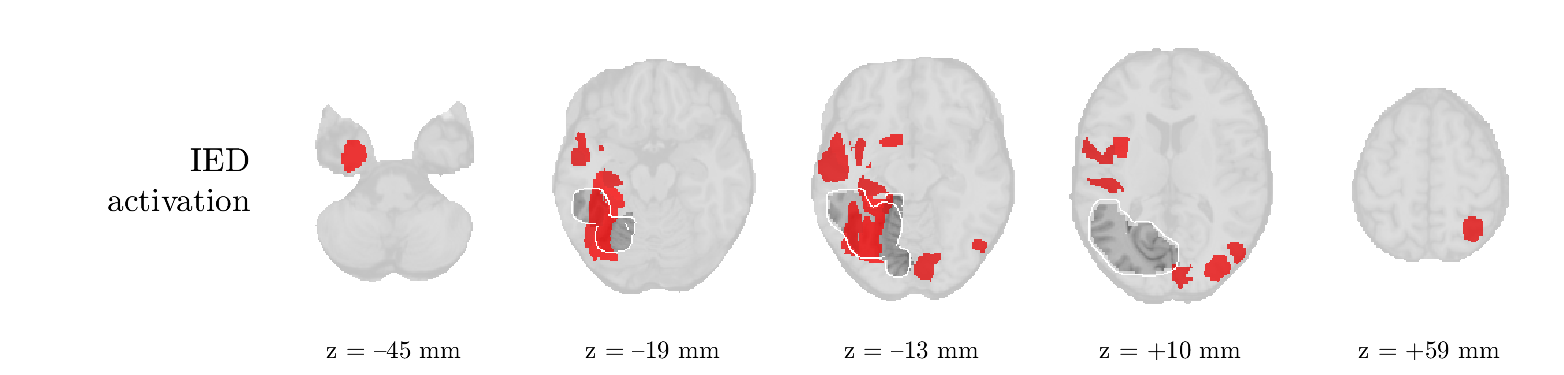}
	\includegraphics[width=1\linewidth]{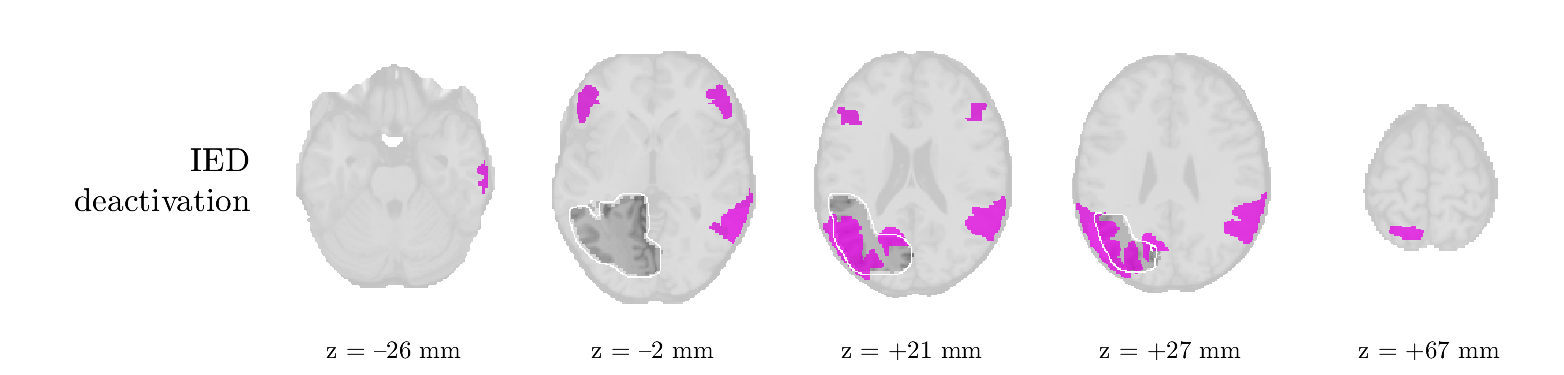}
	\includegraphics[width=1\linewidth]{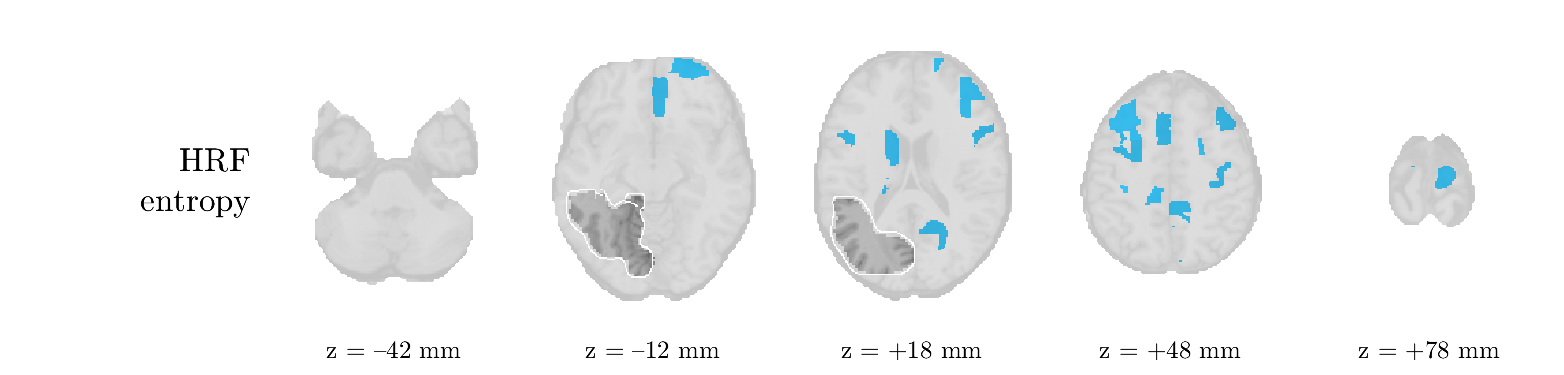}
	\includegraphics[width=1\linewidth]{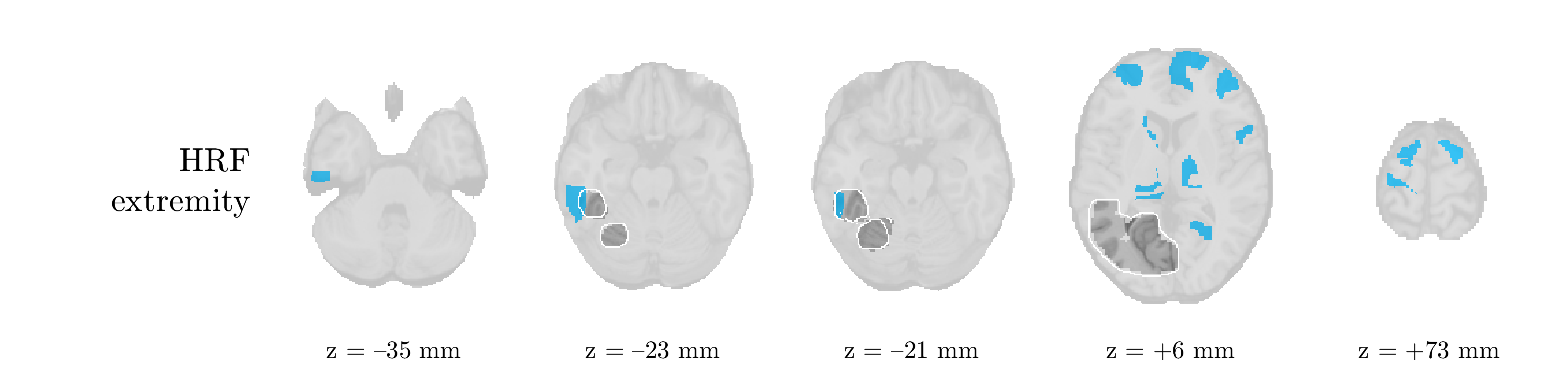}
	\caption{Patient 9's statistical nonparametric maps and HRF entropy/extremity maps. The ground truth ictal onset zone is highlighted in dark gray with a white contour.}
	\label{fig:p11fmri}
\end{figure*}

We analyze the solution with $\hat{R}=2$ sources, and show the results in Figure~\ref{fig:p11eeghrf} and \ref{fig:p11fmri}.
In this patient, there is only a moderate correlation of a component with the MWF reference time course.
This component's topography (left occipital) agrees with the clinical description, however. 
The HRF extremity (and not the entropy) is high in a small part of the IOZ.
Both the significant IED activation and deactivation allow correct localization as well.
The second source seemingly captured high-frequency oscillatory activity in the sensorimotor network, similar to the previous patient.

\subsection*{Patient 11}
\begin{figure*}
	\centering
	\begin{subfigure}[b]{0.6\textwidth}
		\includegraphics[width=1\linewidth]{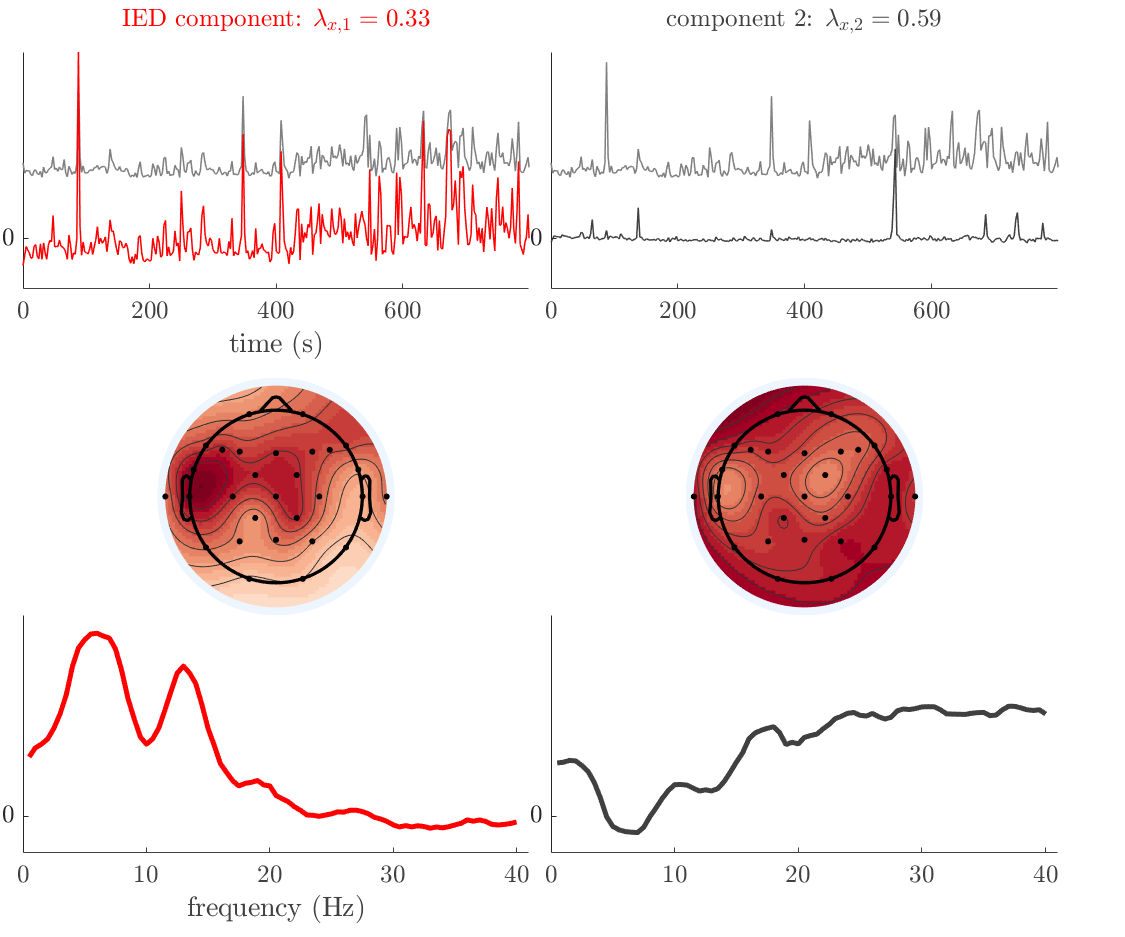}
		\caption{Temporal ($\vect{s}_r$, top), spatial ($\vect{m}_r$, middle), and spectral ($\vect{g}_r$, bottom) profiles of the 2 sources in the EEG domain, and reference IED time course ($\vect{s}_{\text{ref}}$, in grey).}
		\label{fig:p13eeg}
	\end{subfigure}
	
	\begin{subfigure}[b]{0.8\textwidth}
		\includegraphics[width=\textwidth]{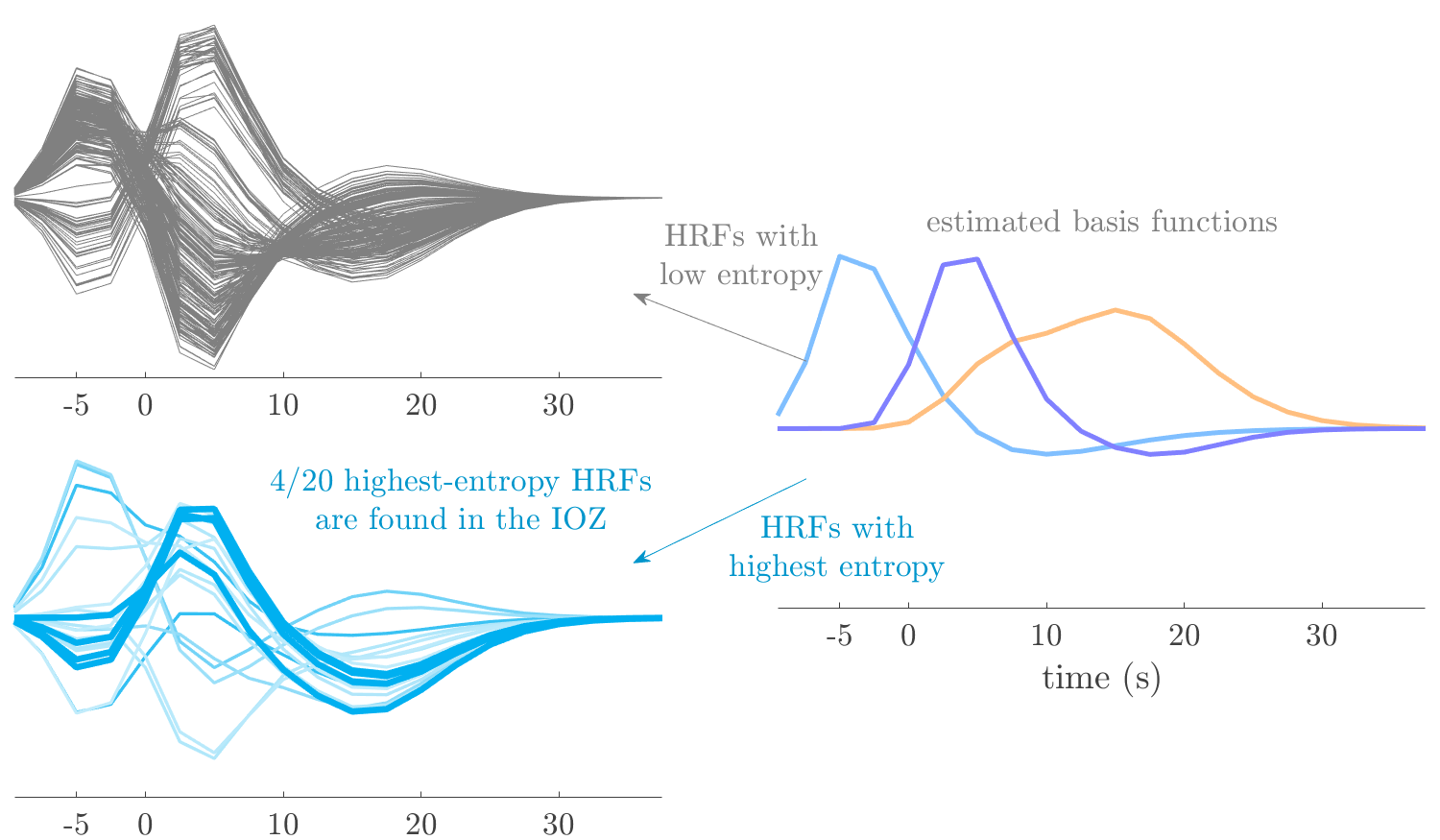}
		\caption{Estimated HRFs with low and high entropy}
		\label{fig:p13hrf}
	\end{subfigure}
	\caption{Patient 11's estimated sources and neurovascular coupling parameters. \subfigcap{b} Four of the ROIs with the highest-entropy HRFs belong to the ictal onset zone (bold line, $p=0.01$).}
	\label{fig:p13eeghrf}
\end{figure*}

\begin{figure*}
	\centering
	\includegraphics[width=1\linewidth]{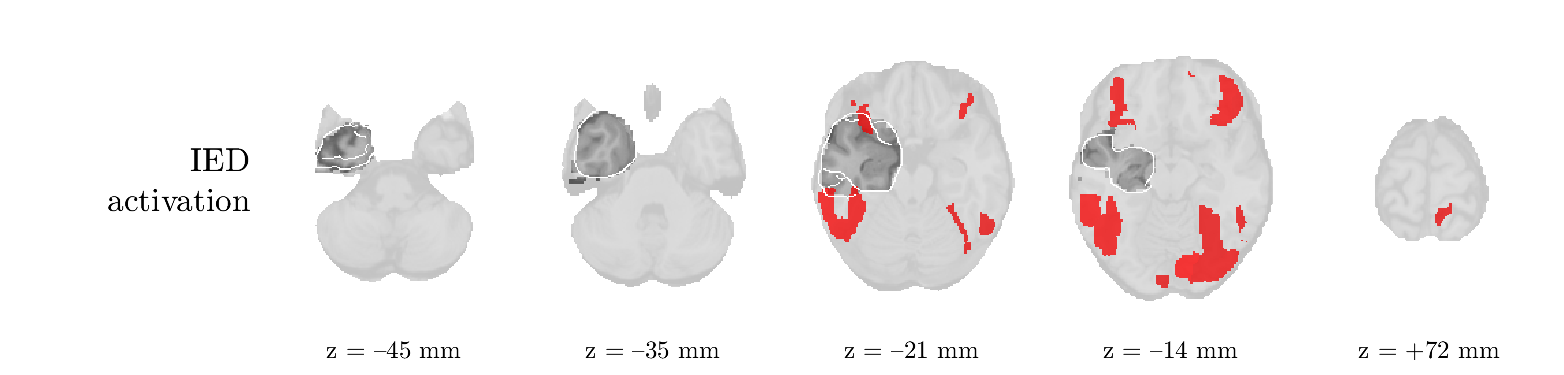}
	\includegraphics[width=1\linewidth]{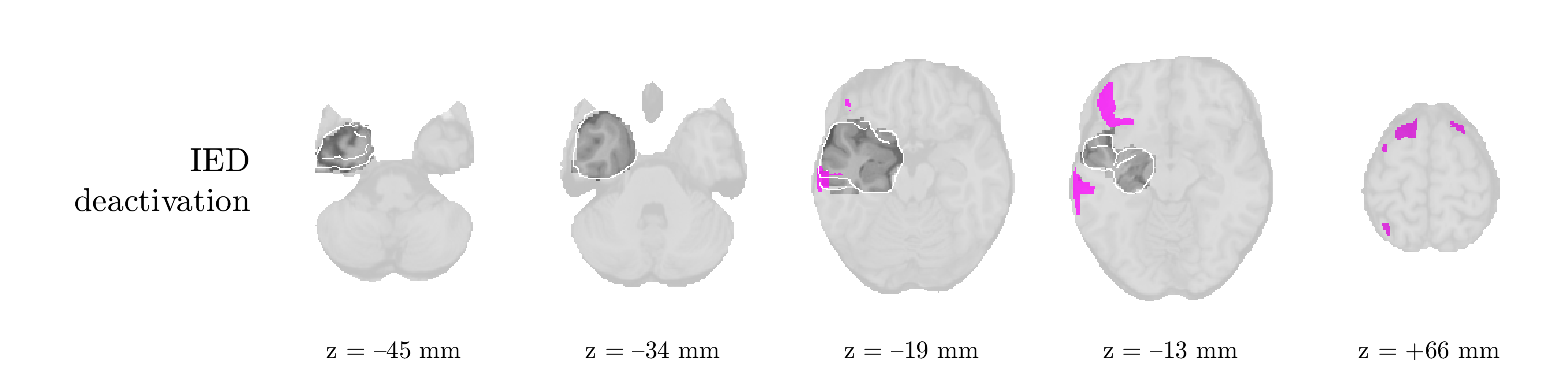}
	\includegraphics[width=1\linewidth]{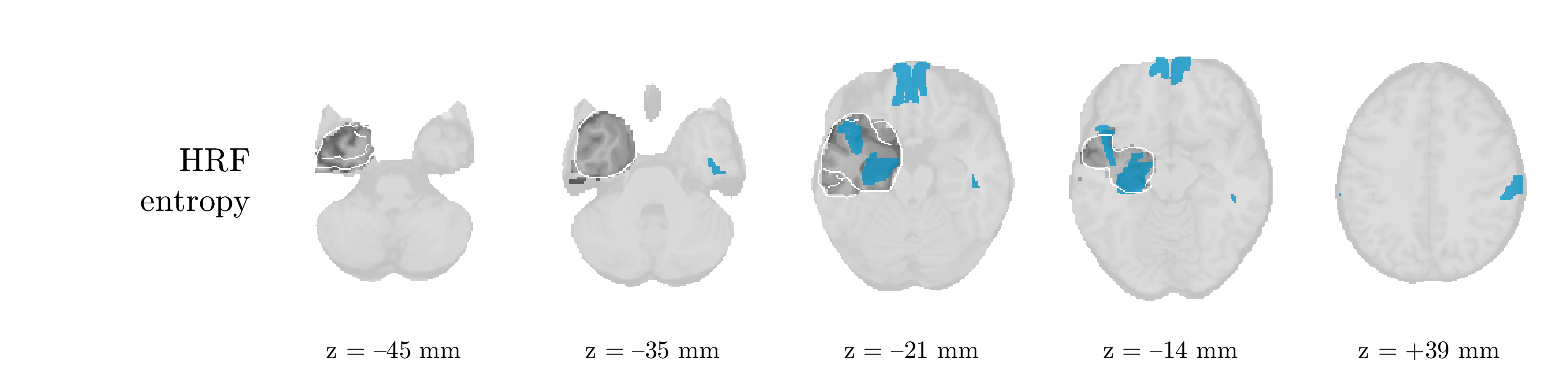}
	\includegraphics[width=1\linewidth]{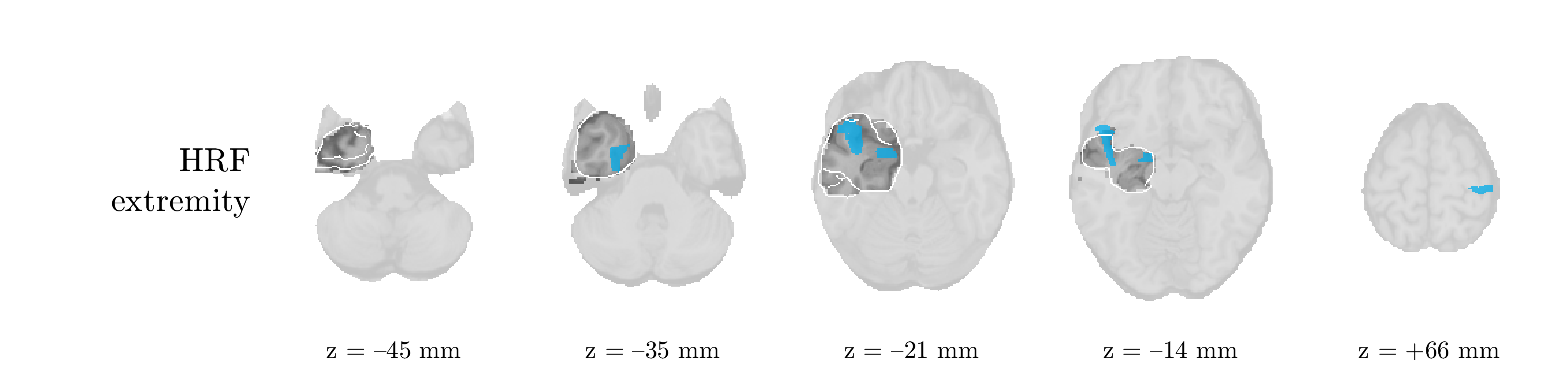}
	\caption{Patient 11's statistical nonparametric maps and HRF entropy/extremity maps. The ground truth ictal onset zone is highlighted in dark gray with a white contour.}
	\label{fig:p13fmri}
\end{figure*}

We analyze the solution with $\hat{R}=2$ sources, and show the results in Figure~\ref{fig:p13eeghrf} and \ref{fig:p13fmri}.
The IED-related source had a high correlation with the MWF reference, but an odd bimodal spectrum. 
Its EEG topography is very consistent with the clinical description.
Both HRF extremity and entropy are useful biomarkers for the IOZ.
The IED activation and deactivation maps each had a very small overlap with the IOZ.
The second source is temporally sparse and captures high-frequency EEG variations, which we identified as muscle artifacts.

\subsection*{Patient 12}
\begin{figure*}
	\centering
	\begin{subfigure}[b]{0.6\textwidth}
		\includegraphics[width=1\linewidth]{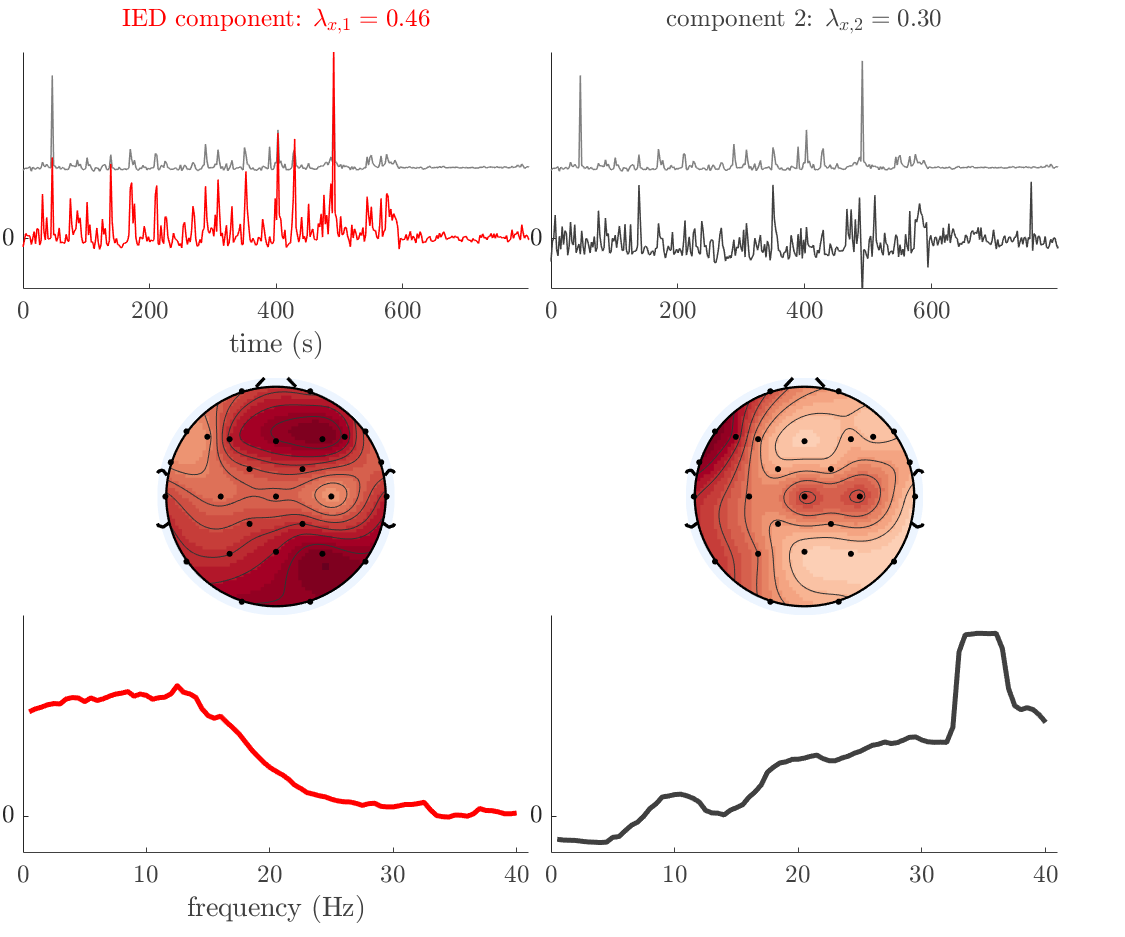}
		\caption{Temporal ($\vect{s}_r$, top), spatial ($\vect{m}_r$, middle), and spectral ($\vect{g}_r$, bottom) profiles of the 3 sources in the EEG domain, and reference IED time course ($\vect{s}_{\text{ref}}$, in grey).}
		\label{fig:p14eeg}
	\end{subfigure}
	
	\begin{subfigure}[b]{0.8\textwidth}
		\includegraphics[width=\textwidth]{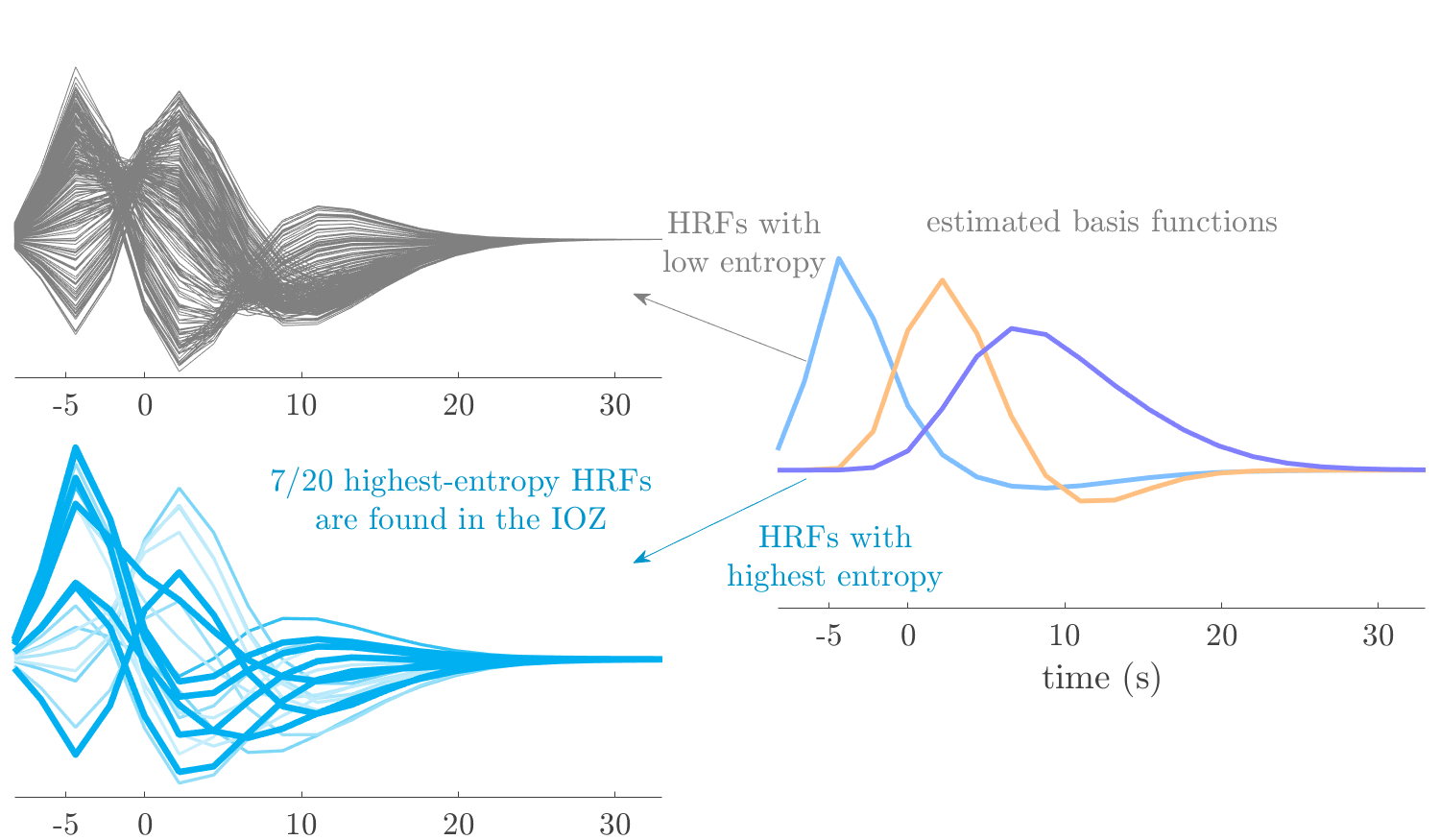}
		\caption{Estimated HRFs with low and high entropy}
		\label{fig:p14hrf}
	\end{subfigure}
	\caption{Patient 12's estimated sources and neurovascular coupling parameters. \subfigcap{b} Seven of the ROIs with the highest-entropy HRFs belong to the ictal onset zone (bold line, $p<10^{-3}$).}
	\label{fig:p14eeghrf}
\end{figure*}

\begin{figure*}
	\centering
	\includegraphics[width=1\linewidth]{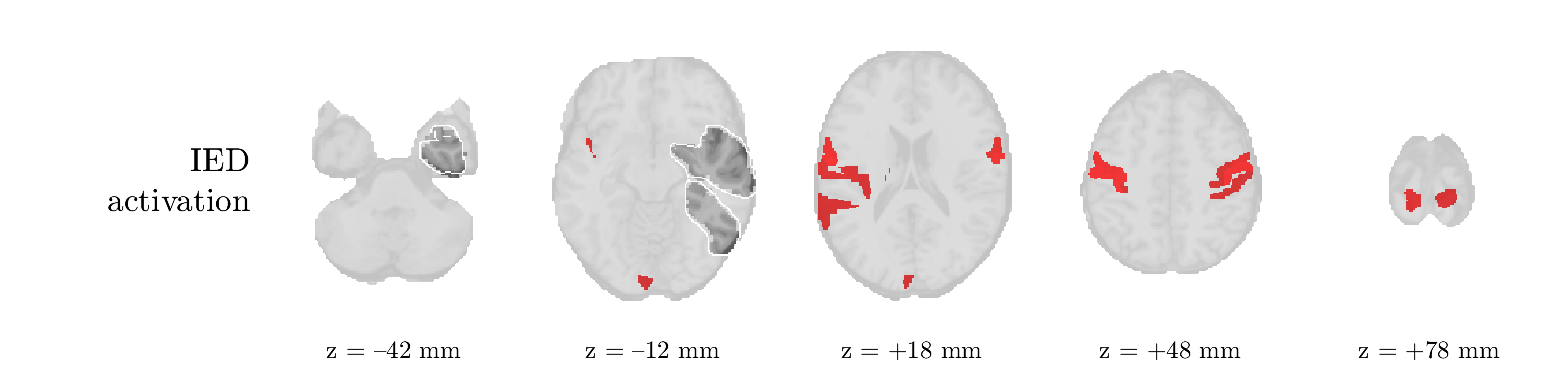}
	\includegraphics[width=1\linewidth]{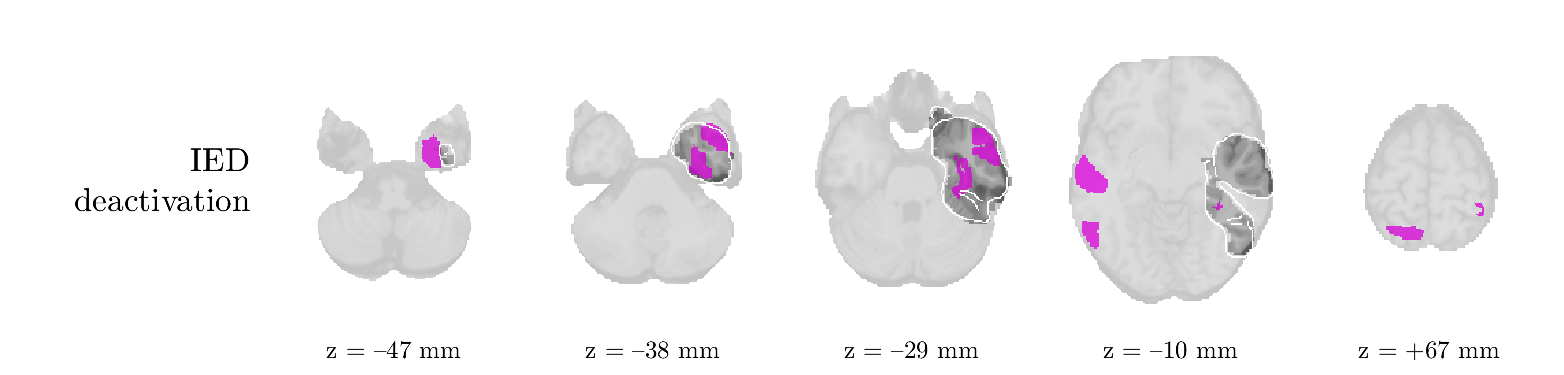}
	\includegraphics[width=1\linewidth]{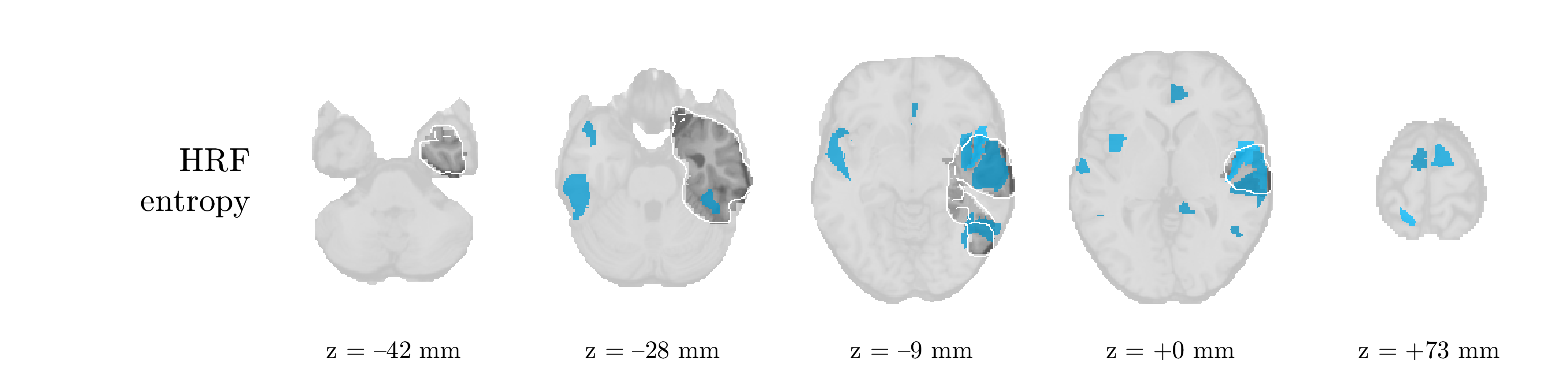}
	\includegraphics[width=1\linewidth]{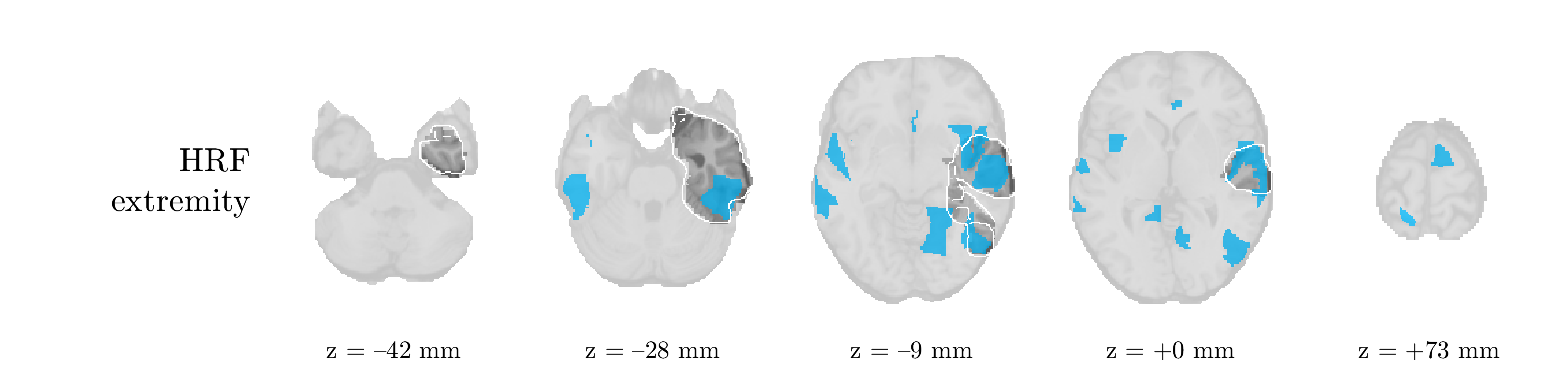}
	\caption{Patient 12's statistical nonparametric maps and HRF entropy/extremity maps. The ground truth ictal onset zone is highlighted in dark gray with a white contour.}
	\label{fig:p14fmri}
\end{figure*}

We analyze the solution with $\hat{R}=2$ sources, and show the results in Figure~\ref{fig:p14eeghrf} and \ref{fig:p14fmri}.
Again we observe an IED-related source and a seemingly artifactual source with a spectral peak near 34 Hz.
Many of the high-entropy HRFs are highly noncausal, and are associated to ROIs inside the IOZ.
Hence, with both HRF metrics, the highest-scoring ROIs provides good localization of the HRF.
While there are no significantly active ROIs in the IOZ, there are several significantly deactivated ROIs, which may indicate that the sign standardization was not done flawlessly (cfr. also some of the negative-peaking HRFs for patient 10).
Surprisingly, the second source had one significantly active ROI, which overlaps with the IOZ, but which did not match its EEG topography. 
Hence, the nature of this source remains ambiguous.

